\documentclass[12pt]{article}
\usepackage{epsfig}
\usepackage{axodraw}
\usepackage{color}
\usepackage{a4}
\usepackage{latexsym}
\usepackage{cite}

\usepackage{pslatex}
\usepackage[latin1]{inputenc}
\usepackage[T1]{fontenc}

\usepackage{color,colordvi}
\def\colour4colour#1{\Blue{#1}}

\renewcommand{\theequation}{\thesection.\arabic{equation}}

\newcommand{\equal}{\:\: = \:\:}

\newcommand{\hspn}{{\hspace{-4mm}}}
\newcommand{\hspp}{{\hspace{4mm}}}
\newcommand{\beq}{\begin{equation}}
\newcommand{\eeq}{\end{equation}}
\newcommand{\bea}{\begin{eqnarray}}
\newcommand{\eea}{\end{eqnarray}}
\newcommand{\nn}{\nonumber}
\newcommand{\MSb}{$\overline{\mbox{MS}}$}
\newcommand{\ra}{\rightarrow}
\newcommand{\DD}{{\cal D}}

\newcommand{\als}{\alpha_{\rm s}}
\newcommand{\ars}{a_{\rm s}}
\newcommand{\muS}{\mu^{\,2}}
\newcommand{\ep}{\varepsilon}

\begin{document}
\setlength{\parskip}{0.25cm}
\setlength{\baselineskip}{0.53cm}

\def\z#1{{\zeta_{#1}}}
\def\zss{\zeta_2^{\,2}}

\def\ca{{C^{}_A}}
\def\cas{{C^{\,2}_A}}
\def\cat{{C^{\,3}_A}}
\def\cf{{C^{}_F}}
\def\cfs{{C^{\, 2}_F}}
\def\cft{{C^{\, 3}_F}}
\def\nf{{n^{}_{\! f}}}
\def\nfs{{n^{\,2}_{\! f}}}

\def\DNnO{D_0}
\def\DNmO{D_{-1}}
\def\DNpO{D_{1}}
\def\DNppO{D_{2}}
\def\DNn#1{D_0^{\:#1}}
\def\DNm#1{D_{-1}^{\:#1}}
\def\DNp#1{D_1^{\:#1}}
\def\DNpp#1{D_2^{\:#1}}

\def\etaD#1{\eta^{\,#1}}
\def\nuD#1{\nu^{\,#1}}

\def\as(#1){{\alpha_{\rm s}^{\:#1}}}
\def\ar(#1){{a_{\rm s}^{\:#1}}}

\def\x1{{(1 \! - \! x)}}
\def\LntO{\ln(1\!-\!x)}
\def\Lnt(#1){\ln^{\,#1}(1\!-\!x)}

\def\muRs{{\mu_R^{\,2}}}
\def\Qs{{Q^{\, 2}}}

\def\dabc2n{{{d^{abc}d_{abc}}\over{n_c}}}
\def\S(#1){{{S}_{#1}}}
\def\Ss(#1,#2){{{S}_{#1,#2}}}
\def\Sss(#1,#2,#3){{{S}_{#1,#2,#3}}}
\def\Ssss(#1,#2,#3,#4){{{S}_{#1,#2,#3,#4}}}
\def\Sssss(#1,#2,#3,#4,#5){{{S}_{#1,#2,#3,#4,#5}}}

\def\frct#1#2{\mbox{\large{$\frac{#1}{#2}$}}}

\def\H(#1){{\rm{H}}_{#1}}
\def\Hh(#1,#2){{\rm{H}}_{#1,#2}}
\def\Hhh(#1,#2,#3){{\rm{H}}_{#1,#2,#3}}
\def\Hhhh(#1,#2,#3,#4){{\rm{H}}_{#1,#2,#3,#4}}

\def\dpqq{\Delta p_{\rm qq}}
\def\dpgg{\Delta p_{\rm gg}}

\def\dpqg{\Delta p_{\rm qg}}
\def\dpgq{\Delta p_{\rm gq}}

\def\dPqqL(#1){\Delta P_{{\rm qq},\,L}^{\,(#1)}}
\def\dPqgL(#1){\Delta P_{{\rm qg},\,L}^{\,(#1)}}
\def\dPgqL(#1){\Delta P_{{\rm gq},\,L}^{\,(#1)}}
\def\dPggL(#1){\Delta P_{{\rm gg},\,L}^{\,(#1)}}

\def\dPnsp(#1){\Delta P_{\rm ns}^{\,+(#1)}}
\def\dPnsm(#1){\Delta P_{\rm ns}^{\,-(#1)}}
\def\dPps(#1){\Delta P_{\rm ps}^{\,(#1)}}
\def\dPqq(#1){\Delta P_{\rm qq}^{\,(#1)}}
\def\dPqg(#1){\Delta P_{\rm qg}^{\,(#1)}}
\def\dPgq(#1){\Delta P_{\rm gq}^{\,(#1)}}
\def\dPgg(#1){\Delta P_{\rm gg}^{\,(#1)}}

\def\Pnsm(#1){P_{\rm ns}^{\,-(#1)}}

\def\Zqq(#1){z_{\rm qq}^{\,(#1)}}
\def\Zqg(#1){z_{\rm qg}^{\,(#1)}}
\def\Zgq(#1){z_{\rm gq}^{\,(#1)}}
\def\Zgg(#1){z_{\rm gg}^{\,(#1)}}

\def\P{\overline{P}}
\def\GG{\overline{G}}

\begin{titlepage}
\noindent
DESY 14-157 \hfill September 2014\\
NIKHEF 14-033 \\ 
LTH 1023 \\
\vspace{1.2cm}
\begin{center}
\Large
{\bf The Three-Loop Splitting Functions in QCD:} \\
\vspace{0.15cm}
{\bf The Helicity-Dependent Case} \\
\vspace{1.6cm}
\large
S. Moch$^{\, a}$, J.A.M. Vermaseren$^{\, b}$ and A. Vogt$^{\, c}$\\
\vspace{1.2cm}
\normalsize
{\it $^b$II.~Institute for Theoretical Physics, Hamburg University\\
\vspace{0.1cm}
D-22761 Hamburg, Germany}\\
\vspace{0.5cm}
{\it $^b$Nikhef Theory Group \\
\vspace{0.1cm}
Science Park 105, 1098 XG Amsterdam, The Netherlands} \\
\vspace{0.5cm}
{\it $^c$Department of Mathematical Sciences, University of Liverpool\\
\vspace{0.1cm}
Liverpool L69 3BX, United Kingdom}\\
\vspace{2.6cm}
\large
{\bf Abstract}
\vspace{-0.2cm}
\end{center}
We present the next-to-next-to-leading order $\:$(NNLO)$\:$ contributions to 
the main splitting functions for the evolution of longitudinally polarized 
parton densities of hadrons in perturbative~QCD. 
The quark-quark and gluon-quark splitting functions have been obtained by
extending our previous all Mellin-$N$ calculations to the structure function 
$g_1^{}$ in electromagnetic deep-inelastic scattering (DIS).
Their quark-gluon and gluon-gluon counterparts have been derived using third-%
order \mbox{fixed-$N$} calculations of structure functions in graviton-exchange
DIS, relations to the unpolarized case and mathematical tools for systems of
Diophantine equations.
The NNLO corrections to the splitting functions are small outside the region
of small momentum fractions $x$ where they exhibit a large double-logarithmic 
enhancement, yet the corrections to the evolution of the parton densities can 
be unproblematic down to at least $x \approx 10^{\,-4}$.
\vfill
\end{titlepage}
%
%
\section{Introduction}
\label{sec:intro}
%
 
The splitting functions for the scale dependence (evolution) of parton 
densities \cite{Altarelli:1977zs,Curci:1980uw,Furmanski:1980cm},
or anomalous dimensions of twist-2 operators \cite{Gross:1973rr,Georgi:1974sr,%
Sasaki:1975hk,Ahmed:1976ee,Floratos:1977au,Floratos:1979ny,Hamberg:1992qt}
in the light-cone operator-product expansion (OPE)~\cite{Christ:1972ms}, are 
important universal (process independent) quantities in perturbative QCD.
A~little more than ten years ago, we completed the calculation of the 
third-order (next-to-next-to-leading order, NNLO) corrections 
$P_{ik}^{\,(2)\!}$, $i,\,k =\rm q, g\,$ for the helicity-averaged (unpolarized) 
case \cite{mvvPns,mvvPsg}. 

These calculations were performed in the approach of Ref.~\cite
{Mom3loop1,Mom3loop2} where physical quantities, specifically structure
functions in inclusive deep-inelastic scattering (DIS), are calculated via
forward amplitudes in dimensional regularization \cite
{'tHooft:1972fi,Bollini:1972ui,Ashmore:1972uj,Cicuta:1972jf}.
In order to access also the lower row of the NNLO flavour-singlet 
splitting-function matrix, i.e., $P_{\rm gq}^{\,(2)}$ and $P_{\rm gg}^{\,(2)}$,
in a third-order calculation, this procedure requires the inclusion of a 
process other than standard gauge-boson exchange DIS.
The method of choice, cf.~Ref.~\cite{FP82}, was to include DIS via a scalar
$\phi$ coupling directly only to gluons via 
$\phi \, G_{a}^{\,\mu\nu} \, G_{a,\,\mu\nu}$,
where $G_{a}^{\,\mu\nu}$ is the gluon field strength tensor, as realized in 
the Standard Model by the Higgs boson in the limit of a heavy top quark and 
five massless flavours \cite{HGGeff1,HGGeff2}.

A corresponding calculation was performed six years ago for the structure
function $g_1^{}$ in polarized photon-exchange DIS, which is sufficient to 
extend the determination of the helicity-dependent (polarized) splitting 
functions \cite{MvN95,WVdP1a,WVdP1b} to NNLO for the upper-row quantities
$\Delta P_{\rm qq}$ and $\Delta P_{\rm qg}$. 
Since we had no access to the corresponding lower-row splitting functions,
these results were only briefly discussed in Ref.~\cite{mvvLL08}.
There is no helicity-sensitive analogue to the above Higgs-boson exchange 
in the Standard Model or an effective theory derived from it 
(initially a pseudo\-scalar $\chi$ with a $\,\chi\,\ep_{\mu\nu\rho\sigma}^{} 
\, G_a^{\,\mu\nu} \, G_a^{\,\rho\sigma}$ coupling to gluons was tried, which
however cannot probe spin information either, as also $\chi$ is a scalar under 
the rotation group).

This leaves only working in supersymmetry, as in Ref.~\cite{GdGGN05} for the 
determination of the NNLO quark-gluon antenna function, or considering DIS by 
graviton exchange.
We have chosen to adopt the second option, which is easier to implement in our
setup and offers additional information and checks by accessing all four 
splitting functions $\Delta P_{ik}$ as well as their unpolarized counterparts
and a full set of physical evolution kernels for both the unpolarized and the 
polarized case.

The basic formalism for graviton-exchange DIS has been developed in 
Ref.~\cite{LamLi}; for a recent application see also Ref.~\cite{SVgDIS}.
There are three structure functions $H_{\,1,\,2,\,3}$ in the unpolarized case,
of which three combinations can be formed which are analogous to $F_{\,2}$
(no gluon contribution at order $\as(0)$), $F_{\,\phi}$ (no quark contribution 
at order $\as(0)$) and $F_{\,L}$ (neither) in gauge-boson and scalar DIS.
In the polarized case there are two structure functions, $H_4$ and $H_6$,
where $H_{\bar{4}} = H_4 - H_6$ and $H_6$ involve only the quark and gluon
distributions, respectively, at the leading order, in perfect analogy with the
system $(F_{\,2},\,F_{\,\phi})$ that we employed for obtaining the unpolarized 
splitting functions.

We have performed complete second-order calculations of all these quantities.
At three loops, however, gravition exchange leads to a large number of 
integrals with a  higher numerator complexity than encountered in the 
calculations for Refs.~\cite{mvvPns,mvvPsg,mvvLL08}. 
Hence repeating the step from fixed-$N$ Mellin moments 
\cite{Mom3loop1,Mom3loop2} to all-$N$ results would require a lot of time 
and$/$or considerably improved algorithms. We have therefore resorted to 
calculating $\Delta P_{\rm gq}^{\,(2)}$ and $\Delta P_{\rm gg}^{\,(2)}$ for 
fixed (odd) values of~$N$. 
Substantial improvement in our diagram handling and in the {\sc Form} 
\cite{FORM3,TFORM,FORM4} implementation of the {\sc Mincer} program 
\cite{MINCER1,MINCER2}, see Ref.~\cite {jvLL14}, together with the availability
of sufficient computing resources, have enabled us to completely determine
$\Delta P_{\rm gq}^{\,(2)}(N)$ for $3 \leq N \leq 27$ and
$\Delta P_{\rm gg}^{\,(2)}(N)$ for $3 \leq N \leq 25$ 
(the $N=1$ moments are not accessible in this calculation \cite{LamLi}),
and both for specific colour factors up to $N=29$.

Initially the extension to high moments was intended to facilitate approximate 
$x$-space results, analogous to but much more accurate than those obtained in 
Ref.~\cite{NVappr} based on the moments of Ref.~\cite{Mom3loop3} for the 
unpolarized case, which would suffice at all $x$-values relevant to `spin 
physics' in the foreseeable future. 
Similar to the somewhat simpler case of transverse polarization in Ref.~\cite
{Veliz}, however, it turned out that it is possible to reach values of $N$ for
which even the most complicated parts could be determined completely from the 
moments and additional endpoint information, in particular the suppression of 
$\,P_{ik}(x) - \Delta P_{ik}(x)\,$ by two powers of $\x1$ in the threshold
limit $x\ra 1$ in a suitable factorization scheme. 
The crucial step in this determination is the solution of systems of 
Diophantine equations for which we have, besides in-house tools coded in 
{\sc Form}, made use of a publicly available program \cite{axbWeb} using the 
LLL-based~\cite{LLL} algorithm described in Ref.~\cite{axbAlg}.

Consequently we are now in the position to present the complete NNLO 
contributions $\Delta P_{ik}^{\,(2)}$ to the helicity-difference splitting 
functions in perturbative QCD. 
The remainder of this article is organized as follows:
In Section~2 we set up our notations and discuss aspects of the second-order 
calculations and results relevant to our determination of the third-order 
corrections which we turn to in Section 3. 
Our \mbox{$N$-space} results for $\Delta P_{ik}^{\,(2)}$ are presented in 
Section~4, and the corresponding $x$-space expressions in Section~5, where we
also briefly illustrate the numerical size of the NNLO contributions to the 
evolution of polarized parton densities. 
We summarize our results in Section~6.
Some additional information on scheme transformations and graviton-exchange DIS
is collected in the Appendix.  A brief account of this research has been 
presented before in Ref.~\cite{mvvLL14}.
 
%
\section{Notations and second-order results} 
\label{sec:2loop}
%
The unpolarized and polarized parton densities of a longitudinally
polarized nucleon are given~by$\!$
\beq
\label{fiUnp}
  \; f_{i}^{}(x,\muS) \equal
  f_{i}^{\,+}(x,\muS) \,+\, f_{i}^{\,-}(x,\muS)
\eeq
and
\beq
\label{fiPol}
 \Delta f_{i}(x,\muS) \equal
  f_{i}^{\,+}(x,\muS) \,-\, f_{i}^{\,-}(x,\muS)
\eeq
where
$f_i^{\,+}$ and $f_i^{\,-}$ represent the number distributions of the parton 
type $i$ with positive and negative helicity, respectively, in a nucleon with
positive helicity. Here $x$ denotes the fraction of the nucleon's momentum 
carried by the parton, 
and $\mu$ the mass-factorization scale which can be identified with the 
coupling-constant renormalization scale without loss of information.

The scale dependence of the quantities in Eqs.~(\ref{fiUnp}) and 
Eqs.~(\ref{fiPol}) is governed by the renor\-malization-group evolution 
equations
\beq
\label{evol}
  \frac{d}{d \ln \muS} \: (\Delta) f_{i\,}^{}(x,\muS) \equal
  \left[ {(\Delta) P^{}_{ik}(\als(\muS))}
  \,\otimes\, (\Delta) f_k^{}(\muS) \right]\! (x) 
\eeq
where $\otimes$ stands for the Mellin convolution in the momentum variable,
given by
\beq
\label{Mconv}
  [ a \otimes b ](x) \equal \int_x^1 \! \frac{dy}{y} \; a(y)\:
  b\left(\frac{x}{y}\right) 
\eeq
if no $1/\x1_+$-distribution are involved.
The splitting functions $(\Delta) P_{ik}^{}$ in Eq.~(\ref{evol}) admit an
expansion in powers of the strong coupling constant $\als$ which we write as
\beq
\label{Pexp}
  (\Delta) P^{}_{ik}(x,\muS) \equal
  \sum_{\,n=0} \: \ar(n+1) (\Delta) P^{\,(n)}_{ik}(x)
\eeq
with
\beq
\label{aDef}
  \ars \:\:\equiv\:\: \frac{ \als(\muS) }{ 4\:\!\pi }
\;\: .
\eeq

Using symmetries, the system (\ref{evol}) of $2\nf+\!1$ coupled 
integro-differential equations, where $\nf$ denotes the numbers of effectively 
massless flavours, can be reduced to $2\nf -\!1$ scalar flavour non-singlet 
equations and the $2 \times 2$ system
\beq
\label{Sevol}
  \frac{d}{d \ln\muS}\,
  \left( \begin{array}{c} \! \Delta f_{\rm q}^{} \! 
                  \\[1mm] \! \Delta f_{\rm g}^{} \! \end{array} \right)
  \:\: = \:\! \left( \begin{array}{cc} 
      \! \Delta P_{\rm qq} & \Delta P_{\rm qg} \! \\[1mm]
      \! \Delta P_{\rm gq} & \Delta P_{\rm gg} \! \end{array} \right) \otimes
  \left( \begin{array}{c} \! \Delta f_{\rm q}^{} \! 
                  \\[1mm] \! \Delta f_{\rm g}^{} \!  \end{array} \right)
  \:\:\equiv\:\: \Delta P \,\otimes\, \Delta f
\eeq
for the polarized gluon density $\Delta f_g^{}(x,\muS)$ and the flavour-singlet 
quark distribution
\beq
\label{qS}
  \Delta f_{\rm q}^{}(x,\muS) \:\: = \:\:
  \sum_{i=1}^{\nf} \left\{ \Delta f_{q_i^{}}(x,\muS) 
                         + \Delta f_{{\bar q}_i^{}}(x,\muS) \right\}
\:\: .
\eeq
The quark-quark splitting function $\Delta P_{\rm qq}$ in Eq.~(\ref{Sevol})
can be decomposed as
\beq
\label{DPqq}
  \Delta P_{\rm qq}^{\,(n)}(x) \:\: =\:\: 
  \Delta P_{\rm ns}^{\,+(n)}(x) + \Delta P_{\rm ps}^{\,(n)}(x)
\eeq
into non-singlet and pure singlet components. The former is related by
$\Delta P_{\rm ns}^{\,+} = P_{\rm ns}^{\,-}\,$ to an unpolarized quantity
calculated in Ref.~\cite{mvvPns}, the latter starts only at $n=1$ and is specific
to the present polarized case. 
It is often convenient to consider the Mellin transforms of all quantities, 
given by
\beq
\label{Mtrf}
  a(N) \:\: = \:\:
  \int_0^1 \!dx\; x^{\,N-1}\: a(x)
\eeq
and an obvious generalization for plus-distributions, since the convolutions
(\ref{Mconv}) correspond to simple products in $N$-space, $[a \otimes b](N)
\,=\, a(N) \, b(N)$.

The complete next-to-leading order (NLO) contributions 
$\Delta P^{\,(1)}_{ik}$ for the quantities in Eq.~(\ref{Sevol}) have been
derived almost 20 years ago in Ref.~\cite{MvN95} in $N$-space using the OPE and
in Refs.~\cite{WVdP1a,WVdP1b} in $x$-space, using the lightlike axial-gauge 
approach of Refs.~\cite{Curci:1980uw,Furmanski:1980cm}. 
Some years ago, we have checked these results, and obtained 
$\Delta P^{\,(2)}_{\rm qq}$ and $\Delta P^{\,(2)}_{\rm qg}$, by extending 
the calculations for Refs.~\cite{mvvPsg,mvvF2L} to the structure function 
$g_{1}^{}$ in polarized DIS which was first addressed beyond the first order 
in Ref.~\cite{ZvNpol}.
All these calculations used dimensional regularization, and 
thus needed to address the issue of the Dirac matrix $\gamma_{5}^{}$ in 
$D \neq 4$ dimensions which enters via the quark helicity-difference projector.
 
The calculations in Ref.~\cite{MvN95} used the `reading-point' scheme for
$\gamma_{5}^{}$ \cite{readP}; those in Refs.~\mbox{\cite{WVdP1a,WVdP1b}} 
were carried out primarily with the `t Hooft$/$Veltman prescription 
\cite{g5HV,g5BM}, but included checks also using the so-called Larin scheme
\cite{g5L1,g5L2},
\beq
\label{g5Larin}
   p \hspace*{-2mm}/ \: \gamma_{5,L}^{} \equal
  {1 \over 6}\; \ep_{\mu\nu\rho\sigma}\; 
  p^{\,\mu} \: \gamma^{\,\nu}\, \gamma^{\,\rho}\, \gamma^{\,\sigma} 
\:\: ,
\eeq
where the resulting contractions of two $\ep$-tensors are evaluated in terms of
the $D$-dimensional metric. 
All our calculations have been carried out using the Larin scheme which is 
equivalent to the \mbox{`t Hooft}$/$Veltman prescription for the present 
massless case.

Quantities calculated using Eq.~(\ref{g5Larin}) need to be subjected to a 
factorization scheme transformation in order to arrive at expressions in the 
standard \MSb\ scheme \cite{'tHooft:1973mm,Bardeen:1978yd}, for example
\beq
\label{g1trf}
  g_1^{} \; = \; {C}_{g_{1\!}^{},\,L} \; {\Delta  _L^{}} \nn\\
         \; = \; ( {C}_{g_{1\!}^{},\,L} \:\: {Z}^{\,-1} ) \; 
                 ( {Z} \:\: {\Delta f}_{\!L}^{} )
         \; = \; {C}_{g_{1\!}^{}} \:\:{\Delta f}
\eeq
where we have switched to a matrix notation in $N$-space and suppressed all 
function arguments. Denoting the perturbative expansion of the transformation
matrix by
\beq
\label{Zexp}
  Z(x,\muS) \:\: = \:\: 1 \,+\, \sum_{n=1} \ar(n)\: Z^{\,(n)}(x) 
  \:\: = \:\:
  1 \,+\, \sum_{n=1} \ar(n) \left( \begin{array}{rr} 
     \! z_{\,\rm qq}^{\,(n)}(x) \! & z_{\,\rm qg}^{\,(n)}(x) \! \\[2mm] 
     \! z_{\,\rm gq}^{\,(n)}(x) \! & z_{\,\rm gg}^{\,(n)}(x) \!
  \end{array} \right)
\:\: ,
\eeq
the transformation (\ref{g1trf}) of the coefficient functions ${C}_{g_{1\!}^{}}$
and the parton densities $\Delta f$ leads to 
\bea
\label{Ptrf}
  \Delta P &\! =\! & \;\;
        \ars \, \Delta P^{\,(0)}
\nn \\[1mm] & & \mbox{\hspn} 
  +\: \ar(2) \, \Big\{ \Delta P_L^{\,(1)} + [Z^{\,(1)},\,\Delta P^{\,(0)}] 
                  - \beta_0\, Z^{\,(1)} \Big\}
\nn \\[1mm] & & \mbox{\hspn} 
  +\: \ar(3) \, \Big\{ \Delta P_L^{\,(2)} + [Z^{\,(2)},\,\Delta P^{(0)}]
    + [Z^{\,(1)},\,\Delta P_L^{\,(1)}] - [Z^{(1)},\,\Delta P^{(0)}] \: Z^{(1)}
\nn \\ & & \mbox{\hspp} 
    + \beta_0\, \left( ( Z^{\,(1)} )^2 - 2\, Z^{\,(2)} \right)
    - \beta_1\, Z^{\,(1)} \Big\}
  \;\; + \;\; {\cal O}(\ar(4))
\eea
for the splitting functions in the \MSb\ scheme, where $[a ,\, b]$ denotes the 
standard matrix commutator. Here $\beta_0$ and $\beta_1$ are the leading two
coefficients in the expansion of the beta function of QCD, 
\beq
\label{aRun}
  \frac{d \ars}{d \ln \muS} 
  \:\: = \:\: \beta(\ars) 
  \:\: = \:\:
  - \sum_{\ell=0} \ar(\ell+2) \: \beta_\ell^{} 
\:\: ,
\eeq
which to NNLO is given by \cite
{Gross:1973rr,Georgi:1974sr,beta1a,beta1b,beta2a,beta2b}
\bea
\label{b012}
  \beta_0 &\! =\! & 
                \frct{11}{3}\: \* \ca 
          \,-\, \frct{2}{3}\: \* \nf
\:\: ,\nn \\[1mm]
  \beta_1 &\! =\! & 
                \frct{34}{3}\: \* \cas 
          \,-\, \frct{10}{3}\: \* \ca\, \* \nf
          \,-\, 2\, \* \cf\, \* \nf
\:\: , \\[1mm]
  \beta_2  &\! =\! &
               \frct{2857}{54} \: \* \cat
         \,-\, \frct{1415}{54} \: \* \cas \* \, \nf
         \,-\, \frct{205}{18} \: \* \cf \, \* \ca \, \* \nf
         \,+\, \cfs \, \* \nf
         \,+\, \frct{79}{54} \: \ca \, \* \nfs
         \,+\, \frct{11}{9} \: \cf \, \* \nfs
\nn
\eea
with $C_A \,=\, n_c \,=\, 3$ and $C_F \,=\, (n_c^{\,2}-1)/(2\:\! n_c)\,=\, 4/3$ 
in $SU(n_c=3)$. $\beta_0$ and $\beta_1$ are scheme-%
independent in massless perturbative QCD; $\,\beta_2$ is given in the \MSb\
scheme adopted in this article.
 
The transformation matrix has been determined to NNLO in Ref.~\cite{MSvN98} as
\beq
\label{ZikM}
  Z_{\,\rm ik} \equal \delta_{\,\rm ik} \:+\: 
   \delta_{\,\rm iq} \,\delta_{\,\rm kq}\, \Big(
     \ars \: z_{\,\rm ns}^{\,(1)}
   \,+\, \ar(2)  \left\{ z_{\,\rm ns}^{\,(2)} + z_{\,\rm ps}^{\,(2)} \right\} 
   \!\Big)
   \; + \; {\cal O}(\ar(3))
\:\: .
\eeq
Its non-singlet entries can be fixed by the relation between the corresponding 
coefficient functions for $g_1^{}$ and the structure function $F_3^{}$ which is
known to order $\as(3)$ \cite{mvvF3}; 
the critical part is the pure-singlet part for which, as far as we know, only 
that one calculation has been performed so far.
For~the convenience of the reader the results are included in Appendix A.
For $\,z_{\rm qg}^{\,(n)} = z_{\rm gg}^{\,(n)} = 0$, Eq.~(\ref{Ptrf}) leads to 
the following transformations of the NLO and NNLO splitting functions:
\bea
\label{P1trf}
  \dPqq(1) &\! =\!& \dPqqL(1)
   \,-\, \beta_0 \, \* \Zqq(1)
   \,-\, \dPqg(0) \, \* \Zgq(1) 
\:\: ,\nn \\[1.5mm]
  \dPqg(1) &\! =\!& \dPqgL(1)
    \,+\, \dPqg(0)\, \* \Zqq(1) 
\:\: ,\nn \\[1.5mm]
  \dPgq(1) &\! =\!& \dPgqL(1)
   \,-\, \dPgq(0) \, \* \Zqq(1) 
   \,+\, \left( \dPqq(0) - \dPgg(0) - \beta_0 \right) \* \Zgq(1) 
\:\: ,\nn \\[1.5mm]
  \dPgg(1) &\! =\!& \dPggL(1)
   \,+\, \dPqg(0) \, \* \Zgq(1) 
\eea
and
\bea
\label{P2trf}
  \dPqq(2) &\! =\!& \dPqqL(2)
   \,+\, \beta_0 \* \left( \! ( \Zqq(1) )^2 - 2\,\* \Zqq(2) \right)
   \,-\, \beta_1 \, \* \Zqq(1)
   \,-\, \dPqgL(1) \, \* \Zgq(1) 
   \,-\, \dPqg(0) \, \* \Zgq(2) 
\:\: ,\nn \\[2mm]
  \dPqg(2) &\! =\!& \dPqgL(2)
   \,+\, \dPqgL(1) \, \*\Zqq(1) 
   \,+\, \dPqg(0) \, \*\Zqq(2) 
\:\: ,\nn \\[2mm]
  \dPgq(2) &\! =\!& \dPgqL(2)
   - \left( \dPgqL(1) - \dPgq(0)\, \*\Zqq(1) \right) \* \Zqq(1)
   - \dPgq(0)\, \*\Zqq(2)
\nn \\[1mm] & & \mbox{}
   + \left( \dPgg(0) - \dPqq(0) + \beta_0 \right) \* \Zqq(1)\, \* \Zgq(1)
   - \left( \dPgg(0) - \dPqq(0) + 2\,\* \beta_0 \right) \* \Zgq(2)
\nn \\[1mm] & & \mbox{}
   + \left( \dPqqL(1) - \dPggL(1) - \beta_1 - \dPqg(0)\, \* \Zgq(1) 
     \right) \* \Zgq(1)
\:\: ,\nn \\[2mm]
  \dPgg(2) &\! =\!& \dPggL(2)
   \,+\, \dPqgL(1) \, \* \Zgq(1)
   \,+\, \dPqg(0) \, \* \Zgq(2) 
\:\: .
\eea
These expressions are reduced to the standard scheme transformation of 
Refs.~\cite{MvN95,WVdP1a,WVdP1b,MSvN98} by dropping all contributions with 
$\Zgq(1)$ or $\Zgq(2)$; 
it will become clear below why these terms have been included in 
Eqs.~(\ref{P1trf}) and (\ref{P2trf}).

\begin{figure}[p]
\vspace*{-1mm}
\centerline{\hspace*{-1mm}\epsfig{file=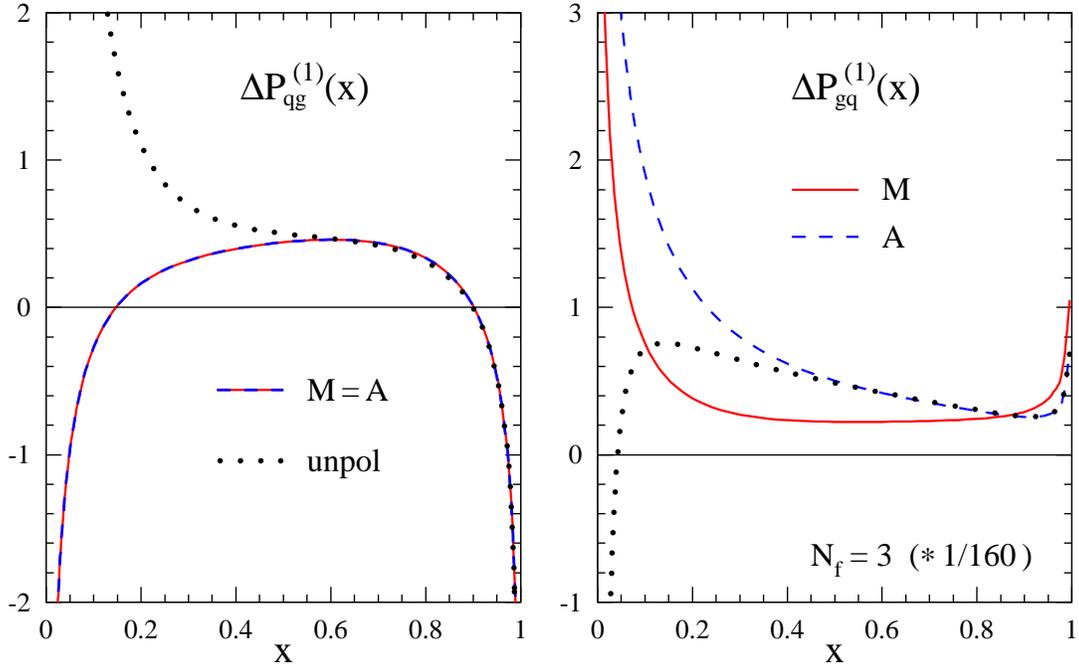,width=15cm}}
\vspace{-3.5mm}
\caption{\label{dPij1MAU}
The NLO contributions to the off-diagonal splitting functions in 
Eq.~(\ref{Sevol}), compared to their unpolarized counterparts. The polarized 
results are shown as published in Refs.~\cite{MvN95,WVdP1a,WVdP1b} (`M') and 
after including an additional term 
$\,z_{\,\rm gq}^{\,(1)} \,=\, -\,\Delta P^{\,(0)}_{\,\rm gq}$
in the transformation (\ref{Ptrf}) from the Larin scheme (`A'), which removes 
all $(1-x)^{\,0,\,1}$ terms from the quantity $\,\delta_{\,\rm gq}^{\,(1)}(x)$
in Eq.~(\ref{dPgq1xto1}).
}
\end{figure}

It is instructive to consider the $x \ra 1$ threshold limit of the splitting
functions. It is expected that the physical probability of a helicity flip is
suppressed by two powers in $\x1$ in this limit \cite{BBSsuppr}. Hence the
differences
\beq
\label{delta}
 \delta_{ik}^{\,(n)} \:\:\equiv\:\: P^{\,(n)}_{ik} - \Delta P^{\,(n)}_{ik}  
\eeq
should be suppressed, in a `physical' factorization scheme, by a factor of 
$\x1^2$, or $1/N^{\,2}$ in \mbox{$N$-space}, relative to the respective sums 
which behave (modulo logarithms) as $\x1^{-1}$ or $N^{\,0}$ for 
$\rm ik = qq,\:gg$ and $\x1^{0}$ or $N^{\,-1}$ for $\rm ik = qg,\:gq$.
For the scheme-independent leading-order (LO) splitting functions, the 
differences (\ref{delta}) read
\bea
  \delta_{\,\rm qq}^{\,(0)}(x) &\! =\! & 0 
\:\: , \nn \\[1mm] 
  \delta_{ik}^{\,(0)}(x) &\! =\! & \! 
  \mbox{ const}\, \cdot \x1^2 \,+\:\ldots
  \quad \mbox{for} \quad ik = {\rm qg,\: gq,\: gg} \; .
\label{dPij0xto1}
\eea
The corresponding NLO results for the \MSb\ splitting functions 
\cite{MvN95,WVdP1a,WVdP1b} are given by
\bea
\label{dPij1xto1}
  \delta_{\,ik}^{\,(1)}(x) &\! =\!&
  {\cal O} \left( \x1^a \right)
  \quad \mbox{ for} \quad 
  ik = {\rm qq,\: gg}\; (a = 1),\; 
       {\rm qg}\;(a = 2)
\:\: , \\[1mm]
\label{dPgq1xto1}
  \delta_{\,\rm gq}^{\,(1)}(x) &\! =\!&
  8\, \* \cf \* (\ca\!-\!\cf)\: \* \ln \x1
  \:+\: \frct{44}{3}\:\* \cf \* \ca - 6\, \* \cfs - \frct{8}{3}\: \* \cf \nf
\nn \\[1mm] & & \mbox{}
  - \, \x1 \Big\{
        8\, \* \cf \* (\ca\!-\!\cf)\, \* \ln \x1
    \,+\, \Big( \, \frct{20}{3}\: \* \cf \ca + 2\, \* \cfs
        - \frct{8}{3}\: \* \cf \nf \Big) \! \Big\}
\quad \\[1mm] & & \mbox{}
  \,+\, {\cal O} \left( (1\!-\!x)^2 \, \right) 
\nn \:\: .
\eea
Interestingly, as already noted in Ref.~\cite{mvvLL08}, all 10 terms in 
Eq.~(\ref{dPgq1xto1}) can be removed by including the simple additional term
$\Zgq(1) \,=\, -\dPgq(0)$ in the NLO scheme transformation (\ref{P1trf}).
The splitting functions $\dPqg(1)(x)$ and $\dPgq(1)(x)$ are shown,
together with their unpolarized counterparts,
 in Fig.~\ref{dPij1MAU} in the standard scheme, from now on denoted by `M' 
wherever required, that uses only Eq.~(\ref{ZikM}) and an alternative scheme 
(`A') that also includes this additional term. 

\begin{figure}[p]
\vspace*{-1mm}
\centerline{\hspace*{-1mm}\epsfig{file=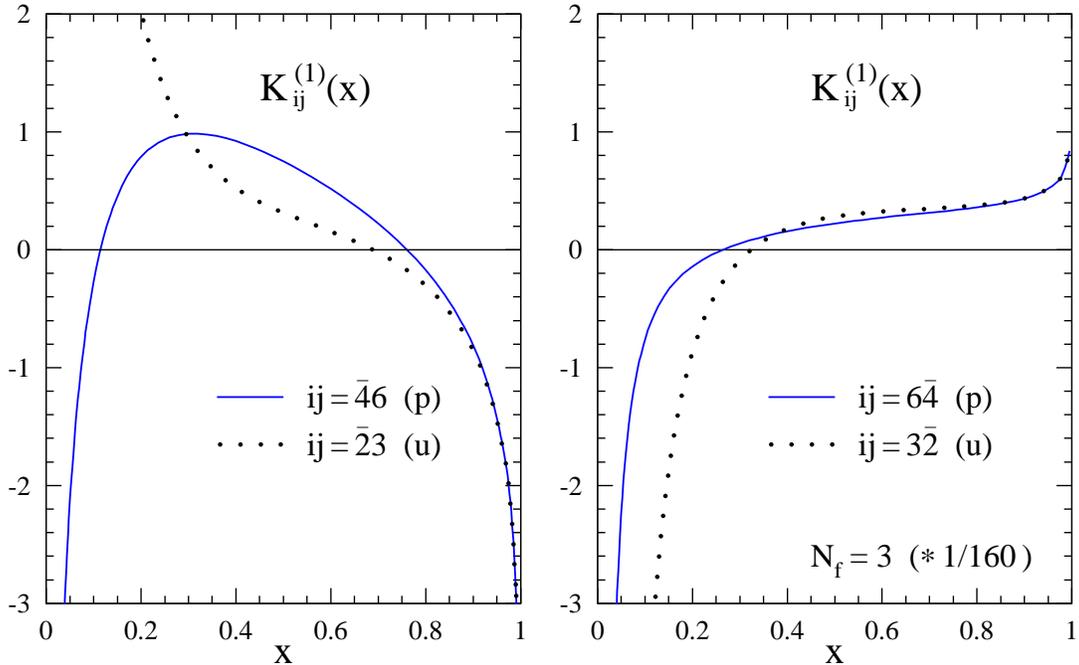,width=15cm}}
\vspace{-3.5mm}
\caption{\label{Kij1PU}
The NLO contributions to the off-diagonal elements of the physical-kernel
matrices for the systems $\,(H_{\bar{2}},\, H_3^{})\,$ and 
$\,(H_{\bar{4}},\, H_6^{})\,$ of structure functions in unpolarized and 
polarized graviton-exchange DIS \cite{LamLi} as defined in and below 
Eqs.~(\ref{Hunp}) and (\ref{Hpol}). 
The factor $1/160 \,\simeq\, 1/(4 \pi)^2$ approximately converts the results 
from our small expansion parameter (\ref{aDef}) to a series in $\als$.
}
\vspace*{-1mm}
\end{figure}

The issue of the physical large-$x$ behaviour of the helicity-dependent 
quark-gluon splitting can be addressed by studying suitable flavour-singlet
physical evolution kernels (or physical anomalous dimensions) for structure 
functions in unpolarized and polarized DIS. Graviton-exchange DIS, for~which the
basic formalism was worked out in Ref.~\cite{LamLi}, provides a sufficiently
large set of structure functions. It is convenient to combine and normalize
four of these functions as
\beq
\label{Hunp}
  H_{\rm u} \equal 
  \left( \begin{array}{c}  \! H_{\bar{2}}^{} \!
                  \\[1mm]  \! H_3^{}         \! \end{array} \right)
 \equal
  \left( 
  \begin{array}{cc} \! C_{\,\bar{2},\rm q}\! & C_{\,\bar{2},\rm g} \!
           \\[1mm]  \! C_{\,3,\rm q} \!      & C_{\,3,\rm g} \!   \end{array} 
  \right)
  \left( \begin{array}{c}  \! f_{\rm q}^{} \!
                  \\[1mm]  \! f_{\rm g}^{} \! \end{array} \right)
  \:\:\equiv\:\:
  C_{\rm u} \, f 
\eeq
with $\,H_{\bar{2}}^{} = H_2 - 4\, H_3\,$ in the unpolarized case, and
\beq
\label{Hpol}
  H_{\rm p} \equal
  \left( \begin{array}{c}  \! H_{\bar{4}}^{} \!
                  \\[1mm]  \! H_6^{}         \! \end{array} \right)
 \equal
  \left(
  \begin{array}{cc} \! C_{\,\bar{4},\rm q}\! & C_{\,\bar{4},\rm g} \!
           \\[1mm]  \! C_{\,6,\rm q} \!      & C_{\,6,\rm g} \!   \end{array}
  \right)
  \left( \begin{array}{c}  \! f_{\rm q}^{} \!
                  \\[1mm]  \! f_{\rm g}^{} \! \end{array} \right)
  \:\:\equiv\:\:
  C_{\rm p} \, \Delta f
\eeq
with $\,H_{\bar{4}}^{} = 2 ( H_4 - H_6 )\,$ in the polarized case, 
where we have changed the $x^{\,n}$ prefactors relative to Eq.~(31) of 
Ref.~\cite{LamLi} such that $(C_{\rm u})_{\rm ij} = (C_{\rm p})_{\rm ij} 
= \delta_{\,\rm ij}$ at LO. The corresponding NLO coefficient functions
can be found in Appendix B.
The physical-kernel matrices $K_a\,$, $\,a= \rm u,\:p$ (for~the renormalization 
scale $\,\muRs \,=\, \Qs\,$) are obtained from the coefficient functions, the 
beta function (\ref{aRun}) and the respective unpolarized ($P_{\rm u}\,=\, P\,$)
and polarized ($P_{\rm p} \,=\, \Delta P\,$) splitting functions, 
cf.~Eq.~(\ref{Sevol}),~by
\beq
\label{PhysK}
  \frac{d H_a}{d \ln Q^{\,2}} \equal 
  \Big( \beta(\ars)\; \frac{d\, C_a}{d \ars}\, + C_a\, P_a \Big) \, 
  C_a^{\,-1} H_a \:\: \equiv \:\:  K_a \: H_a 
\:\: .
\eeq
The expansion of this result to order $\ar(3)$ can be read off from 
Eq.~(\ref{Ptrf}) for $\,Z \,=\, C_a\,$.

We have performed complete two-loop calculations of these structure functions, 
recovering both the unpolarized and polarized NLO flavour-singlet splitting 
functions from graviton-exchange DIS, and used these results to obtain the NLO 
physical kernels $K_{\,\rm u}^{\,(1)\!}(x)$ and $K_{\,\rm p}^{\,(1)\!}(x)$. 
The respective off-diagonal elements for the systems (\ref{Hunp}) and 
(\ref{Hpol}) are compared in Fig.~\ref{Kij1PU}.
It is clear, also from the corresponding analytical results, that also the 
large-$x$ limits of the kernels $K_{\,3\bar{2}}^{\,(1)\!}(x)$ and
$K_{\,6\bar{4}}^{\,(1)\!}(x)$ corresponding to the splitting functions
$(\Delta) P_{\rm gq}^{\,(1)}$ are consistent with the expectation of 
Ref.~\cite{BBSsuppr}; hence Eq.~(\ref{dPgq1xto1}) is indeed a unphysical
feature of the standard transformation to the \MSb\ scheme.

%
\section{Determination of the third-order corrections}
\label{sec:3calc}
\setcounter{equation}{1}
%
As before, we have calculated inclusive DIS via the optical theorem, which 
relates the \mbox{probe$\,$($q$)$\,$-} $\,$parton$\,$($p$) total cross sections 
(with $Q^{\,2} = - q^{\,2} > 0$ and $p^{\,2} = 0\,$) to forward amplitudes, 
and a dispersion relation in $x\,$ that provides the $N$-th moments from the 
coefficient of $(2p \cdot q)^N$ \mbox{\cite{Mom3loop1,Mom3loop2}}.
For~the splitting functions $\dPqq(2)$ and $\dPqg(2)$ we have extended the 
three-loop all-$N$ calculations of Refs.~\cite{mvvPns,mvvPsg} to the 
photon-exchange structure function $g_{1}^{}$. 
As discussed in Ref.~\cite{mvvLL08}, a large number of additional integrals, 
arising from a fairly small set of top-level integrals with higher numerator 
powers, had to be calculated for this extension; their determination took 
several months.

The situation is far worse in the case of graviton-exchange DIS, which is our
means to access also $\dPgq(2)$ and $\dPgg(2)$, in terms of both the complexity 
and the number of new top-level integrals.  We have therefore not tried a 
direct all-$N$ calculation in this case,     but managed to set up
a two-step procedure with the same result.  The first step is a 
calculation of fixed-$N$ moments for the structure functions in polarized 
graviton-exchange DIS, as in Refs.~\cite{Mom3loop1,Mom3loop2} using the 
{\sc Mincer} program \cite{MINCER1,MINCER2}, 
but up to much higher moments in particular for $H_6$, cf.~Eq.~(\ref{Hpol}).
The~second step is the determination of the all-$N$ expressions for $\dPgq(2)$
and $\dPgg(2)$ from the moments calculated in the first step together with
insight into the structure of these functions.
 
In order to drive the first step to a point where the second became possible,
and its results could be verified by one or two yet higher moments, improvements
had to be made in our diagram preparation and the {\sc Mincer} code, see also
Ref.~\cite{jvLL14}.
The diagrams were generated, as before, with a special version of {\sc Qgraf}
\cite{QGRAF}. Unlike in our previous calculations, however, the diagrams with 
the same group-invariant colour factor, the same topology and subtopology 
(see below), 
and the same flavour structure have been combined in the `diagram' files which 
are managed, as before, using the database program {\sc Minos} \cite{MINOS}.
In this way the number of third-order diagrams has been reduced from 5176 to 
1142 and from 15208 to 1249 for the quark and gluon contributions, respectively,
to $H_4$ and $H_6$. The combined diagrams take roughly as much time as the most 
difficult individual diagram in the set, which leads to an overall gain in 
speed by a factor of three to five.

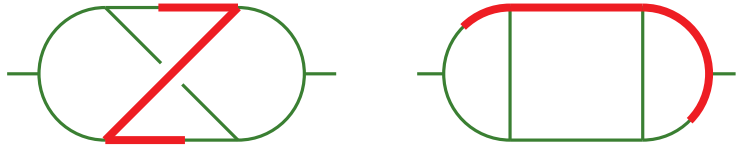
\begin{figure}[b]
\vspace*{-8mm}
\begin{center}
\begin{picture}(150,100)(0,0)
\SetColor{OliveGreen}
\SetScale{1.0}
\SetWidth{1.2}
       \Line(35,70)(85,70) \Line(35,20)(85,20)
       \Line(35,70)(56,49) \Line(64,41)(85,20)
       \Line(85,70)(35,20)
       \CArc(35,45)(25,90,270) \CArc(85,45)(25,270,90)
       \Line(10,45)(-2,45) \Line(110,45)(122,45)
       {\SetColor{Red} \SetWidth{3}
       \Line(55,70)(85,70)
       \Line(35,20)(65,20) \Line(85,70)(35,20)}
\SetColor{Black}
\end{picture}
\begin{picture}(150,100)(0,0)
\SetColor{OliveGreen}
\SetScale{1.0}
\SetWidth{1.2}
       \Line(35,70)(85,70) \Line(35,20)(85,20)
       \Line(35,70)(35,20) \Line(85,70)(85,20)
        \CArc(35,45)(25,90,270) \CArc(85,45)(25,270,90)
        \Line(10,45)(0,45) \Line(110,45)(120,45)
        {\SetColor{Red} \SetWidth{3}
        \CArc(35,45)(25,90,135) \CArc(85,45)(25,-45,90)
        \Line(35,70)(85,70)}
\SetColor{Black}
\end{picture}
\end{center}
\vspace*{-8mm}
\caption{\label{Diagrams} 
The NO$_{25}$ (left) and LA$_{14}$ (right) subtopologies for the forward 
probe-parton amplitudes. The momentum $q$ of the probe, with $q^{\,2} < 0$,
enters the diagram from the right and leaves on the~left. The parton momentum 
$p$, with $p^{\,2} = 0$, flows through the fat (in the coloured version: red) 
lines.
}
\vspace*{-1mm}
\end{figure}

The overall most demanding subtopology, in terms of execution time and required
disk space, is NO$_{25}$ (see Fig.~\ref{Diagrams}), i.e, the  most difficult
$p$-flow in the most difficult three-loop topology.
Also notable are the LA$_{14}$ (also shown in Fig.~\ref{Diagrams}), O4$_{57}$,
O2$_{26}$ cases, where the momentum $p$ flows through four internal lines, and
the three-line BE$_{57}$ and BE$_{28}$ `Benz' cases.
The largest diagram calculated took about $10^{\,7}$ CPU seconds and required
6.7 TB of disk space for the projection on $N$.
The~results for $3 \leq N \leq 25$ were employed for obtaining the all-$N$
expressions for $\dPgq(2)$ and $\dPgg(2)$. For checking these expressions,
the quark case was computed completely at $N=27$ and in the `planar limit'
$\ca -2\,\cf \ra 0$ at $N=29$, and the gluon case for the $\cat$ terms
at $N=27$ and~$N=29$.
The latter was possible since most of the slowest and largest diagrams do not
contribute to this colour factor, which is the most complicated one in terms
of the structure of the splitting function.

Most of the diagram calculations were performed on the {\tt ulgqcd} cluster
in Liverpool, using {\sc Tform} \cite{TFORM,FORM4} with 16 workers on more than 
200 cores; the hardest diagrams at the highest values of $N$ were calculated on 
a new high-end computer at {\sc Nikhef}. For the previous optimization of 
{\sc Mincer} we were also able to use a multi-core workstation at DESY-Zeuthen.

As an example, we show the non-$\z3$ parts of the moments $3 \leq N \leq 25$ of the $\cft$ part of
$\,\Delta P_{\rm gq}^{\,(2)}$ in the Larin scheme, i.e., before the
transformation of the output of the mass factorization to \MSb$\,$:
\\[2mm] \noindent
{\small
\hspace*{2mm}
~N = ~3: ~~$ 186505/
( 3^{\,5}\, 2^{\,5} ) $
\\[0.2mm]
\hspace*{2mm}
~N = ~5: ~~$ 9473569/
( 5^{\,5}\, 3^{\,5}\, 2^{\,2} ) $
\\[0.2mm]
\hspace*{2mm}
~N = ~7: ~-$ 509428539731/
( 7^{\,5}\, 5^{\,4}\, 3^{\,2}\, 2^{\,11} ) $
\\[0.2mm]
\hspace*{2mm}
~N = ~9: ~-$ 266884720969207/
( 7^{\,4}\, 5^{\,5}\, 3^{\,10}\, 2^{\,7} ) $
\\[0.2mm]
\hspace*{2mm}
~N = 11: -$ 3349566589170829651/
( 11^{\,5}\, 7^{\,4}\, 5^{\,4}\, 3^{\,9}\, 2^{\,7} ) $
\\[0.2mm]
\hspace*{2mm}
~N = 13: -$ 751774767290148022507/
( 13^{\,5}\, 11^{\,4}\, 7^{\,3}\, 5^{\,3}\, 3^{\,7}\, 2^{\,8} ) $
 \hfill {\normalsize (3.1)}
\\[0.2mm]
\hspace*{2mm}
~N = 15: -$ 23366819019913026454180147/
( 13^{\,4}\, 11^{\,4}\, 7^{\,4}\, 5^{\,5}\, 3^{\,9}\, 2^{\,16} ) $
\\[0.2mm]
\hspace*{2mm}
~N = 17: -$ 305214227818628090680174170947/
( 17^{\,5}\, 13^{\,4}\, 11^{\,4}\, 7^{\,4}\, 5^{\,4}\, 3^{\,10}\, 2^{\,10} ) $
\\[0.2mm]
\hspace*{2mm}
~N = 19: -$ 570679648684656807578199791973487/
( 19^{\,5}\, 17^{\,4}\, 13^{\,4}\, 11^{\,4}\, 7^{\,3}\, 5^{\,5}\, 3^{\,7}\, 
 2^{\,9} ) $
\\[0.2mm]
\hspace*{2mm}
~N = 21: -$ 2044304092089235762279148843319979/
( 19^{\,4}\, 17^{\,4}\, 13^{\,4}\, 11^{\,4}\, 7^{\,5}\, 5^{\,3}\, 3^{\,9}\, 
 2^{\,11} ) $
\\[0.2mm]
\hspace*{2mm}
~N = 23: -$ 289119840113761409530260333250139823739/
( 23^{\,5}\, 19^{\,4}\, 17^{\,4}\, 13^{\,4}\, 11^{\,4}\, 7^{\,4}\, 5\, 
 3^{\,9}\, 2^{\,13} ) $
\\[0.2mm]
\hspace*{2mm}
~N = 25: -$ 1890473255283802937678830745102921869938637/
( 23^{\,4}\, 19^{\,4}\, 17^{\,4}\, 13^{\,5}\, 11^{\,4}\, 7^{\,4}\, 5^{\,10}\, 
 3^{\,5}\, 2^{\,12} ) $
}

\vspace*{1mm}
In order to obtain, with certainty, the analytical forms of 
$\,\Delta P_{\rm gq}^{\,(2)}(N)$ and $\,\Delta P_{\rm gg}^{\,(2)}(N)$ from only 
12~moments, we need to make use of additional constraints on the structure of 
these functions. 
At~least up to NNLO, the splitting functions can be expressed in terms of 
harmonic sums \cite{Hsums}, see also Ref.~\cite{BKurth}, which can be 
recursively defined by
\beq
\label{Hsum1}
  S_{\pm m}(N) \equal \sum_{i=1}^{N}\: \frac{(\pm 1)^i}{i^{\, m}}
\eeq
and
\beq
\label{Hsum2}
  S_{\pm m_1^{},m_2^{},\ldots,m_k^{}}(N) \equal \sum_{i=1}^{N}\:
  \frac{(\pm 1)^{i}}{i^{\, m_1^{}}}\: S_{m_2^{},\ldots,m_k^{}}(i) 
\:\: .
\eeq
The sum of the absolute values of the indices $m_k$ defines the weight
of the harmonic sum. Assigning a weight $m$ to the un-summed denominators
\beq
\label{Ddef}
  D_k^{\:m} \:\:\equiv\:\: \frac{1}{(N+k)_{}^m}
\eeq
which can be expressed as differences of two harmonic sums of weight $m$, 
the N$^n$LO splitting functions include terms up to weight $2n+\!1$. 
For example, the $\cfs \nf$ contribution to $\dPqg(2)(N)$~reads
\bea
\label{dPqg2cf2nf}
  && \hspn 
  \frct{1}{8}\left.\Delta P^{\,(2)}_{\,\rm qg}(N) \right|_{\cfs\nf}
  \: = \;
          2\, \* \colour4colour{ \dpqg } \* (
          - \, \S(-4)
          + 2\, \* \Ss(-2,-2)
          + 4\, \* \Ss(1,-3)
          + 2\, \* \Ssss(1,1,1,1)
          - \Sss(1,1,2)
          - 5\, \* \Sss(1,2,1)
\quad \nn \\[-2.5mm] && \hspace*{4.4cm}
          + \, 4\, \* \Ss(1,3)
          + 2\, \* \Ss(2,-2)
          - 6\, \* \Sss(2,1,1)
          + 6\, \* \Ss(2,2)
          + 7\, \* \Ss(3,1)
          - 3\, \* \S(4)
          )
\nn \\[2mm] && \mbox{}
        + 4\, \* \colour4colour{ \S(-3) }\, \*  (
            \DNn2
          - 2\, \* \DNnO
          + 2\, \* \DNpO
          )
    \:+\: 8\, \* \colour4colour{ \Ss(1,-2) }\, \*  (
            2\, \* \DNp2
          - \DNnO
          + \DNpO
          )
\nn \\[1mm] && \mbox{}
        + \colour4colour{ \Sss(1,1,1) }\, \*  (
            5\, \* \DNn2
          - 2\, \* \DNp2
          - 21/2\, \* \DNnO
          + 12\, \* \DNpO 
          )
    \:-\: 2\, \* \colour4colour{ \Ss(1,2) }\, \*  (
            2\, \* \DNn2
          - 2\, \* \DNp2
          - 5\, \* \DNnO
          + 5\, \* \DNpO
          )
\nn  \\[1mm] && \mbox{}
        - 2\, \* \colour4colour{ \Ss(2,1) }\, \*  (
            4\, \* \DNn2
          + 2\, \* \DNp2
          - 11\, \* \DNnO
          + 11\, \* \DNpO
          )
    \:+\: 2\, \* \colour4colour{ \S(3) }\, \*  (
            3\, \* \DNn2
          + 6\, \* \DNp2
          - 11\, \* \DNnO
          + 11\, \* \DNpO
          )
\nn  \\[1mm] && \mbox{}
       -  3\, \* \colour4colour{ \z3 } \, \*  (
            2\, \* \DNn2
          + 4\, \* \DNp2
          - 9\, \* \DNnO
          + 12\, \* \DNpO
          )
    \:-\: 6\, \* \colour4colour{ \DNppO }\, \*  ( \S(-2) + 1 ) 
 \\[2.5mm] && \mbox{}
         +  2\, \* \colour4colour{ \S(-2) }\, \*  (
            8\, \* \DNp3
          - 5\, \* \DNn2
          - 6\, \* \DNp2
          + 10\, \* \DNnO
          - 9\, \* \DNpO
          )
    \:-\: \colour4colour{ \Ss(1,1) }\, \*  (
            10\, \* \DNn3
          + 6\, \* \DNp3
          - 35/2\, \* \DNn2
\nn \\[1mm] && \hspp \mbox{}
          - 5\, \* \DNp2
          + 29\, \* \DNnO
          - 36\, \* \DNpO
          )
    \:+\: 2\, \* \colour4colour{ \S(2) }\,  \*  (
            4\, \* \DNn3
          + 6\, \* \DNp3
          - 10\, \* \DNn2
          - 4\, \* \DNp2
          + 17\, \* \DNnO
          - 22\, \* \DNpO
          )
\nn \\[2.5mm] && \mbox{}
       +  \colour4colour{ \S(1) }\, \*  (
            7\, \* \DNn4
          + 4\, \* \DNp4
          - 43/2\, \* \DNn3
          - 15\, \* \DNp3
          + 99/2\, \* \DNn2
          + 18\, \* \DNp2
          - 78\, \* \DNnO
          + 329/4\, \* \DNpO
          )
\nn \\[1mm] && \mbox{}
          + 32\, \* \DNp5
          - 15/2\, \* \DNn4
          - 3\, \* \DNp4
          + 59/8\, \* \DNn3
          + 53/4\, \* \DNp3
          + 77/8\, \* \DNn2
          + 213/8\, \* \DNp2
\nn \\[1mm] && \hspp \mbox{}
          - 1357/32\, \* \DNnO
          + 777/16\, \* \DNpO
\nn 
\eea
in the standard \MSb\ scheme \cite{MSvN98}, where all harmonic sums are 
understood to be taken at argument~$N$. Here we have also made used of the 
first of the abbreviations
\beq
\label{dpij0}
  \Delta p_{\rm qg} \equal 2\,D_1^{} - D_0^{}
\;\; , \quad
  \Delta p_{\rm gq} \equal 2\,D_0^{} - D_1^{}
\eeq
for the $N$-dependence of the lowest-order splitting functions, 
cf.~Eq.~(\ref{dPij0N}) below.

If the unpolarized counterpart of Eq.~(\ref{dPqg2cf2nf}) is written down in the
same notation, the first two lines are the same except for the replacement of
$\Delta p_{\rm qg}$ by $p_{\rm qg} \,=\, 2\,D_2^{} - 2\,D_1^{} + D_0^{}$.
The same holds for the $\ca \cf \nf$ and $\cas \nf$ contributions. 
As in other results in massless perturbative QCD, the number of harmonic sums 
is reduced by the absence of sums with index $-1$. This leaves seven sums of 
weight 3, of which one is missing in Eq.~(\ref{dPqg2cf2nf}) but not the
corresponding $\ca \cf \nf$ and $\cas \nf$ expressions. Half of their in 
principle 28 coefficients with $D_{0,1}^{}$ and $D_{0,1}^{\,2}$ are fixed by 
the $1/N^{\,2}$ suppression of the difference $\delta_{\rm qg}^{\,(2)}$ in
Eq.~(\ref{delta}), which is found to hold separately for each harmonic sum.
Taking into account the lower-weight sums, this large-$N$ behaviour relates 
as many as 24 coefficients to the unpolarized result for each of the three 
non-$\nf$ colour factors.

Another crucial feature of Eq.~(\ref{dPqg2cf2nf}) and all other available 
results for splitting functions is that all coefficients are integer in a 
suitable normalization. 
E.g., after eliminating all terms linear in $D_0^{}$ and $D_1^{}$ using the 
$1/N^3$ large-$N$ behaviour, the remaining coefficients in 
Eq.~(\ref{dPqg2cf2nf}) are integers once factors of $2^{\,w-3}$ have been 
bracketed out of the terms with sums of weight $w < 3$.
Consequently the equations relating the remaining coefficients to fixed-$N$ 
moments are Diophantine equations, and far less that $n$ equations are required 
to determine $n$ unknown coefficients.
While there are a few additional constraints, on the coefficient of the
$D_{0,1}^{\,5}$ and $D_1^{\,4\,}$ terms corresponding to the $\ln^{\,5} x$ and 
$x \ln^{\,5,\,4} x$ small-$x$ logarithms and the remaining coefficients of 
$\Sss(1,1,1)(N)$, see below, it is clear that it is vital for the determination
of $\,\Delta P_{\rm gg}^{\,(2)}(N)$ to have an extension of the $A$-scheme of 
Fig.~\ref{dPij1MAU} to NNLO, in order not to miss out on those 24 large-$N$ 
constraints.

\begin{figure}[tbh]
\centerline{\hspace*{-1mm}\epsfig{file=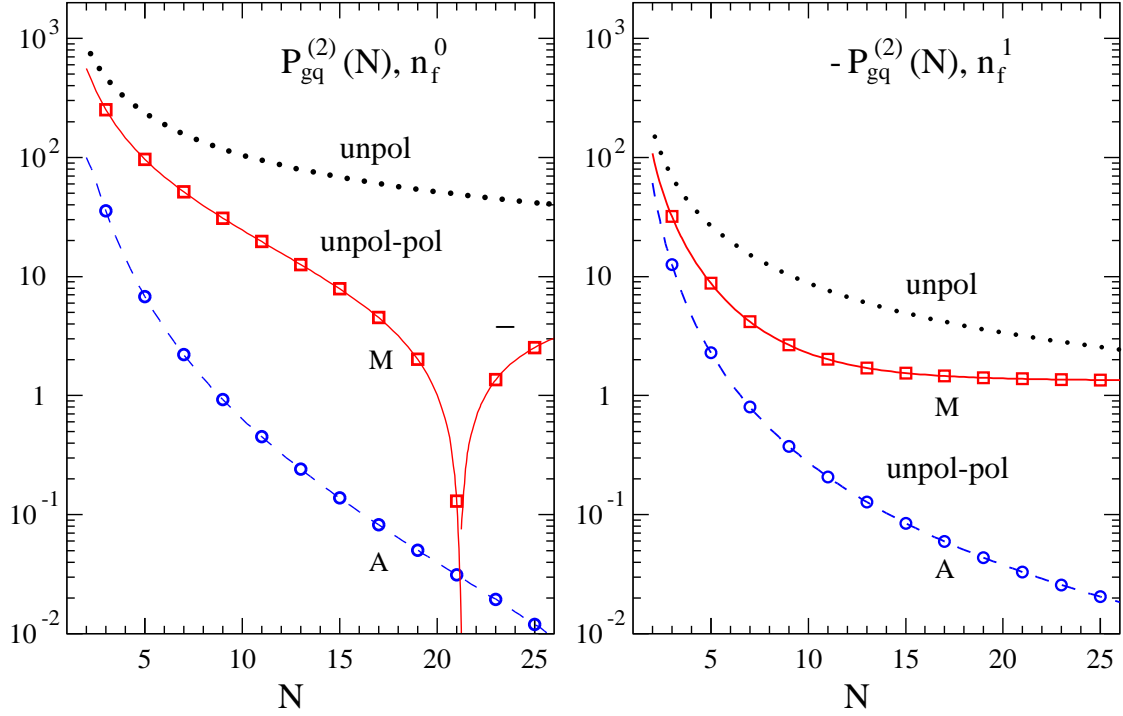,width=15.4cm}}
\vspace{-2.5mm}
\caption{\label{dPgq2mom}
The NNLO differences $\delta_{\rm gq}^{\,(2)}(N) \,=\, P^{\,(2)}_{\rm gq}(N) - 
\Delta P^{\,(2)}_{\rm gq}(N)$ for the non-$\nf$ and $n_{\! f}^{\:1}$ terms in 
the M and A schemes for $\ca = 3$ and $\cf = 4/3$, compared to the unpolarized 
result. The symbols show moments calculated using {\sc Mincer}, the solid 
and dashed lines the exact all-$N$ results presented below. As at NLO, 
cf.~Fig.~\ref{dPij1MAU}, the M-scheme difference turns negative at large $N$.
}
\vspace*{-1.5mm}
\end{figure}

The double-logarithmic $\Sss(1,1,1)$ and $\Sss(1,1,1,1)$ contributions
to $\dPgq(2)(N)$ can be derived from the calculations of polarized graviton-%
exchange DIS, without any reference to the unpolarized results, from the
single-log threshold enhancement of the physical kernel $K_{\rm p}$ in 
Eq.~(\ref{PhysK}), cf.~Ref.~\cite{SMVV}. An additional scheme transformation 
that removes those contributions to $\delta_{\rm gq}$ is found to be
\beq
\label{ZgqA}
  Z_{\,{\rm gq},\,A} \equal 
     - \: \ars\: \Delta P^{\,(0)}_{\,\rm gq} 
  \: - \: \frct{1}{2}\: \ar(2)\: \Delta P^{\,(1)}_{\,{\rm gq},\,L}
  \: + \: {\cal O}(\ar(3))
\:\: .
\eeq
The assumption that this remarkably simple transformation leads to 
$\delta_{\rm gq}(x) = {\cal O}(\x1^2)$ is consistent with the results for 
$N \leq 25$ is illustrated in Fig.~4 for the $n_{\! f}^{\,0}$ and 
$n_{\! f}^{\,1}$ contributions in QCD.

The physical kernels for the system $(H_{\bar{4}},\,H_6)$ also allow to
settle another issue observed in Ref.~\cite{mvvLL08}, the apparent partial
disagreement of the leading small-$x$ logarithm of $\dPqg(2)(x)$ with the old
resummation result of Ref.~\cite{BVpol}: the $\ln^{\,4}x$ contribution to
$K_{\bar{4}6}$ agrees perfectly with that prediction, which clarifies its 
proper interpretation, see also Refs.~\cite{Bartels:1996wc,KL83,BVns}.
Consequently it should be possible to use the prediction of Ref.~\cite{BVpol}, 
via $K_{6\bar{4}\,}$, also for $\dPgq(2)(x)$. Furthermore the $\,x\ln^{\,5}x$ 
and $x\ln^{\,4}x$ terms of this function can be fixed by extending the analysis 
of the small-$x$ limits of the unfactorized expressions in Ref.~\cite{AV11xto0} 
to the present case, see also Ref.~\cite{avLL12}.

Finally we need to briefly address the issue of denominators other than 
$D_0^{}$ and $D_1^{}$, as occurring in the sixth line of Eq.~(\ref{dPqg2cf2nf}),
and with sums to weight 3 in its $\ca \cf \nf$ counterpart. Due to the 
different leading-order structure, there are far fewer such terms here than in 
the unpolarized case. 
Terms with $D_2^{}$ in $\dPqg(2)(N)$, $D_{-1}^{}$ in $\dPgq(2)(N)$ and 
$D_{-1}^{}\, D_2^{}$ in $\dPgg(2)(N)$ do neither affect the prime-number 
decomposition of the denominators of the odd-$N$ moments, e.g., the $N=17$ 
moments do not involve a factor $1/19$, cf.~Eq.~(3.1), nor can they lead to 
an overall pole at $N=1$.

We are now ready to turn to the determination of the all-$N$ expressions.
The structure of the critical $\cft$, $\ca \cfs$ and $\cas\, \cf$ parts of 
$\dPgq(2)$ is analogous to Eq.~(\ref{dPqg2cf2nf}) discussed in detail above. 
With the coefficients of the weight-4 sums fixed by the unpolarized result 
\cite{mvvPsg}, we are left with 2$\times$32 coefficients of sums at weight 3 
and below combined with powers of $D_0^{}$ and $D_1^{}$, recall 
Eq.~(\ref{Ddef}), plus at most 11~sums combined with $D_{-1}^{}$.
The large-$N$ suppression of $\delta_{\rm gq}^{\,(2)}$ in the \mbox{A-scheme}
and the other endpoint constraints fix 29 or 30 of these coefficients 
(depending whether or not $D_{-1}^{}\,\Sss(1,1,1)(N)$ is included in the basis 
set), leaving up to 45 unknown integer parameters.

We have developed {\sc Form} tools for analyzing the prime-number structure
of the moments, see Eq.~(3.1), and deriving relations between the remaining 
parameters using the Chinese remainder theorem~\cite{ChinRem}. 
These tools have proved sufficient, sometimes together with a brute-force scan 
of a few variables, for simpler cases. It~is however not easy
to derive more than about ten relations for the three difficult 
$n_{\! f}^{\:0}$ parts of $\dPgq(2)$.
For these cases we have employed the program provided in Ref.~\cite{axbWeb},
see also Refs.~\cite{LLL,axbAlg} to solve the remaining system of linear
Diophantine equations.
Since this program looks for short vectors, it is best for our purposes to
eliminate 4 to 6 `unpleasant' coefficients, in particular those of low-weight 
combinations such as $D_{0}^{\,2}$, $D_{1}^{\,2}$, $D_{0}^{\,2} S_1$, 
$D_{1}^{\,2} S_1$, using the moments to $N = 9\,$ or $N = 13\,$, and work with 
the remaining 6 to 8 equations. 
 
For example, using the moments (3.1) this procedure leads to the result 
\bea
\label{dPgq2cf3}
  && \hspn
  \frct{1}{8}\left.\Delta P^{\,(2)}_{\,\rm gq}(N) \right|_{\cft}
  \: = \;
          2\, \* \colour4colour{ \dpgq } \* (
          - \, \S(-4)
          + 6\, \* \Ss(-2,-2)
          + 4\, \* \Ss(1,-3)
          + 2\, \* \Ssss(1,1,1,1)
          + \Sss(1,1,2)
\quad \nn \\[-1.5mm] && \hspace*{4.0cm} \mbox{}
          + 3\, \* \Sss(1,2,1)
          - 3\, \* \Ss(1,3)
          + 2\, \* \Ss(2,-2)
          + 2\, \* \Sss(2,1,1)
          - 2\, \* \Ss(2,2)
          )
\nn \\[2mm] && \mbox{}
        - 4 \, \* \colour4colour{ \S(-3) }\, \*  (
            2\, \* \DNn2
          - \DNnO
          + \DNpO
          )
  \: - \: 8 \, \* \colour4colour{ \Ss(1,-2) }\, \*  (
            \DNp2
          - 2\, \* \DNnO
          + 2\, \* \DNpO
          )
\nn \\[1mm] && \mbox{}
        + \colour4colour{ \Sss(1,1,1) }\, \*  (
            2\, \* \DNn2
          - 5\, \* \DNp2
          - 6\, \* \DNnO
          - 3/2\, \* \DNpO
          )
  \: - \: 2\, \* \colour4colour{ \Ss(1,2) }\, \*  (
            \DNp2
          + 4\, \* \DNnO
          - \DNpO
          )
\nn \\[1mm] && \mbox{}
        - \colour4colour{ \Ss(2,1) }\, \*  (
            4\, \* \DNn2
          + 4\, \* \DNp2
          - 4\, \* \DNnO
          + 7\, \* \DNpO
          )
   \: + \: \colour4colour{ \S(3) }\,  \*  (
            2\, \* \DNn2
          + \DNp2
          + 6\, \* \DNnO
          - 3/2\, \* \DNpO
          )
\nn \\[1mm] && \mbox{}
       +  6\, \* \colour4colour{ \z3 } \* ( 2\,\* \DNnO - \DNpO ) \* 
         ( 2\, \* \S(1) - 3 )
  \: - \: 6 \, \* \colour4colour{ \DNmO }\, \*  ( \S(-2) + 1 )
\\[2.5mm] && \mbox{}
        - \colour4colour{ \S(-2) }\, \*  (
            8\, \* \DNp3
          + 4\, \* \DNn2
          + 18\, \* \DNp2
          - 26\, \* \DNnO
          + 24\, \* \DNpO
          )
  \: - \: \colour4colour{ \Ss(1,1) }\, \*  (
            6\, \* \DNn3
          + 6\, \* \DNp3
          + 4\, \* \DNn2
\nn \\[1mm] && \hspp \mbox{}
          + 5\, \* \DNp2
          + 2\, \* \DNnO
          - 7/4\, \* \DNpO
          )
   \: + \: 2 \, \* \colour4colour{ \S(2) }\, \*  (
            \DNp3
          + 2\, \* \DNp2
          + 10\, \* \DNnO
          - 4\, \* \DNpO
          )
\nn \\[2.5mm] && \mbox{}
        - \colour4colour{ \S(1) }\, \*  (
            6\, \* \DNn4
          + 7\, \* \DNp4
          + 4\, \* \DNn3
          + 23/2\, \* \DNp3
          - 27/2\, \* \DNn2
          + 39/4\, \* \DNp2
          - 8\, \* \DNnO
          + 23/4\, \* \DNpO
          )
\nn \\[1mm] && \mbox{}
          - 8\, \* \DNn5
          - 12\, \* \DNp5
          + 23\, \* \DNn4
          - 28\, \* \DNp4
          - 39/4\, \* \DNn3
          - 427/8\, \* \DNp3
          - 341/8\, \* \DNn2
\nn \\[1mm] && \mbox{}
          - 767/8\, \* \DNp2
          + 2427/16\, \* \DNnO
          - 4547/32\, \* \DNpO
\nn
\eea
in the standard (M) definition of the \MSb\ scheme \cite{MSvN98}, where we have
again used the abbreviations (\ref{Ddef}) and (\ref{dpij0}) and suppressed the 
argument $N$ of the harmonic~sums.
The corresponding expressions for the $\ca \cfs$ and $\cas\, \cf$ parts are
somewhat longer, see below. The $\nf$-dependent terms are much shorter; 
their determination does not require the $N=23$ and $N=25$ moments.

Note the simplicity of the coefficients in Eq.~(\ref{dPgq2cf3}), 
in particular those of the terms with overall weights of 5 and 4 and sums of 
weight 2 or higher, which strongly indicates that the result is correct even 
without further checks.
In fact, if any erroneous information is entered for an externally fixed 
parameters, e.g., a wrong coefficient of $D_1^{\,5}$, or if the set of 
functions is too small, e.g., by omitting the term with $D_{-1}$,
then either no solution exists for the system of Diophantine equations, or only
solutions with nonsensically large coefficients (also) for the high-weight
terms.

Nevertheless it is, of course, necessary to validate the resulting all-$N$ 
formulae. For this purpose their predictions at higher values of $N$ have been
compared to additional {\sc Mincer} moments such~as
{\small
\bea
  - \Delta P^{\,(2)}_{\,{\rm gq},\,L}(N\!=\!27) &\,=\,&
  4609770383587605432813291530849726335264810727/ 
\quad \nn \\[-0.5mm]
& &
  \left( 23^{\,4}\, 19^{\,4}\, 17^{\,4}\, 13^{\,4}\, 11^{\,4}\, 7^{\,5}\, 
  5^{\,8}\, 3^{\,15}\, 2^{\,13} \right) \: \* \cft \; + \;\ldots
\eea
}
The diagram calculations for the corresponding result at $N=29$ have been
carried out only in the planar limit $\,\ca\! - 2\,\cf \ra 0\,$ at $\nf = 0$.
As this result combines the three difficult all-$N$ expressions for the 
$\cft$, $\ca \cfs$ and $\cas\,\cf$ colour factors, which have been obtained
independently from each other, it provides another strong check of all these 
results including Eq.~(\ref{dPgq2cf3}). 
Perfect agreement is found for the not entirely trivial fractions at both
values of $N$.

The overall most difficult case was the $\nf$-independent, i.e., $\cat$ 
part of $\dPgg(2)$. Also here the harmonic sums beyond weight 3 can be 
determined from the unpolarized case; the same holds for all terms not 
involving any un-summed denominators: these contribute to either the $1/\x1_+$ 
of the $\delta \x1$ terms the large-$x$ limit which are the same for 
$P_{\rm gg}$ and $\Delta P_{\rm gg}$. 
This reduces the problem to the same basis set as in the case of $\dPgq(2)$ at
$\nf=0$. 
The $1/N^{\,2}$ suppression of $\delta_{\,gg}$ with respect to $P_{\rm gg}$, 
however, only removes one instead two coefficients for each harmonic sum up to 
weight 3.
 
Taking into account our additional knowledge of the coefficients of $D_0^{\,5}$
from Ref.~\cite{BVpol} (this coefficient is the same for $K_{66}^{\,(2)}$ and 
$\dPgg(2)$, unlike for the off-diagonal cases),
$D_1^{\,5}$ and $D_1^{\,4}$, cf.~Refs.\ \mbox{\cite{AV11xto0,avLL12}}, 
and of $\Sss(1,1,1)$, cf.~Ref.~\cite{SMVV}, 
this leaves 49 terms with $D_0^{}$ and $D_1^{}$ plus the functions with the
`extra' denominator $D_{-1}\,D_2$ corresponding to $D_{-1}$ in the previous
case of $\dPgq(2)$.
The non-$C_F$ parts of $\Delta P_{\rm gg}$ are non-singlet like quantities, 
e.g., they are not affected by scheme transformations with $z_{\rm gg} = 0$, 
see Eqs.~(\ref{P1trf}) and (\ref{P2trf}). Hence we could use some non-singlet 
heuristics, see Ref.~\cite{Veliz}, to reduce the overall basis to 52 functions,
which we were able to determine using our own programs and,
in the final step, Ref.~\cite{axbWeb} with 8 equations at $11 \leq N \leq 25$ 
for 41 unknowns. 

Quite a few of the resulting coefficients are far less simple than those in 
Eq.~(\ref{dPgq2cf3}), see Eq.~(\ref{dPgg2N}) below; on the other hand seven 
coefficients put in are zero, and there are some expected relations. 
The result has been checked against the {\sc Mincer} calculations at $N=27$ 
and $N=29$ which were finished only after we had obtained $\dPgg(2)(N)$. 
Another important check is the first moment which is not accessible directly
\cite{LamLi}, but can be obtained by Mellin-inverting to $x$-space expressions
in terms of harmonic polylogarithms \cite{Hpols} from which arbitrary moments
can be calculated. The results is
\beq
\label{dPgg2N1}
  \dPgg(2)(N\!=\!1) \equal \beta_{\,2}^{\:\mbox{\scriptsize \MSb }}
\eeq
see Eq.~(\ref{b012}), as expected from the two previous orders.
This result is the same in all factorization schemes considered here also for 
the $\,C_F$ terms due to $\,\dPqg(n)(N\!=\!1) \,=\, 0\,$ in Eq.~(\ref{P2trf}),
cf.~Ref.~\cite{AltLam90}.

%
\setcounter{equation}{0}
\section{The NNLO splitting functions in Mellin space}
\label{sec:Nres}
%
%
The analytical odd-$N$ expressions of the splitting functions to NNLO can be 
written in terms of harmonic sums \cite{Hsums} as recalled in 
Eqs.~(\ref{Hsum1}) and (\ref{Hsum2}) above. Our notation is different from 
section~3 of Refs.~\cite{mvvPns,mvvPsg}: here all sums are taken at 
argument $N$ (which we usually suppress), for~the additional un-summed
denominators we employ the abbreviations (\ref{Ddef}), (\ref{dpij0}) and 
\beq
\label{EtaNu}
  \eta \:\:\equiv \:\: \{ N (N+1) \}^{-1}
\:\: , \quad
  \nu \:\:\equiv \:\: \{ (N-1)(N+2) \}^{-1}
\:\: .
\eeq
In this notation the leading-order (LO) contributions 
\cite{Altarelli:1977zs,Sasaki:1975hk,Ahmed:1976ee} to Eq.~(\ref{Sevol}), see
also Eq.~(\ref{DPqq}), read 
\bea
\label{dPij0N}
  \Delta P^{\,(0)}_{\,\rm ns}(N) & \! = \! &  
         \colour4colour{\cf}  \* \left(
          - 4\,\* \S(1) 
          + 2\,\* \eta
          + 3
          \right)
\:\: , \nn \\[0.5mm]
  \dPps(0)(N) & \! = \! & 0  
\:\: , \nn \\[0.5mm]
  \dPqg(0)(N) & \! = \! &  
  2\, \* \colour4colour{\nf}\,  \* \dpqg
\:\: , \nn \\[1mm]
  \dPgq(0)(N) & \! = \! &  
  2\, \* \colour4colour{\cf}\, \* \dpgq 
\:\: , \nn \\[1mm]
  \dPgg(0)(N) & \! = \! &  
         \colour4colour{\ca}  \* \left(
          - 4\, \* \S(1) 
          + 8\, \* \eta 
          + 11/3
          \right)
          - 2/3 \: \* \colour4colour{\nf}
\:\: ,  
\eea
and their next-to-leading order (NLO) counterparts of 
Refs.~\cite{MvN95,WVdP1a,WVdP1b} are given by 
\bea
  \Delta P^{\,+(1)}_{\,\rm ns}(N) \!\! &\! =\! &
  4\, \* \colour4colour{\cfs}\, \*  \Big(
       - 4\,\* ( \S(-3)
       - 2\, \* \Ss(1,-2)
       - \Ss(1,2)
       - \Ss(2,1) )
       - 3\, \* \S(2)
       + 3/8
       - 4\, \* \eta\, \* \S(-2)
\nn\\[-1.5mm] && \mbox{}
       - 2\, \* \eta\, \* \S(2) 
       + 2\, \* (
            2\, \* \eta
          + \etaD2
          - 2\, \* \DNn2
       )\, \* \S(1)
       - \eta
       - 11\, \* \etaD2
       - 5\, \* \etaD3
       + \DNn2
       + 2\, \* \DNn3
  \Big)
\nn\\ && \mbox{\hspn}
   + 4\, \* \colour4colour{\ca \* \cf}\, \*  \Big(
         2\, \* ( \S(-3) - \S(3) ) 
       - 4\, \* \S(1,-2)
       + 11/3\: \* \S(2)
       - 67/9\: \* \S(1)
       + 17/24
\nn\\[-1.5mm] && \mbox{}
       + 2\, \* \eta\, \* \S(-2)   
       + 217/18\: \* \eta
       + 35/6\: \* \etaD2
       + 2\, \* \etaD3
       - 11/3\: \* \DNn2
  \Big)
\nn\\ && \mbox{\hspn}
   + 4/9\: \* \colour4colour{\cf \* \nf}\, \*  \Big(
       - 6\, \* \S(2)
       + 10\, \* \S(1)
       - 3/4
       - 17\, \* \eta
       - 3\, \* \etaD2
       + 6\, \* \DNn2
  \Big)
\:\: , \\[1mm]
\label{dPps1N}
  \dPps(1)(N) &\! =\! &
  4\, \* \colour4colour{\cf \* \nf}\:  \*  \Big(
          - 5\, \* \eta
          + 3\, \* \etaD2
          + 2\, \* \etaD3
          + 4\, \* \DNn2
          - 4\, \* \DNn3
          \Big)
\:\: , \\[1mm]
\label{dPqg1N}
  \dPqg(1)(N) &\! =\! &
     4\, \* \colour4colour{\cf \* \nf}\, \*  \Big( 
          2\, \* \dpqg \,\*  (
            \Ss(1,1)
          - \S(2)
          )
       - 2\, \* (
            2\, \* \DNnO
          - \DNn2
          - 2\, \* \DNpO
          )\: \* \S(1)
\nn\\[-1.5mm] && \mbox{} 
       - 11\, \* \DNnO
          + 9/2\: \* \DNn2
          - \DNn3
          + 27/2\: \* \DNpO
          + 4\, \* \DNp2
          - 2\, \* \DNp3
       \Big)
\nn\\ && \mbox{\hspn}
   + 4\, \* \colour4colour{\ca \* \nf} \*  \Big( 
          - 2\, \* \dpqg \,\*  (
            \S(-2)
          + \Ss(1,1)
          )
       + 4\, \* (
            \DNnO
          - \DNpO
          - \DNp2
          )\: \* \S(1)
\nn\\[-1.5mm] && \mbox{}
       + 12\, \* \DNnO
          - \DNn2
          - 2\, \* \DNn3
          - 11\, \* \DNpO
          - 12\, \* \DNp2
          - 12\, \* \DNp3
       \Big)
\:\: , \\[1mm]
\label{dPgq1N}
  \dPgq(1)(N) &\! =\! &
   4\, \* \colour4colour{\cfs}\, \*  \Big( 
          - \dpgq\, \* (
             2\, \* \Ss(1,1)
           - \S(1)
           )
       + 2\, \* (
            \DNpO
          + \DNp2
          )\: \* \S(1)
\nn\\[-1.5mm] && \mbox{}
       - 17/2\: \* \DNnO 
          + 2\, \* \DNn2
          + 2\, \* \DNn3
          + 4\, \* \DNpO
          + 1/2\: \* \DNp2
          + \DNp3
       \Big)
\nn\\ && \mbox{\hspn}
   + 4\, \* \colour4colour{\ca \* \cf}\, \*  \Big(
          2\, \* \dpgq\, \*  (
            \Ss(1,1)
          - \S(-2)
          - \S(2)
          )
       - (
            10/3\: \* \DNnO
          + 4\, \* \DNn2
          + 1/3\: \* \DNpO
          )\: \* \S(1)
\nn\\[-1mm] && \mbox{}
       + 41/9\: \* \DNnO
          - 4\, \* \DNn2
          + 4\, \* \DNn3
          + 35/9\: \* \DNpO
          + 38/3\: \* \DNp2
          + 6\, \* \DNp3
       \Big)
\nn\\ && \mbox{\hspn}
   + 8/9\: \* \colour4colour{\cf \* \nf}\, \*  \Big(
         3\, \*  \dpgq\, \* \S(1)
       - 4\, \* \DNnO
          - \DNpO
          - 3\, \* \DNp2
       \Big)
\:\: , \\[1mm]
\label{dPgg1N}
  \dPgg(1)(N) &\! =\! &
  4\, \* \colour4colour{\cas}\, \*  \Big(
         4\, \* (  
            \Ss(1,-2)
          + \Ss(1,2)
          + \Ss(2,1) )
       - 2\, \* ( 
            \S(3)
          + \S(-3) )
       - 67/9\: \* \S(1)
          + 8/3
\nn\\[-0.5mm] && \mbox{}
       - 8\, \* \eta \* (
           \S(2)
         + \S(-2) 
          )
       + 8\, \* (
            2\, \* \eta
          + \etaD2
          - 2\, \* \DNn2
          )\, \* \S(1)
\nn\\ && \mbox{}
          + 901/18\: \* \eta
          - 149/3\: \* \etaD2
          - 24\, \* \etaD3
          - 32\, \* \DNn2
          + 32\, \* \DNn3     
          \Big)
\nn\\ && \mbox{\hspn}
       + 4/3\: \* \colour4colour{\ca \* \nf}\, \*  \Big(
         10/3\: \* \S(1)
          - 2
          - 26/3\: \* \eta
          + 2\, \* \etaD2
          \Big)
\nn\\ &&  \mbox{\hspn}
       + 4\, \* \colour4colour{\cf \* \nf}\, \*  \Big(
          - 1/2
          - 7\, \* \eta
          + 5\, \* \etaD2
          + 2\, \* \etaD3
          + 6\, \* \DNn2
          - 4\, \* \DNn3
          \Big)
\:\: .  
\eea

For completeness also including the non-singlet contribution, which is 
identical to the function $P_{\,\rm ns}^{\,-(2)}(N)$ given (in a different
notation) already in Eq.~(3.8) of Ref.~\cite{mvvPns}, the polarized 
next-to-next-to-leading (NNLO) quark-quark splitting function $\dPqq(2)(N)$ 
is the sum of
\bea
\label{dPns2N}
  && \hspn \Delta P^{\,+(2)}_{\,\rm ns}(N) \equal 
    16 \, \* \colour4colour{\cft}  \, \*  \Big(
       - 12 \, \* \S(-5)
       + 24 \, \* \Ss(-4,1)
       + 4 \, \* \Ss(-3,-2)
       + 4 \, \* \Ss(-3,2)
       + 12 \, \* \Ss(-2,-3)
\nn\\[-0.5mm] && \mbox{}
       - 24 \, \* \Sss(-2,1,-2)
       + 44 \, \* \Ss(1,-4)
       - 64 \, \* \Sss(1,-3,1)
       + 16 \, \* \Sss(1,-2,-2)
       - 8 \, \* \Sss(1,-2,2)
       - 80 \, \* \Sss(1,1,-3)
\nn\\[0.5mm] && \mbox{}
       + 96 \, \* \Ssss(1,1,-2,1)
       - 16 \, \* \Sss(1,2,-2)
       - 8 \, \* \Sss(1,2,2)
       - 16 \, \* \Sss(1,3,1)
       - 8 \, \* \Ss(1,4)
       + 52 \, \* \Ss(2,-3)
       - 56 \, \* \Sss(2,-2,1)
\nn\\[0.5mm] && \mbox{}
       - 16 \, \* \Sss(2,1,-2)
       - 8 \, \* \Sss(2,1,2)
       - 8 \, \* \Sss(2,2,1)
       + 4 \, \* \Ss(2,3)
       + 12 \, \* \Ss(3,-2)
       - 8 \, \* \Sss(3,1,1)
       + 8 \, \* \Ss(3,2)
       + 4 \, \* \Ss(4,1)
\nn\\[0.5mm] && \mbox{}
       + 4 \, \* \S(5)
       - \S(-4)\, \* (9 + 22 \, \* \eta) 
       - \Ss(-3,1) \, \* (6 - 32 \, \* \eta) 
       + 4 \, \* \eta \, \* 
       ( \Ss(-2,2) 
         - 2 \, \* \Ss(-2,-2) 
         + 2 \, \* \Ss(3,1)
         + \S(4) 
       )
\nn\\[0.5mm] && \mbox{}
       + 2 \, \* \Ss(1,-3)\, \* (3 + 20 \, \* \eta) 
       + 4 \, \* \Sss(1,-2,1) \, \* (3 - 12 \, \* \eta) 
       + 12 \, \* \Ss(1,3)
       + 2 \, \* \Ss(2,-2) \, \* (3 + 4 \, \* \eta) 
\nn\\[0.5mm] && \mbox{}
       + 2 \, \* \Ss(2,2) \, \* (3 + 2 \, \* \eta) 
       + \S(-3) \, \* (
         + 3 \, \* \eta
         - 4 \, \* \etaD2 
         - 12 \, \* \DNn2
         ) 
       + 2 \, \* \Ss(-2,1) \, \* (
           5 \, \* \eta
         + 10 \, \* \etaD2 
         + 4 \, \* \DNn2
         ) 
\nn\\[0.5mm] && \mbox{}
       - 4 \, \* \Ss(1,-2) \, \* (
           7 \, \* \eta 
         + 5 \, \* \etaD2 
         - 4 \, \* \DNn2
         )
       - 4 \, \* (\Ss(1,2)+\Ss(2,1)) \, \* (
           2 \, \* \eta 
         + \etaD2 
         - 2 \, \* \DNn2
         ) 
\nn\\[0.5mm] && \mbox{}
       - \S(3) \, \* (13/4 
         - 4 \, \* \eta 
         - 5 \, \* \etaD2 
         + 4 \, \* \DNn2) 
       - \S(2) \, \* (3/8 
         - 2 \, \* \eta 
         - 17/2 \, \* \etaD2 
         - 4 \, \* \etaD3 
         + 2 \, \* \DNn2 
         + 4 \, \* \DNn3) 
\nn\\[0.5mm] && \mbox{}
       - \S(-2) \, \* (3 
         - 12 \, \* \z3 
         + 2 \, \* \eta 
         - 14 \, \* \etaD2 
         - 6 \, \* \etaD3 
         - 2 \, \* \DNn2 
         + 8 \, \* \DNn3) 
       + 4 \, \* \Ss(1,1) \, \* (3 \, \* \etaD2 + \etaD3) 
\nn\\[0.5mm] && \mbox{}
       - \S(1) \, \* (
           47/2 \, \* \eta 
         + 53/4 \, \* \etaD2 
         + 48 \, \* \etaD3 
         + 13 \, \* \etaD4 
         - 18 \, \* \DNn2 
         + 18 \, \* \DNn3 
         - 24 \, \* \DNn4) 
\nn\\[0.5mm] && \mbox{}
       + \z3 \, \* ( 15/2 
         + 6 \, \* \eta
         + 6 \, \* \etaD2 
       )
       + 29/32 
       - 215/8 \, \* \eta 
       + 26 \, \* \etaD2 
       + 45 \, \* \etaD3 
       + 49 \, \* \etaD4 
       + 11 \, \* \etaD5 
\nn\\[-0.5mm] && \mbox{}
       + 175/8 \, \* \DNn2
       - 43/2 \, \* \DNn3 
       + 15/2 \, \* \DNn4 
       - 16 \, \* \DNn5
       \Big)
\nn\\ && \mbox{\hspn}
  + 8 \, \* \colour4colour{\cfs \, \* \ca}  \, \*  \Big(
         20 \, \* \S(-5)
       - 40 \, \* \Ss(-4,1)
       + 4 \, \* \Ss(-3,-2)
       - 4 \, \* \Ss(-3,2)
       - 20 \, \* \Ss(-2,-3)
       - 16 \, \* \Sss(-2,-2,1)
\nn\\[-0.5mm] && \mbox{}
       + 56 \, \* \Sss(-2,1,-2)
       - 68 \, \* \Ss(1,-4)
       + 128 \, \* \Sss(1,-3,1)
       - 64 \, \* \Sss(1,-2,-2)
       + 8 \, \* \Sss(1,-2,2)
       + 144 \, \* \Sss(1,1,-3)
\nn\\[0.5mm] && \mbox{}
       - 224 \, \* \Ssss(1,1,-2,1)
       - 32 \, \* \Sss(1,1,3)
       + 16 \, \* \Sss(1,2,-2)
       + 32 \, \* \Sss(1,3,1)
       + 44 \, \* \Ss(1,4)
       - 84 \, \* \Ss(2,-3)
       + 120 \, \* \Sss(2,-2,1)
\nn\\[0.5mm] && \mbox{}
       + 16 \, \* \Sss(2,1,-2)
       + 20 \, \* \Ss(2,3)
       - 20 \, \* \Ss(3,-2)
       + 4 \, \* \Ss(3,2)
       + 4 \, \* \Ss(4,1)
       - 20 \, \* \S(5)
       + (89/3 + 34 \, \* \eta) \, \* \S(-4)
\nn\\[0.5mm] && \mbox{}
       + 268/9 \, \* (\Ss(1,2) + \Ss(2,1) + 2 \, \* \Ss(1,-2) - \S(-3))
       + 2 \, \* \Ss(-3,1) \, \* (31/3 - 32 \, \* \eta) 
       + 4 \, \* \Ss(-2,-2) \, \* (3 + 8 \, \* \eta) 
\nn\\[0.5mm] && \mbox{}
       - 4 \, \* \eta \, \* \Ss(-2,2)
       - 2 \, \* \Ss(1,-3) \, \* (31/3 + 36 \, \* \eta) 
       - 4 \, \* \Sss(1,-2,1) \, \* (31/3 - 28 \, \* \eta) 
       - 4 \, \* \Ss(1,3) \, \* (31/3 - 4 \, \* \eta) 
\nn\\[0.5mm] && \mbox{}
       - 2 \, \* \Ss(2,-2) \, \* (31/3 + 4 \, \* \eta) 
       - 44/3 \, \* \Ss(2,2)
       - 8 \, \* \Ss(3,1) \, \* (1/3 + 2 \, \* \eta) 
       - \S(4) \, \* (23/3 + 22 \, \* \eta) 
\nn\\[0.5mm] && \mbox{}
       + \S(-3) \, \* (
           37/3 \, \* \eta 
         + 14 \, \* \etaD2 
         + 12 \, \* \DNn2 ) 
       - 2 \, \* \Ss(-2,1) \, \* (
           53/3 \, \* \eta 
         + 30 \, \* \etaD2 
         + 4 \, \* \DNn2 ) 
\nn\\[0.5mm] && \mbox{}
       + 4 \, \* \Ss(1,-2) \, \* (
           13 \, \* \eta 
         + 10 \, \* \etaD2 
         - 4 \, \* \DNn2) 
       + \S(3) \, \* (13 
         + 26/3 \, \* \eta 
         - 10 \, \* \etaD2 
         - 4 \, \* \DNn2 ) 
\nn\\[0.5mm] && \mbox{}
       + \S(-2) \, \* (9 
         - 36 \, \* \z3 
         - 586/9 \, \* \eta 
         - 34/3 \, \* \etaD2 
         + 38/3 \, \* \DNn2 
         + 8 \, \* \DNn3 ) 
\nn\\[0.5mm] && \mbox{}
       - \S(2) \, \* (151/12 
         + 350/9 \, \* \eta 
         + 46/3 \, \* \etaD2 
         + 4 \, \* \etaD3 
         - 44/3 \, \* \DNn2 )
\nn\\[0.5mm] && \mbox{}
       + \S(1) \, \* (
           715/9 \, \* \eta 
         + 494/9 \, \* \etaD2 
         + 137/3 \, \* \etaD3 
         + 8 \, \* \etaD4 
         - 580/9 \, \* \DNn2 
         + 16 \, \* \DNn3 
         - 24 \, \* \DNn4) 
\nn\\[0.5mm] && \mbox{}
       - \z3 \, \* ( 45/2 
         + 18 \, \* \eta 
         + 18 \, \* \etaD2 )
       + 151/32 
       - 4 \, \* \etaD5 
       - 341/6 \, \* \etaD4 
       - 1805/9 \, \* \etaD3 
\nn\\[-0.5mm] && \mbox{}
       - 3691/18 \, \* \etaD2 
       - 5/18 \, \* \eta 
       + 217/36 \, \* \DNn2 
       + 185/9 \, \* \DNn3 
       + 38 \, \* \DNn4 
       + 16 \, \* \DNn5
       \Big)
\nn\\ && \mbox{\hspn}
  + 8 \, \* \colour4colour{\cf \, \* \cas}  \, \*  \Big(
       - 4 \, \* \S(-5)
       + 8 \, \* \Ss(-4,1)
       - 4 \, \* \Ss(-3,-2)
       + 4 \, \* \Ss(-2,-3)
       + 8 \, \* \Sss(-2,-2,1)
       - 16 \, \* \Sss(-2,1,-2)
\nn\\[-0,5mm] && \mbox{}
       + 12 \, \* \Ss(1,-4)
       - 32 \, \* \Sss(1,-3,1)
       + 24 \, \* \Sss(1,-2,-2)
       - 32 \, \* \Sss(1,1,-3)
       + 64 \, \* \Ssss(1,1,-2,1)
       + 16 \, \* \Sss(1,1,3)
       - 16 \, \* \Sss(1,3,1)
\nn\\[0.5mm] && \mbox{}
       - 12 \, \* \Ss(1,4)
       + 16 \, \* \Ss(2,-3)
       - 32 \, \* \Sss(2,-2,1)
       - 8 \, \* \Ss(2,3)
       + 4 \, \* \Ss(3,-2)
       + 4 \, \* \Ss(4,1)
       + 4 \, \* \S(5)
       -  \S(-4) \, \* (31/3 + 6 \, \* \eta) 
\nn\\[0.5mm] && \mbox{}
       - (11/3 - 8 \, \* \eta) \, \* (2 \, \* \Ss(-3,1) + \Ss(3,1) 
         - 4 \, \* \Sss(1,-2,1))
       - 6 \, \* \Ss(-2,-2) \, \*  (1 + 2 \, \* \eta) 
       + 2 \, \* \Ss(1,-3) \, \* (11/3 + 8 \, \* \eta) 
\qquad \nn\\[0.5mm] && \mbox{}
       + \Ss(1,3) \, \* (11 - 8 \, \* \eta) 
       + 22/3 \, \* \Ss(2,-2)
       + \S(4) \, \* (31/3 + 6 \, \* \eta) 
       + \S(-3) \, \* (134/9 
         - 23/3 \, \* \eta
         - 5 \, \* \etaD2 ) 
\nn\\[0.5mm] && \mbox{}
       - 4  \, \* \Ss(1,-2) \, \* (67/9 
         + 3 \, \* \eta 
         + 5/2 \, \* \etaD2 ) 
       - 1/2 \, \* \S(3) \, \* (389/9 
         + \eta
         - 9 \, \* \etaD2 ) 
       + 1043/54 \, \* \S(2)
\nn\\[0.5mm] && \mbox{}
       + \Ss(-2,1) \, \* (
           38/3 \, \* \eta
         + 20 \, \* \etaD2 ) 
       - \S(-2) \, \* (3 
         - 12 \, \* \z3 
         - 302/9 \, \* \eta 
         + 4/3 \, \* \etaD2 
         + 3 \, \* \etaD3
         + 22/3 \, \* \DNn2 ) 
\nn\\[0.5mm] && \mbox{}
       - \S(1) \, \* (245/12 
         + 6 \, \* \eta
         + 7/6 \, \* \etaD2 
         + 11/6 \, \* \etaD3 
         - 1/2 \, \* \etaD4 ) 
       + \z3 \, \* ( 15/2 + 6 \, \* \eta + 6 \, \* \etaD2 )
\nn\\[0.5mm] && \mbox{}
       - 1657/288 
       + 20521/216 \, \* \eta 
       + 4819/54 \, \* \etaD2 
       + 261/4 \, \* \etaD3 
       + 11/3 \, \* \etaD4 
       - 3 \, \* \etaD5 
\nn\\[-0.5mm] && \mbox{}
       - 2759/54 \, \* \DNn2 
       + 44/3 \, \* \DNn3 
       - 22 \, \* \DNn4
       \Big)
\nn\\ && \mbox{\hspn}
  + 8/3 \, \* \colour4colour{\cfs \, \* \nf}  \, \*  \Big(
       - 8 \, \* \S(-4)
       - 8 \, \* \Ss(-3,1)
       + 8 \, \* \Ss(1,-3)
       + 16 \, \* \Sss(1,-2,1)
       + 16 \, \* \Ss(1,3)
       + 8 \, \* \Ss(2,-2)
       + 8 \, \* \Ss(2,2)
\nn\\[-0.5mm] && \mbox{}
       + 8 \, \* \Ss(3,1)
       - 4 \, \* \S(4)
       - 80/3 \, \* \Ss(1,-2)
       - 40/3 \, \* \Ss(1,2)
       - 40/3 \, \* \Ss(2,1)
       + 4 \, \* \S(-3) \, \* (10/3 - \eta) 
       - 8 \, \* \eta \, \* \Ss(-2,1)
\nn\\[0.5mm] && \mbox{}
       - \S(3) \, \* (6 + 8 \, \* \eta) 
       + 4 \, \* \S(-2) \, \* (
           22/3 \, \* \eta
         + \etaD2 
         - 2 \, \* \DNn2 ) 
       + \S(2) \, \* (5/2 
         + 56/3 \, \* \eta 
         + 4 \, \* \etaD2
         - 8 \, \* \DNn2 ) 
\nn\\[0.5mm] && \mbox{}
       + \S(1) \, \* (55/4 
         - 64/3 \, \* \eta
         - 92/3 \, \* \etaD2 
         - 8 \, \* \etaD3
         - 12 \, \* \z3 
         + 64/3 \, \* \DNn2 ) 
       + \z3 \, \* ( 9 + + 6 \, \* \eta )
\nn\\[-0.5mm] && \mbox{}
       - 69/8 
       + 83/24 \, \* \eta 
       + 457/6 \, \* \etaD2 
       + 278/3 \, \* \etaD3 
       + 19 \, \* \etaD4 
       - 71/6 \, \* \DNn2 
       + 10/3 \, \* \DNn3 
       - 24 \, \* \DNn4
       \Big)
\nn\\ && \mbox{\hspn}
  + 8/3 \, \* \colour4colour{\cf \, \* \ca \, \* \nf}  \, \*  \Big(
         4 \, \* \S(-4)
       + 4 \, \* \Ss(-3,1)
       - 4 \, \* \Ss(1,-3)
       - 8 \, \* \Sss(1,-2,1)
       - 6 \, \* \Ss(1,3)
       - 4 \, \* \Ss(2,-2)
       + 2 \, \* \Ss(3,1)
       - 4 \, \* \S(4)
\nn\\[-0.5mm] && \mbox{}
       + 4 \, \* \eta \, \* \Ss(-2,1)
       + 40/3 \, \* \S(1,-2)
       - 2 \, \* \S(-3) \, \* (10/3 - \eta) 
       + \S(3) \, \* (14 + 3 \, \* \eta) 
       - 167/9 \, \* \S(2)
\nn\\[0.5mm] && \mbox{}
       - 2 \, \* \S(-2) \, \* ( 
           22/3 \, \* \eta 
         + \etaD2 
         - 2 \, \* \DNn2 ) 
       + \S(1) \, \* (209/18 
         + 2 \, \* \etaD2 
         + \etaD3 
         + 12 \, \* \z3 ) 
       - \z3 \, \* ( 9 + 6 \, \* \eta )
\nn\\[-0.5mm] && \mbox{}
       + 15/2 
       - 943/12 \, \* \eta 
       - 953/18 \, \* \etaD2 
       - 121/3 \, \* \etaD3 
       - 8 \, \* \etaD4 
        + 389/9 \, \* \DNn2 
        - 8 \, \* \DNn3 
        + 12 \, \* \DNn4
       \Big)
\nn\\ && \mbox{\hspn}
  + 8/9 \, \* \colour4colour{\cf \, \* \nfs}  \, \*  \Big(
         2/3 \, \* \S(1)
       + 10/3 \, \* \S(2)
       - 2 \, \* \S(3)
       - 17/8 
       + 34/3 \, \* \eta 
       + 20/3 \, \* \etaD2 
       + \etaD3 
       - 22/3 \, \* \DNn2
       \Big)
%
\eea
and
\bea
\label{dPps2N}
  && \hspn \dPps(2)(N) \equal 
   8 \, \* \colour4colour{\ca \, \* \cf \, \* \nf}  \, \*  \Big(
       - \S(-3)  \, \*  (
            5 \, \* \eta
          - 6 \, \* \etaD2
          )
       - ( \Ss(1,-2) - \Sss(1,1,1) + \Ss(1,2) - 3 \, \* \z3 )  \, \*  (
            2 \, \* \eta
          - 4 \, \* \etaD2
          )
\qquad \nn\\[-0.5mm] && \mbox{}
       + 4 \, \* \eta \, \* \Ss(-2,1)
       - \S(3)  \, \*  (
            5/2 \: \* \eta
          - 7 \, \* \etaD2
          )
       + \S(-2)  \, \*  (
            21 \, \* \eta
          - 13 \, \* \etaD2
          - 14 \, \* \etaD3
          - 20 \, \* \DNn2
          + 16 \, \* \DNn3
          )
\nn\\[0.5mm] && \mbox{}
       - \Ss(1,1)  \, \*  (
            11/6 \: \* \eta
          + 1/3 \: \* \etaD2
          - 2 \, \* \etaD3
          )
       + \S(2)  \, \*  (
            5 \, \* \eta
          - \etaD2
          - 4 \, \* \etaD3
          - 5 \, \* \DNn2
          + 4 \, \* \DNn3
          )
       + \S(1)  \, \*  (
            203/9 \: \* \eta
\nn\\[0.5mm] && \mbox{}
          - 115/9 \: \* \etaD2
          - 3/2 \, \* \etaD3
          - \etaD4
          - 41/3 \: \* \DNn2
          + 34/3 \: \* \DNn3
          + 2 \, \* \DNn4
          )
        + 1268/27 \, \* \eta
          - 107/54 \: \* \etaD2
\nn\\[-0.5mm] && \mbox{}
          + 93 \, \* \etaD3
          - 283/3 \: \* \etaD4
          - 38 \, \* \etaD5
          - 575/9 \: \* \DNn2
          + 1367/18 \, \* \DNn3
          - 83 \, \* \DNn4
          + 32 \, \* \DNn5
       \Big)
\nn\\ && \mbox{\hspn}
  + 8 \, \* \colour4colour{\cfs \, \* \nf}  \, \*  \Big(
       - ( 2 \, \* \Sss(1,1,1) - 2 \, \* \Ss(1,2) - \S(3) + 6\, \* \z3 )\, \* (
              \eta
          - 2 \, \* \etaD2
          )
       + \Ss(1,1)  \, \*  (
            3/2 \: \* \eta
          - 2 \, \* \etaD2
          - 2 \, \* \etaD3
          )
\nn\\[-0.5mm] && \mbox{}
       + 2\, \* \S(2)  \, \*  (
            7 \, \* \eta
          - 4 \, \* \etaD2
          - 2 \, \* \etaD3
          - 6 \, \* \DNn2
          + 6 \, \* \DNn3
          )
       - 2\, \* \S(1)  \, \*  (
            45/4 \: \* \eta
          - 3 \, \* \etaD2
          - 21 \, \* \etaD3
          - 7 \, \* \etaD4
          - 6 \, \* \DNn2
\nn\\[-0.5mm] && \mbox{}
          + 3 \, \* \DNn3
          + \DNn4
          )
       + 5 \, \* \eta
          + 3 \, \* \etaD2
          - 75/2 \: \* \etaD3
          - 39 \, \* \etaD4
          - 8 \, \* \etaD5
          + 7 \, \* \DNn2
          - 29/2 \: \* \DNn3
          + 9 \, \* \DNn4
       \Big)
\nn\\ && \mbox{\hspn}
  + 8/3 \, \* \colour4colour{\cf \, \* \nfs}  \, \*  \Big(
         \Ss(1,1)  \, \*  (
            \eta
          - 2 \, \* \etaD2
          )
       - \S(1)  \, \*  (
            44/3 \: \* \eta
          - 31/3 \: \* \etaD2
          - 6 \, \* \etaD3
          - 11 \, \* \DNn2
          + 10 \, \* \DNn3
          )
\nn\\[-1mm] && \mbox{}
       + 160/9 \: \* \eta
          - 53/9 \: \* \etaD2
          - 30 \, \* \etaD3
          - 8 \, \* \etaD4
          - 34/3 \: \* \DNn2
          + 17/3 \: \* \DNn3
          + 6 \, \* \DNn4
       \Big)
\:\: .
\eea
In $N$-space the off-diagonal NNLO entries of the matrix (\ref{Sevol}) are 
given by
\bea
 \label{dPqg2N}
   && \hspn \dPqg(2)(N) \equal
     8 \, \* \colour4colour{\cfs \, \* \nf} \, \*  \Big(
          2 \, \* \dpqg \, \*  (
          - \S(-4)
          + 2 \, \* \Ss(-2,-2)
          + 4 \, \* \Ss(1,-3)
          + 2 \, \* \Ssss(1,1,1,1)
          - \Sss(1,1,2)
          - 5 \, \* \Sss(1,2,1)
\nn\\[-0.5mm] && \mbox{}
          + 4 \, \* \Ss(1,3)
          + 2 \, \* \Ss(2,-2)
          - 6 \, \* \Sss(2,1,1)
          + 6 \, \* \Ss(2,2)
          + 7 \, \* \Ss(3,1)
          - 3 \, \* \S(4)
          )
       - 4 \, \* \S(-3)  \, \*  (
            2 \, \* \DNnO
          - \DNn2
          - 2 \, \* \DNpO
          )
\nn\\[0.5mm] && \mbox{}
       - 8 \, \* \Ss(1,-2)  \, \*  (
            \DNnO
          - \DNpO
          - 2 \, \* \DNp2
          )
       - \Sss(1,1,1)  \, \*  (
            21/2 \, \* \DNnO
          - 5 \, \* \DNn2
          - 12 \, \* \DNpO 
          + 2 \, \* \DNp2
          )
\nn\\[0.5mm] && \mbox{}
       + 2 \, \* \Ss(1,2)  \, \*  (
            5 \, \* \DNnO
          - 2 \, \* \DNn2
          - 5 \, \* \DNpO
          + 2 \, \* \DNp2
          )
       + 2 \, \* \Ss(2,1)  \, \*  (
            11 \, \* \DNnO
          - 4 \, \* \DNn2
          - 11 \, \* \DNpO
          - 2 \, \* \DNp2
          )
\nn\\[0.5mm] && \mbox{}
       - 2 \, \* \S(3)  \, \*  (
            11 \, \* \DNnO
          - 3 \, \* \DNn2
          - 11 \, \* \DNpO
          - 6 \, \* \DNp2
          )
       - 6 \, \* \DNppO  \, \*  ( \S(-2) + 1 )
\nn\\[0.5mm] && \mbox{}
       + 2 \, \* \S(-2)  \, \*  (
            10 \, \* \DNnO
          - 5 \, \* \DNn2
          - 9 \, \* \DNpO
          - 6 \, \* \DNp2
          + 8 \, \* \DNp3
          )
       - \Ss(1,1)  \, \*  (
            29 \, \* \DNnO
          - 35/2 \, \* \DNn2
          + 10 \, \* \DNn3
\nn\\[0.5mm] && \mbox{}
          - 36 \, \* \DNpO
          - 5 \, \* \DNp2
          + 6 \, \* \DNp3
          )
       + 2 \, \* \S(2)  \, \*  (
            17 \, \* \DNnO
          - 10 \, \* \DNn2
          + 4 \, \* \DNn3
          - 22 \, \* \DNpO
          - 4 \, \* \DNp2
          + 6 \, \* \DNp3
          )
\nn\\[0.5mm] && \mbox{}  
       + \S(1)  \, \*  (
          - 78 \, \* \DNnO
          + 99/2 \, \* \DNn2
          - 43/2 \, \* \DNn3
          + 7 \, \* \DNn4
          + 329/4 \, \* \DNpO
          + 18 \, \* \DNp2
          - 15 \, \* \DNp3
          + 4 \, \* \DNp4
          )
\nn\\[0.5mm] && \mbox{}
       + 3 \, \* \z3  \, \*  (
            9 \, \* \DNnO
          - 2 \, \* \DNn2
          - 12 \, \* \DNpO
          - 4 \, \* \DNp2
          )
       - 1357/32 \, \* \DNnO
          + 77/8 \, \* \DNn2
          + 59/8 \, \* \DNn3
\nn\\[-0.5mm] && \mbox{}
          - 15/2 \, \* \DNn4
          + 777/16 \, \* \DNpO
          + 213/8 \, \* \DNp2
          + 53/4 \, \* \DNp3
          - 3 \, \* \DNp4
          + 32 \, \* \DNp5
       \Big)
\nn\\ && \mbox{\hspn}
   + 8 \, \* \colour4colour{\ca \, \* \cf \, \* \nf} \, \*  \Big(
         2 \, \* \dpqg \, \*  (
          - 11/2 \, \* \S(-4)
          + 6 \, \* \Ss(-3,1)
          - 3 \, \* \Ss(-2,-2)
          - 2 \, \* \Sss(-2,1,1)
          + 2 \, \* \Ss(-2,2)
          + 6 \, \* \Ss(1,-3)
\nn\\[-0.5mm] && \mbox{}
          - 6 \, \* \Sss(1,-2,1)
          - 6 \, \* \Sss(1,1,-2)
          - 4 \, \* \Ssss(1,1,1,1)
          - 3 \, \* \Sss(1,1,2)
          + 3 \, \* \Sss(1,2,1)
          + \Ss(1,3)
          + 3 \, \* \Ss(2,-2)
          + 6 \, \* \Sss(2,1,1)
\nn\\[0.5mm] && \mbox{}
          - 6 \, \* \Ss(3,1)
          + 3/2 \, \* \S(4)
          + 3 \, \* \z3 \, \* \S(1)
          )
       - 3 \, \* \DNppO \, \*  (
            2 \, \* \S(-3)
          - 2 \, \* \Ss(-2,1)
          - 2 \, \* \S(1)
          - 2 \, \* \Ss(1,-2)
          - \S(-2)
          - 1
          )
\nn\\[0.5mm] && \mbox{}
       - \S(-3)  \, \*  (
            15 \, \* \DNnO
          - 6 \, \* \DNn2
          - 18 \, \* \DNpO
          - 8 \, \* \DNp2
          )
       + 2 \, \* \Ss(-2,1)  \, \*  (
            5 \, \* \DNnO
          - 2 \, \* \DNn2
          - 8 \, \* \DNpO
          )
\nn\\[0.5mm] && \mbox{}
       + \Sss(1,1,1)  \, \*  (
            37/3 \, \* \DNnO
          - \DNn2
          - 47/3 \, \* \DNpO
          - 2 \, \* \DNp2
          )
       + 8 \, \* \Ss(1,-2)  \, \*  (
            13/4 \, \* \DNnO
          - \DNn2
          - 4 \, \* \DNpO
          - 2 \, \* \DNp2
          )
\nn\\[0.5mm] && \mbox{}
       + 3 \, \* \Ss(1,2)  \, \*  (
            11/2 \, \* \DNnO
          - 4 \, \* \DNn2
          - 5 \, \* \DNpO
          - 4 \, \* \DNp2
          )
       - 3 \, \* \Ss(2,1)  \, \*  (
            11/2 \, \* \DNnO
          - 5 \, \* \DNpO
          - 4 \, \* \DNp2
          )
\nn\\[0.5mm] && \mbox{}
       + \S(3)  \, \*  (
            61/3 \, \* \DNnO
          - \DNn2
          - 59/3 \, \* \DNpO
          - 18 \, \* \DNp2
          )
       + \S(-2)  \, \*  (
            8 \, \* \DNnO
          - 2 \, \* \DNn2
          + 2 \, \* \DNn3
          - 11 \, \* \DNpO
          - 4 \, \* \DNp3
          )
\nn\\[0.5mm] && \mbox{}
       + \Ss(1,1)  \, \*  (
            317/9 \, \* \DNnO
          - 41/6 \, \* \DNn2
          - 6 \, \* \DNn3
          - 313/9 \, \* \DNpO
          - 31 \, \* \DNp2
          - 2 \, \* \DNp3
          )
\nn\\[0.5mm] && \mbox{}
       + 2 \, \* \S(2)  \, \*  (
            17/18 \, \* \DNnO
          - 5 \, \* \DNn2
          + 6 \, \* \DNn3
          - 23/9 \, \* \DNpO
          + 10 \, \* \DNp2
          + 6 \, \* \DNp3
          )
       + \S(1)  \, \*  (
            1195/27 \, \* \DNnO
\nn\\[0.5mm] && \mbox{}
          - 29/9 \, \* \DNn2
          - 11 \, \* \DNn3
          + 8 \, \* \DNn4
          - 1595/27 \, \* \DNpO
          - 67/2 \, \* \DNp2
          + 3 \, \* \DNp3
          + 34 \, \* \DNp4
          )
\nn\\[0.5mm] && \mbox{}
       - 6 \, \* \z3  \, \*  (
            18 \, \* \DNnO
          - 5 \, \* \DNn2
          - 21 \, \* \DNpO
          - 10 \, \* \DNp2
          )
       + 69407/288 \, \* \DNnO
          - 15259/216 \, \* \DNn2
\nn\\[0.5mm] && \mbox{}
          - 701/72 \, \* \DNn3
          + 89/6 \, \* \DNn4
          - 4 \, \* \DNn5
          - 34927/144 \, \* \DNpO
          - 36461/216 \, \* \DNp2
          - 3359/36 \, \* \DNp3
\nn\\[-0.5mm] && \mbox{}
          - 1/3 \, \* \DNp4
          + 8 \, \* \DNp5
       \Big)
\nn\\ && \mbox{\hspn}
   + 8 \, \* \colour4colour{\cas \, \* \nf} \, \*  \Big(
          2 \, \* \dpqg \, \*  (
          - 3/2 \, \* \S(-4)
          + 2 \, \* \Ss(-3,1)
          + 3 \, \* \Ss(-2,-2)
          + 2 \, \* \Sss(-2,1,1)
          + 2 \, \* \Ss(1,-3)
          - 2 \, \* \Sss(1,-2,1)
          - \Ss(1,3)
\nn\\[-0.5mm] && \mbox{}
          + 6 \, \* \Sss(1,1,-2)
          + 2 \, \* \Ssss(1,1,1,1)
          + 4 \, \* \Sss(1,1,2)
          + 2 \, \* \Sss(1,2,1)
          - \Ss(2,-2)
          - 2 \, \* \Ss(2,2)
          + 3 \, \* \Ss(3,1)
          - 5/2 \, \* \S(4)
          - 3 \, \* \z3 \, \* \S(1)
          )
\nn\\[0.5mm] && \mbox{}
       - \S(-3)  \, \*  (
            104/3 \, \* \DNnO
          - 13 \, \* \DNn2
          - 115/3 \, \* \DNpO
          - 14 \, \* \DNp2
          )
       + 4 \, \* \Ss(-2,1)  \, \*  (
            2 \, \* \DNnO
          - \DNn2
          - 2 \, \* \DNpO
          )
\nn\\[0.5mm] && \mbox{}
       - 6 \, \* \Ss(1,-2)  \, \*  (
            7 \, \* \DNnO
          - 5/3 \, \* \DNn2
          - 7 \, \* \DNpO
          - 6 \, \* \DNp2
          )
       - \Sss(1,1,1)  \, \*  (
            11/6 \, \* \DNnO
          + 4 \, \* \DNn2
          - 11/3 \, \* \DNpO
          - 4 \, \* \DNp2
          )
\nn\\[0.5mm] && \mbox{}
       - \Ss(2,1)  \, \*  (
            35/6 \, \* \DNnO
          - 23/3 \, \* \DNpO
          - 8 \, \* \DNp2
          )
       - \S(3)  \, \*  (
            106/3 \, \* \DNnO
          - 25/2 \, \* \DNn2
          - 223/6 \, \* \DNpO
          - 17 \, \* \DNp2
          )
\nn\\[0.5mm] && \mbox{}
       - \Ss(1,2)  \, \*  (
            157/6 \, \* \DNnO
          - 8 \, \* \DNn2
          - 73/3 \, \* \DNpO
          - 24 \, \* \DNp2
          )
       + 3 \, \* \z3  \, \*  (
            27 \, \* \DNnO
          - 8 \, \* \DNn2
          - 30 \, \* \DNpO
          - 16 \, \* \DNp2
          )
\nn\\[0.5mm] && \mbox{}
      - 3 \, \* \DNppO \, \*  ( \S(-2) + 1 )
       - \S(-2)  \, \*  (
            776/9 \, \* \DNnO
          - 21 \, \* \DNn2
          - \DNn3
          - 709/9 \, \* \DNpO
          - 69 \, \* \DNp2
          - 62 \, \* \DNp3
          )
\nn\\[0.5mm] && \mbox{}
       - 2 \, \* \Ss(1,1)  \, \*  (
            1/9 \, \* \DNnO
          + 7 \, \* \DNn2
          - 4 \, \* \DNn3
          + 65/18 \, \* \DNpO
          - 71/6 \, \* \DNp2
          - 12 \, \* \DNp3
          )
\nn\\[0.5mm] && \mbox{}
       - \S(2)  \, \*  (
            36 \, \* \DNnO
          - 12 \, \* \DNn2
          - 35 \, \* \DNpO
          - 61/3 \, \* \DNp2
          - 16 \, \* \DNp3
          )
       + \S(1)  \, \*  (
            2515/54 \, \* \DNnO
          - 91/2 \, \* \DNn2
\nn\\[0.5mm] && \mbox{}
          + 35/2 \, \* \DNn3
          + 9/2 \, \* \DNn4
          - 4555/108 \, \* \DNpO
          - 59/9 \, \* \DNp2
          + 233/6 \, \* \DNp3
          + 49 \, \* \DNp4
          )
\nn\\[0.5mm] && \mbox{}
       - 16099/36 \, \* \DNnO
          + 2867/27 \, \* \DNn2
          - 75/2 \, \* \DNn3
          + 82/3 \, \* \DNn4
          - 15 \, \* \DNn5
          + 8227/18 \, \* \DNpO
\nn\\[-0.5mm] && \mbox{}
          + 8941/27 \, \* \DNp2
          + 2143/9 \, \* \DNp3
          + 691/3 \, \* \DNp4
          + 158 \, \* \DNp5
       \Big)
\nn\\ && \mbox{\hspn}
   + 8/9 \, \* \colour4colour{\cf \, \* \nfs} \, \*  \Big(
          3\, \* \dpqg  \, \*  (
            \Sss(1,1,1)
          - 2 \, \* \S(3)
          )
       + \Ss(1,1)  \, \*  (
            4 \, \* \DNnO
          + 3 \, \* \DNn2
          - 14 \, \* \DNpO
          )
       - \S(2)\, \*  (
            11 \, \* \DNnO
          - 16 \, \* \DNpO
          )
\nn\\[-0.5mm] && \mbox{}
       + \S(1)  \, \*  (
            14/3 \, \* \DNnO
          - 4 \, \* \DNn2
          + 19/6 \, \* \DNpO
          )
       + 4193/16 \, \* \DNnO
          - 3217/12 \, \* \DNn2
          + 901/4 \, \* \DNn3
\nn\\[-0.5mm] && \mbox{}
          - 129 \, \* \DNn4
          + 36 \, \* \DNn5
          - 2113/8 \, \* \DNpO
          + 97/12 \, \* \DNp2
          + 151/2 \, \* \DNp3
          - 42 \, \* \DNp4
          - 72 \, \* \DNp5
       \Big)
\nn\\ && \mbox{\hspn}
   + 8/9 \, \* \colour4colour{\ca \, \* \nfs} \, \*  \Big(
          3\, \* \dpqg  \, \*  (
          - 2 \, \* \S(-3)
          - \Sss(1,1,1)
          + \Ss(1,2)
          - \Ss(2,1)
          - \S(3)
          )
       - 2 \, \* \S(-2)  \, \*  (
            2\, \* \DNnO
          - 7 \, \* \DNpO
          )
\\[-0.5mm] && \mbox{}
       - 2 \, \* \Ss(1,1)  \, \*  (
            2 \, \* \DNnO
          - 7 \, \* \DNpO
          + 3 \, \* \DNp2
          )
       + 6  \, \* \DNp2 \, \* \S(2)
       - \S(1)  \, \*  (
            23/3 \, \* \DNnO
          - 4/3 \, \* \DNpO
          - 17 \, \* \DNp2
          + 12 \, \* \DNp3
          )
\nn\\[-0.5mm] && \mbox{}
       + 118 \, \* \DNnO
          - 1067/12 \, \* \DNn2
          + 99/2 \, \* \DNn3
          - 527/4 \, \* \DNpO
          - 46/3 \, \* \DNp2
          + 65 \, \* \DNp3
          - 12 \, \* ( \DNn4 + \DNp4 )
       \Big)
\;\; , \qquad
\nn \\[1mm] 
%
 \label{dPgq2N}
   && \hspn \dPgq(2)(N) \equal
   8 \, \* \colour4colour{\cft} \, \*  \Big( 
          2 \, \* \dpgq \, \*  (
          - \S(-4)
          + 6 \, \* \Ss(-2,-2)
          + 4 \, \* \Ss(1,-3)
          + 2 \, \* \Ssss(1,1,1,1)
          + \Sss(1,1,2)
          + 3 \, \* \Sss(1,2,1)
\nn\\[-0.5mm] && \mbox{}
          - 3 \, \* \Ss(1,3)
          + 2 \, \* \Ss(2,-2)
          + 2 \, \* \Sss(2,1,1)
          - 2 \, \* \Ss(2,2)
          - 9 \, \* \z3
          + 6 \, \* \z3 \, \* \S(1)
          )
       + 4 \, \* \S(-3)  \, \*  (
            \DNnO
          - 2 \, \* \DNn2
          - \DNpO
          )
\nn\\[0.5mm] && \mbox{}
       + 8 \, \* \Ss(1,-2)  \, \*  (
            2 \, \* \DNnO
          - 2 \, \* \DNpO
          - \DNp2
          )
       - \Sss(1,1,1)  \, \*  (
            6 \, \* \DNnO
          - 2 \, \* \DNn2
          + 3/2 \, \* \DNpO
          + 5 \, \* \DNp2
          )
\nn\\[0.5mm] && \mbox{}
       - 2 \, \* \Ss(1,2)  \, \*  (
            4 \, \* \DNnO
          - \DNpO
          + \DNp2
          )
       + 4 \, \* \Ss(2,1)  \, \*  (
            \DNnO
          - \DNn2
          - 7/4 \, \* \DNpO
          - \DNp2
          )
       - 6 \, \* \DNmO \, \*  ( \S(-2) + 1 )
\nn\\[0.5mm] && \mbox{}
       + \S(3)  \, \*  (
            6 \, \* \DNnO
          + 2 \, \* \DNn2
          - 3/2 \, \* \DNpO
          + \DNp2
          )
       + \S(-2)  \, \*  (
            26 \, \* \DNnO
          - 4 \, \* \DNn2
          - 24 \, \* \DNpO
          - 18 \, \* \DNp2
          - 8 \, \* \DNp3
          )
\nn\\[0.5mm] && \mbox{}
       - \Ss(1,1)  \, \*  (
            2 \, \* \DNnO
          + 4 \, \* \DNn2
          + 6 \, \* \DNn3
          - 7/4 \, \* \DNpO
          + 5 \, \* \DNp2
          + 6 \, \* \DNp3
          )
       + 2 \, \* \S(2)  \, \*  (
            10 \, \* \DNnO
          - 4 \, \* \DNpO
          + 2 \, \* \DNp2
\nn\\[0.5mm] && \mbox{}
          + \DNp3
          )
       + \S(1)  \, \*  (
            8 \, \* \DNnO
          + 27/2 \, \* \DNn2
          - 4 \, \* \DNn3
          - 6 \, \* \DNn4
          - 23/4 \, \* \DNpO
          - 39/4 \, \* \DNp2
          - 23/2 \, \* \DNp3
          - 7 \, \* \DNp4
          )
\nn\\[0.5mm] && \mbox{}
       + 2427/16 \, \* \DNnO
          - 341/8 \, \* \DNn2
          - 39/4 \, \* \DNn3
          + 23 \, \* \DNn4
          - 8 \, \* \DNn5
          - 4547/32 \, \* \DNpO
          - 767/8 \, \* \DNp2
\nn\\[-0.5mm] && \mbox{}
          - 427/8 \, \* \DNp3
          - 28 \, \* \DNp4
          - 12 \, \* \DNp5
       \Big)
\nn\\ && \mbox{\hspn}
   + 8 \, \* \colour4colour{\ca \, \* \cfs} \, \*  \Big( 
          \dpgq \, \*  (
          - 3 \, \* \S(-4)
          - 10 \, \* \Ss(-2,-2)
          + 4 \, \* \Sss(-2,1,1)
          - 8 \, \* \Ss(1,-3)
          + 4 \, \* \Sss(1,-2,1)
         + 12 \, \* \Sss(1,1,-2)
\nn\\[-0.5mm] && \mbox{} 
          - 8 \, \* \Ssss(1,1,1,1)
          + 6 \, \* \Sss(1,1,2)
          + 2 \, \* \Sss(1,2,1)
          + 10 \, \* \Ss(1,3)
          - 6 \, \* \Ss(2,-2)
          + 4 \, \* \Sss(2,1,1)
          - 5 \, \* \S(4)
          - 18 \, \* \z3 \, \* \S(1)
          + 27 \, \* \z3
          )
\nn\\[0.5mm] && \mbox{}
       + 9 \, \* \DNmO \, \*  ( \S(-2) + 1 )
       + 2 \, \* \S(-3)  \, \*  (
            6 \, \* \DNnO
          - 3 \, \* \DNpO
          + \DNp2
          )
       - 2 \, \* \Ss(-2,1)  \, \*  (
            4 \, \* \DNnO
          - \DNpO
          + 2 \, \* \DNp2
          )
\nn\\[0.5mm] && \mbox{} 
       - 4 \, \* \Ss(1,-2)  \, \*  (
            7 \, \* \DNnO
          - 4 \, \* \DNpO
          )
       + \Sss(1,1,1)  \, \*  (
            73/3 \, \* \DNnO
          + 2 \, \* \DNn2
          - 23/3 \, \* \DNpO
          + \DNp2
          )
\nn\\[0.5mm] && \mbox{}
       + \Ss(1,2)  \, \*  (
            35/3 \, \* \DNnO
          + 4 \, \* \DNn2
          - 71/6 \, \* \DNpO
          - 8 \, \* \DNp2
          )
       - \Ss(2,1)  \, \*  (
            5/3 \, \* \DNnO
          - 8 \, \* \DNn2
          + 13/6 \, \* \DNpO
          + 6 \, \* \DNp2
          )
\nn\\[0.5mm] && \mbox{}
       - \S(3)  \, \*  (
            10 \, \* \DNnO
          + 16 \, \* \DNn2
          - 3 \, \* \DNpO
          - 4 \, \* \DNp2
          )
       + \S(-2)  \, \*  (
            6 \, \* \DNnO
          - 6 \, \* \DNn2
          - 4 \, \* \DNn3
          - \DNpO
          + 7 \, \* \DNp2
          + 2 \, \* \DNp3
          )
\nn\\[0.5mm] && \mbox{}
       - \Ss(1,1)  \, \*  (
            31/18 \, \* \DNnO
          - 8/3 \, \* \DNn2
          + 2 \, \* \DNn3
          + 137/9 \, \* \DNpO
          + 22 \, \* \DNp2
          + 14 \, \* \DNp3
          )
\nn\\[0.5mm] && \mbox{}
       - 4/3 \, \* \S(2)  \, \*  (
            10 \, \* \DNnO
          + 5/2 \, \* \DNn2
          + 6 \, \* \DNn3
          - 5/4 \, \* \DNpO
          + 67/8 \, \* \DNp2
          + 3 \, \* \DNp3
          )
\nn\\[0.5mm] && \mbox{}
       + \S(1)  \, \*  (
            293/54 \, \* \DNnO
          - 64/9 \, \* \DNn2
          + 8/3 \, \* \DNn3
          - 8 \, \* \DNn4
          + 613/108 \, \* \DNpO
          + \DNp2
          - 39/2 \, \* \DNp3
\nn\\[0.5mm] && \mbox{}
          - 24 \, \* \DNp4
          )
       - 3343/48 \, \* \DNnO
          + 11093/216 \, \* \DNn2
          + 365/36 \, \* \DNn3
          - 89/3 \, \* \DNn4
          + 16 \, \* \DNn5
\nn\\[-0.5mm] && \mbox{}
          + 11273/288 \, \* \DNpO
          - 3197/216 \, \* \DNp2
          - 701/72 \, \* \DNp3
          + 8/3 \, \* \DNp4
          - 8 \, \* \DNp5
       \Big)
\nn\\ && \mbox{\hspn}
   + 8 \, \* \colour4colour{\cas \, \* \cf} \, \*  \Big( 
          \dpgq  \, \*  (
          - 11 \, \* \S(-4)
          + 16 \, \* \Ss(-3,1)
          + 2 \, \* \Ss(-2,-2)
          - 4 \, \* \Sss(-2,1,1)
          + 4 \, \* \Ss(-2,2)
          + 24 \, \* \Ss(1,-3)
\nn\\[-0.5mm] && \mbox{}
          + 4 \, \* \Ss(1,3)
          - 20 \, \* \Sss(1,-2,1)
          - 12 \, \* \Sss(1,1,-2)
          + 4 \, \* \Ssss(1,1,1,1)
          - 8 \, \* \Sss(1,1,2)
          - 8 \, \* \Sss(1,2,1)
          + 10 \, \* \Ss(2,-2)
\nn\\[0.5mm] && \mbox{}
          - 8 \, \* \Sss(2,1,1)
          + 12 \, \* \Ss(2,2)
          + 8 \, \* \Ss(3,1)
          - 3 \, \* \S(4)
          - 9 \, \* \z3
          + 6 \, \* \z3 \, \* \S(1)
          )
       - 6 \, \* \DNmO \, \*  ( 
            \S(-3) 
          - \Ss(-2,1)
          - \S(1)
\nn\\[0.5mm] && \mbox{}
          - \Ss(1,-2)
          + 3/2 \, \* \S(-2)
          + 3/2
          )
       + 1/3\, \* \S(-3) \, \*  (
            133 \, \* \DNnO
          - 114 \, \* \DNn2
          - 137 \, \* \DNpO
          - 39 \, \* \DNp2
          )
\nn\\[0.5mm] && \mbox{}
       - 4/3 \, \* \Ss(-2,1)  \, \*  (
            10 \, \* \DNnO
          - 12 \, \* \DNn2
          - 11 \, \* \DNpO
          - 3 \, \* \DNp2
          )
       + 2/3 \, \* \Ss(1,-2)  \, \*  (
            53 \, \* \DNnO
          - 6 \, \* \DNn2
          - 40 \, \* \DNpO
\nn\\[0.5mm] && \mbox{}
          - 15 \, \* \DNp2
          )
       - \Sss(1,1,1)  \, \*  (
            55/3 \, \* \DNnO
          + 4 \, \* \DNn2
          - 55/6 \, \* \DNpO
          - 4 \, \* \DNp2
          )
       + 7/6 \, \* \S(3)  \, \*  (
            35 \, \* \DNnO
          - 18 \, \* \DNn2
\nn\\[0.5mm] && \mbox{}
          - 223/7 \, \* \DNpO
          - 9 \, \* \DNp2
          )
       + ( \Ss(1,2) + \Ss(2,1) ) \, \*  (
            7/3 \, \* \DNnO
          + 12 \, \* \DNn2
          + 41/6 \, \* \DNpO
          + 2 \, \* \DNp2
          )
\nn\\[0.5mm] && \mbox{}
       + \S(-2)  \, \*  (
            124/3 \, \* \DNnO
          - 3 \, \* \DNn2
          - 2 \, \* \DNn3
          - 173/3 \, \* \DNpO
          - 202/3 \, \* \DNp2
          - 31 \, \* \DNp3
          )
\nn\\[0.5mm] && \mbox{}
       + \Ss(1,1)  \, \*  (
            25/18 \, \* \DNnO
          + 7 \, \* \DNn2
          + 24 \, \* \DNn3
          + 581/36 \, \* \DNpO
          + 80/3 \, \* \DNp2
          + 12 \, \* \DNp3
          )
\nn\\[0.5mm] && \mbox{}
       + \S(2)  \, \*  (
            5/9 \, \* \DNnO
          + 38/3 \, \* \DNn2
          - 32 \, \* \DNn3
          - 148/9 \, \* \DNpO
          - 79/2 \, \* \DNp2
          - 18 \, \* \DNp3
          )
\nn\\[0.5mm] && \mbox{}
       - 1/3 \, \* \S(1)  \, \*  (
            883/9 \, \* \DNnO
          + 152/3 \, \* \DNn2
          - 29/2 \, \* \DNn3
          + 75 \, \* \DNn4
          - 403/18 \, \* \DNpO
          + 1/4 \, \* \DNp2
\nn\\[0.5mm] && \mbox{}
          + 65 \, \* \DNp3
          + 75/2 \, \* \DNp4
          )
       + 1913/6 \, \* \DNnO
          - 5513/54 \, \* \DNn2
          + 776/9 \, \* \DNn3
          - 47 \, \* \DNn4
\nn\\[-0.5mm] && \mbox{}
          + 30 \, \* \DNn5
          - 3349/12 \, \* \DNpO
          - 17843/108 \, \* \DNp2
          - 7373/36 \, \* \DNp3
          - 629/3 \, \* \DNp4
          - 79 \, \* \DNp5
       \Big)
\nn\\ && \mbox{\hspn}
   + 8/3 \, \* \colour4colour{\cfs \, \* \nf} \, \*  \Big( 
          \dpgq \, \*  (
          - 5 \, \* \Sss(1,1,1)
          - 4 \, \* \Ss(1,2)
          - 2 \, \* \Ss(2,1)
          + 3 \, \* \S(3)
          + 12 \, \* \z3
          )
       - 6 \, \* \DNmO \, \*  ( \S(-2) + 1 )
\nn\\[-0.5mm] && \mbox{}
       - 6 \, \* \S(-2)  \, \*  (
            4 \, \* \DNnO
          - 4 \, \* \DNn2
          - 5 \, \* \DNpO
          - 2 \, \* \DNp2
          )
       + \Ss(1,1)  \, \*  (
            41/3 \, \* \DNnO
          - 2 \, \* \DNn2
          - 4/3 \, \* \DNpO
          + 6 \, \* \DNp2
          )
\nn\\[0.5mm] && \mbox{}
       + \S(2)  \, \*  (
            4 \, \* \DNnO
          + 4 \, \* \DNn2
          + \DNpO
          + 2 \, \* \DNp2
          )
       - 1/9\, \* \S(1)  \, \*  (
            31 \, \* ( \DNnO + \DNpO )
          - 48 \, \* \DNn2
          - 36 \, \* \DNn3
          - 54 \, \* \DNp3
          )
\nn\\[0.5mm] && \mbox{}
       - 1685/8 \, \* \DNnO
          + 3371/36 \, \* \DNn2
          - 337/6 \, \* \DNn3
          + 50 \, \* \DNn4
          - 24 \, \* \DNn5
          + 10043/48 \, \* \DNpO
\nn\\[-0.5mm] && \mbox{}
          + 3769/36 \, \* \DNp2
          + 829/12 \, \* \DNp3
          + 46 \, \* \DNp4
          + 12 \, \* \DNp5
          \Big)
\nn\\ && \mbox{\hspn}
   + 8/3 \, \* \colour4colour{\ca \, \* \cf \, \* \nf} \, \*  \Big( 
          \dpgq \, \*  (
            4 \, \* \S(-3)
          - 4 \, \* \Ss(-2,1)
          - 8 \, \* \Ss(1,-2)
          + 5 \, \* \Sss(1,1,1)
          - 5 \, \* \Ss(1,2)
          - 5 \, \* \Ss(2,1) 
          - 2 \, \* \S(3)
          )
\nn\\[-0.5mm] && \mbox{}
       - 12\, \* \z3\, \* \dpgq
       + 6 \, \* \DNmO \, \*  ( \S(-2) + 1 )
       + 2 \, \* \S(-2)  \, \*  (
            10 \, \* \DNnO
          - 6 \, \* \DNn2
          - 8 \, \* \DNpO
          - \DNp2
          )
\nn\\[0.5mm] && \mbox{}
       - \Ss(1,1)  \, \*  (
            80/3 \, \* \DNnO
          + 6 \, \* \DNn2
          - 37/3 \, \* \DNpO
          - 4 \, \* \DNp2
          )
       + 2/3 \, \* \S(2)  \, \*  (
            5 \, \* \DNnO
          + 6 \, \* \DNn2
          + 5 \, \* \DNpO
          + 9/2 \, \* \DNp2
          )
\nn\\[0.5mm] && \mbox{}
       + \S(1)  \, \*  (
            91/9 \, \* \DNnO
          + 2/3 \, \* \DNn2
          + 8 \, \* \DNn3
          + 118/9 \, \* \DNpO
          + 55/2 \, \* \DNp2
          + 17 \, \* \DNp3
          )
       + 345/4 \, \* \DNnO
\nn\\[-0.5mm] && \mbox{}
          - 248/9 \, \* \DNn2
          - 41/3 \, \* \DNn3
          - 643/6 \, \* \DNpO
          - 2671/36 \, \* \DNp2
          - 59/6 \, \* \DNp3
          + 14 \, \* \DNp4
       \Big)
\nn\\ && \mbox{\hspn}
   + 8/9 \, \* \colour4colour{\cf \, \* \nfs} \, \*  \Big( 
          3 \, \* \dpgq \, \* \Ss(1,1)
       + \S(1)  \, \*  (
          - 4 \, \* \DNnO
          - \DNpO
          - 3 \, \* \DNp2
          )
       - 6 \, \* \DNnO
          + 5 \, \* \DNpO
          - \DNp2
          - 3 \, \* \DNp3
       \Big)
\:\: .
\eea
Finally the polarized third-order gluon-gluon splitting function reads
\bea
\label{dPgg2N}
  && \hspn \dPgg(2)(N) \equal
   16 \, \* \colour4colour{\cat}  \, \*  \Big(
          - 4 \, \* \S(-5)
          + 8 \, \* \Ss(-4,1)
          + 4 \, \* \Ss(-3,-2)
          + 2 \, \* \Ss(-3,2)
          + 4 \, \* \Ss(-2,-3)
          - 4 \, \* \Sss(-2,-2,1)
\nn\\[-0.5mm] && \mbox{}
          - 4 \, \* \Sss(-2,1,-2)
          + 16 \, \* \Ss(1,-4)
          - 16 \, \* \Sss(1,-3,1)
          - 4 \, \* \Sss(1,-2,-2)
          - 4 \, \* \Sss(1,-2,2)
          - 24 \, \* \Sss(1,1,-3)
\nn\\[0.5mm] && \mbox{}
          + 16 \, \* \Ssss(1,1,-2,1)
          - 8 \, \* \Sss(1,1,3)
          - 8 \, \* \Sss(1,2,-2)
          - 8 \, \* \Sss(1,2,2)
          - 8 \, \* \Sss(1,3,1)
          + 8 \, \* \Ss(1,4)
          + 18 \, \* \Ss(2,-3)
          - 12 \, \* \Sss(2,-2,1)
\nn\\[0.5mm] && \mbox{}
          - 8 \, \* \Sss(2,1,-2)
          - 8 \, \* \Sss(2,1,2)
          - 8 \, \* \Sss(2,2,1)
          + 10 \, \* \Ss(2,3)
          + 4 \, \* \Ss(3,-2)
          - 8 \, \* \Sss(3,1,1)
          + 10 \, \* \Ss(3,2)
          + 8 \, \* \Ss(4,1)
          - 4 \, \* \S(5) 
\nn\\[0.5mm] && \mbox{}
          + 11/6 \, \*  (
               2 \, \*  \Ss(-2,-2) 
             - \Ss(1,3) 
             - \Ss(3,1) 
             )
	  - 67/9 \, \*  ( 
               \S(-3) 
             + \S(3) 
             - 2 \, \* \Ss(1,-2) 
             - 2 \, \* \Ss(1,2)
             - 2 \, \* \Ss(2,1) 
             )
\nn\\[0.5mm] && \mbox{}
          + 1/6 \, \* \S(2) 
          - 245/24 \, \* \S(1) 
          + 79/32
       + 8 \, \* \eta  \, \*  (
          - 4 \, \* \S(-4)
          + 4 \, \* \Ss(-3,1)
          + \Ss(-2,-2)
          + \Ss(-2,2)
          + 6 \, \* \Ss(1,-3)
\nn\\[0.5mm] && \mbox{}
          - 4 \, \* \Sss(1,-2,1)
          + 2 \, \* \Ss(1,3)
          + 2 \, \* \Ss(2,-2)
          + 2 \, \* \Ss(2,2)
          + 2 \, \* \Ss(3,1)
          - 2 \, \* \S(4)
          )
       - 11 \, \* \nu  \, \*  ( 
            \S(-3) 
          - \Ss(-2,1) 
\nn\\[0.5mm] && \mbox{}
          - \Ss(1,-2) 
          + \S(-2)
          - \S(1) 
          + 1 
          )
       + \S(-3)  \, \*  (
            33 \, \* \eta
          - 16 \, \* \etaD2
          - 24 \, \* \DNn2
          )
       + \S(3)  \, \*  (
	    86/3 \, \* \eta
          - 6 \, \* \etaD2
          - 24 \, \* \DNn2
          )
\nn\\[0.5mm] && \mbox{}
       - \Ss(1,-2)  \, \*  (
            43 \, \* \eta
          + 32 \, \* \etaD2
          - 32 \, \* \DNn2
          )
       - ( \Ss(1,2) + \Ss(2,1) )  \, \*  (
            32 \, \* \eta
          + 16 \, \* \etaD2
          - 32 \, \* \DNn2
          )
\nn\\[0.5mm] && \mbox{}
       - \Ss(-2,1)  \, \*  (
            23 \, \* \eta
          - 16 \, \* \DNn2
          )
       - \S(-2)  \, \*  (
            802/9 \, \* \eta
          - 338/3 \, \* \etaD2
          - 60 \, \* \etaD3
          - 64 \, \* \DNn2
          + 64 \, \* \DNn3
          )
\nn\\[0.5mm] && \mbox{}
       + \Ss(1,1)  \, \*  (
            48 \, \* \etaD2
          + 16 \, \* \etaD3
          )
       + \S(2)  \, \*  (
	  - 1745/18 \, \* \eta
	  + 173/3 \, \* \etaD2
          + 32 \, \* \etaD3
          + 64 \, \* \DNn2
          - 64 \, \* \DNn3
          )
\nn\\[0.5mm] && \mbox{}
       + \S(1)  \, \*  (
            487/18 \, \* \eta
          - 17/3 \, \* \etaD2
          - 761/3 \, \* \etaD3
          - 74 \, \* \etaD4
          - 365/9 \, \* \DNn2
          - 76/3 \, \* \DNn3
          + 48 \, \* \DNn4
          )
\nn\\[0.5mm] && \mbox{}
       - 1571/54 \, \* \eta
          - 32503/216 \, \* \etaD2
          + 1493/36 \, \* \etaD3
          + 1666/3 \, \* \etaD4
          + 156 \, \* \etaD5
          + 638/9 \, \* \DNn2
\nn\\[-0.5mm] && \mbox{}
          - 644/9 \, \* \DNn3
          + 172 \, \* \DNn4
          - 128 \, \* \DNn5
       \Big)
\nn\\ && \mbox{\hspn}
   + 8 \, \* \colour4colour{\cas \, \* \nf}  \, \*  \Big(
	   2/3 \, \*  ( 
            \Ss(1,3) 
          + \Ss(3,1) 
          - 2 \, \* \Ss(-2,-2) 
          )
	 + 20/9 \, \*  (
            \S(-3) 
          - 2 \, \* \Ss(1,-2) 
          - 2 \, \* \Ss(1,2) 
          - 2 \, \* \Ss(2,1)
          + \S(3) 
          )
\nn\\[-0.5mm] && \mbox{}
        - 1/3 \, \*  \S(2) 
        + 209/54 \, \*  \S(1) 
        - 233/144 
        + 4 \, \* \z3  \, \*  (
            \S(1)  
          - 2 \, \* \eta
          + 3 \, \* \etaD2
          )
        - \nu \, \*  (
            4 \, \* \S(-3)
          - 4 \, \* \Ss(-2,1)
\nn\\[0.5mm] && \mbox{}
          - 4 \, \* \Ss(1,-2)
          - 2 \, \* \S(-2)
          - 4 \, \* \S(1)
          - 2
          )
       + \S(-3)  \, \*  (
            3 \, \* \eta
          + 6 \, \* \etaD2
          )
       - 4 \, \*  \Ss(-2,1)  \, \* 
          \eta
        - 2 \, \* \Ss(1,-2)  \, \*  (
            \eta
          + 6 \, \* \etaD2
          )
\nn\\[0.5mm] && \mbox{}
       - \S(3)  \, \*  (
	    11/6 \, \* \eta
          + 3 \, \* \etaD2
          )
       + \S(-2)  \, \*  (
            77/9 \, \* \eta
          - 13/3 \, \* \etaD2
          + 2 \, \* \etaD3
          )
       + 4/3 \, \* \S(2)  \, \*  (
            23/3 \, \* \eta
          - \etaD2
          )
\nn\\[0.5mm] && \mbox{}
       - \S(1)  \, \*  (
            901/36 \, \* \eta
          + 166/9 \, \* \etaD2
          + 43/6 \, \* \etaD3
          + 3 \, \* \etaD4
          - 232/9 \, \* \DNn2
          + 16/3 \, \* \DNn3
          )
        - 2662/27 \, \* \eta
\nn\\[-0.5mm] && \mbox{}
          + 4375/54 \, \* \etaD2
          + 169/9 \, \* \etaD3
          - 17/3 \, \* \etaD4
          + 2 \, \* \etaD5
          + 716/9 \, \* \DNn2
          - 704/9 \, \* \DNn3
          + 16 \, \* \DNn4
       \Big)
\nn\\ && \mbox{\hspn}
   + 8 \, \* \colour4colour{\ca \, \* \cf \, \* \nf}  \, \*  \Big(
         55/12 \, \* \S(1)
       - 241/144
       - 2 \, \* \z3 \, \*  (
            2 \, \* \S(1)  
          - \eta
          + 12 \, \* \etaD2
          )
       - \S(-3)  \, \*  (
            10 \, \* \eta
          + 8 \, \* \etaD2
          )
\nn\\[-0.5mm] && \mbox{}
       + 8 \, \* \nu  \, \*  (
            \S(-3)
          - \Ss(-2,1)
          - \Ss(1,-2)
          - \S(1)
          - 5/4 \, \*  ( \S(-2) + 1 )
          )
       + 8 \, \* \Ss(-2,1)  \, \*  \eta
       + \Ss(1,-2)  \, \*  (
            4 \, \* \eta
          + 32 \, \* \etaD2
          )
\nn\\[0.5mm] && \mbox{}
       - ( \Sss(1,1,1) - \Ss(1,2) )  \, \*  (
            2 \, \* \eta
          - 4 \, \* \etaD2
          )
       - \S(3)  \, \*  (
            \eta
          - 14 \, \* \etaD2
          )
       + \Ss(1,1)  \, \*  (
	    11/6 \, \* \eta
	  + 1/3 \, \* \etaD2
          - 2 \, \* \etaD3
          )
\nn\\[0.5mm] && \mbox{}
       + \S(-2)  \, \*  (
            33 \, \* \eta
          - 20 \, \* ( \etaD2 + \DNn2 )
          - 16 \, \* ( \etaD3 - \DNn3 )
          )
       + \S(2)  \, \*  (
            40/3 \, \* \eta
          - 29/3 \, \* \etaD2
          - 4 \, \* \etaD3
          - 15 \, \* \DNn2
\nn\\[0.5mm] && \mbox{}
          + 12 \, \* \DNn3
          )
       + \S(1)  \, \*  (
            89/18 \, \* \eta
          + 202/9 \, \* \etaD2
          + 130/3 \, \* \etaD3
          + 14 \, \* \etaD4
          - 3 \, \* \DNn2
          + 2 \, \* \DNn3
          - 2 \, \* \DNn4
          )
\nn\\[0.5mm] && \mbox{}
        - 1483/54 \, \* \eta
          + 3845/54 \, \* \etaD2
          + 169/9 \, \* \etaD3
          - 554/3 \, \* \etaD4
          - 56 \, \* \etaD5
          + 30 \, \* \DNn2
          - 95/6 \, \* \DNn3
\nn\\[-0.5mm] && \mbox{}
          - 35 \, \* \DNn4
          + 32 \, \* \DNn5
       \Big)
\nn\\ && \mbox{\hspn}
   + 8 \, \* \colour4colour{\cfs \, \* \nf}  \, \*  \Big(
         1/8
       + 6 \, \* \z3  \, \*  (
            \eta
          + 2 \, \* \etaD2
          )
       + 8 \, \* ( \S(-3) - 2 \, \* \Ss(1,-2) ) \, \*  \etaD2
       + ( \Sss(1,1,1) - \Ss(1,2) ) \, \*  (
            2 \, \* \eta
          - 4 \, \* \etaD2
          )
\nn\\[-0.5mm] && \mbox{}
       - \S(3)  \, \*  (
            \eta
          + 6 \, \* \etaD2
          )
       + 12 \, \* \nu  \, \* (
            \S(-2) + 1
          )
       - 10 \, \*  \S(-2)  \, \*  \eta
       + \S(2)  \, \*  (
            11\,  \* \eta
          - 10 \, \* \etaD2
          - 4 \, \* \etaD3
          - 8 \, \* \DNn2
          + 4 \, \* \DNn3
          )
\qquad \nn\\[0.5mm] && \mbox{}
       - \Ss(1,1)  \, \*  (
	    3/2 \, \* \eta
          - 2 \, \* \etaD2
          - 2 \, \* \etaD3
          )
       - \S(1)  \, \*  (
	    23/2 \, \* \eta
          + 6 \, \* \etaD2
          - 10 \, \* \etaD3
          - 2 \, \* \etaD4
          - 14 \, \* \DNn2
          + 10 \, \* \DNn3
\nn\\[0.5mm] && \mbox{}
          - 2 \, \* \DNn4
          )
        - 55 \, \* \eta
          + 12 \, \* \etaD2
          - 19/2 \, \* \etaD3
          - 21 \, \* \etaD4
          - 4 \, \* \etaD5
          + 38 \, \* \DNn2
          - 75/2 \, \* \DNn3
          + 15 \, \* \DNn4
       \Big)
\nn\\ && \mbox{\hspn}
   + 2/27 \, \* \colour4colour{\ca \, \* \nfs}  \, \*  \Big(
         87/4
       + \S(1)  \, \*  (
            8
          - 27 \, \* \eta
          + 48 \, \* \etaD2
          )
        - 3 \, \* \eta
          - 16 \, \* \etaD2
          - 24 \, \* \etaD3
       \Big)
\nn\\ && \mbox{\hspn}
   + 8/27 \, \* \colour4colour{\cf \, \* \nfs}  \, \*  \Big(
         33/8
       + ( \Ss(1,1) - 2 \, \* \S(2) ) \, \*  (
          - 9 \, \* \eta
          + 18 \, \* \etaD2
          )
       - \S(1)  \, \*  (
            84 \, \* \eta
          - 51 \, \* \etaD2
          - 18 \, \* \etaD3
\nn\\[-1mm] && \mbox{}
          - 81 \, \* \DNn2
          + 54 \, \* \DNn3
          )      
        - 16 \, \* \eta
          + 65 \, \* \etaD2
          - 120 \, \* \etaD3
          - 36 \, \* \etaD4
          - 45 \, \* \DNn3
          + 54 \, \* \DNn4
       \Big)
%
\:\: .
\eea
All these results refer to the standard transformation to the \MSb\ scheme of
Ref.~\cite{MSvN98}, see Eq.~(\ref{ZikM}).
With the exception of the $C_A \nfs$ part of Eq.~(\ref{dPgg2N}), which was 
derived in Ref.~\cite{BGracey} (see also Ref.~\cite{Gracey}), 
Eqs.~(\ref{dPps2N}) -- (\ref{dPgg2N}) are new results of the present article.

\pagebreak
 
The last two equations include the denominator $\nu$ defined in 
Eq.~(\ref{EtaNu}), and are therefore only valid at $N \geq 3$. 
The first moment of the NNLO quark-gluon splitting function is
\bea
\label{dPgg0N1}
   \dPgq(0)(N\!=\!1) \! &\! = \!&
           3\,\* \cf 
\:\: , \\[2mm]
\label{dPgg1N1}
   \dPgq(1)(N\!=\!1) \! &\! = \!&
          \frac{71}{3}\: \* \cf\, \* \ca 
     \:-\: 9\,\* \cfs
     \:-\: \frac{2}{3}\: \* \cf\, \* \nf 
\:\: , \\[2mm] 
\label{dPgq2N1}
   \dPgq(2)(N\!=\!1) \! &\! = \!&
           \frac{1607}{12}\: \* \cf \* \, \cas
     \:-\: \frac{461}{4}\: \* \cfs \* \, \ca
     \:+\: \frac{63}{2}\: \* \cft
     \:+\: \bigg( \,\frac{41}{3} - 72\, \*\z3 \bigg)\,\* \cf \, \* \ca \, \* \nf
\nn \\[1mm] & &
     \:-\: \bigg( \,\frac{107}{2} - 72\, \* \z3 \bigg) \, \* \cfs \* \, \nf
     \:-\: \frac{13}{3}\: \* \cf \* \nfs
\:\: .
\eea
The corresponding results for the gluon-gluon splitting function are identical
to the coefficients of the beta function recalled in Eq.~(\ref{b012}). 
The NLO and NNLO pure-singlet results are related to Eqs.~(\ref{dPgg0N1}) and 
(\ref{dPgg1N1}) by
\beq
\label{dPpsnN1}
   \dPps(n)(N\! =\! 1) \equal
   - 2\,\nf\, \dPgq(n-1)(N\! =\! 1)
\:\: .
\eeq
 
In the OPE, this relation for the anomalous dimension of the pure-singlet axial
current together with Eq.~(\ref{dPgg2N1}) for the first moment of $\dPgg(2)$ 
has been shown in Ref.~\cite{g5L2} to be a direct consequence of the 
requirement that the axial anomaly \cite{Adler:1969gk,Bell:1969ts} should 
preserve the one-loop character of the operator relation  \cite{Adler:1969er}
\beq
\label{ABJ}
  \partial^{\,\mu\!} j^{\,5}_\mu \equal 
  -2\, \nf\, a_s\, {\widetilde G}_a^{\,\mu\nu} G_{a,\,\mu\nu}
\eeq
in dimensional regularization,
where $j^{\,5}_\mu \,=\, \overline{\psi}\,\gamma_{5}\gamma_{\mu}\psi$ and 
$G^{\,\mu\nu}_a$ ($\,{\widetilde G}^{\,\mu\nu}_a 
\,=\, 1/2\,\epsilon^{\,\mu\nu\alpha\beta}G_{a,\alpha\beta})$ denote the
renormalized axial current and the (dual) gluon field-strength tensor.
In this context
Eqs.~(\ref{dPgg2N1}) and (\ref{dPpsnN1}) are thus consistency requirements
ensuring the correct renormalization of the pure-singlet axial current with
the chosen finite renormalization constants $Z_{\,\rm ik}$, 
see Eq.~(\ref{ZikM}).
Consequently~Eq.~(\ref{dPpsnN1}) for $n=3$, together with Eq.~(\ref{dPgq2N1}) 
and $\dPnsp(n)(N\! =\! 1) \,=\, \dPqg(n)(N\! =\! 1) \,=\, 0$, fixes the first 
moments of the upper-row splitting functions at order $\as(4)$.

The quantities given above do not provide the complete set of third-order 
helicity-difference splitting functions. Additional even-$N$ functions
$\Delta P_{\rm ns}^{\,-,\rm v}$ exist for the quark-antiquark differences 
\bea
\label{eq:qpm}
  \Delta f_{ik}^{\,-} &\! = \!&
  \Delta f_{q_i^{}} - \Delta f_{\bar{q}_i^{}}
  - \left( \Delta f_{q_k^{}} - \Delta f_{\bar{q}_k^{}} \right)
\:\: , \\
  \Delta f^{\,\rm v} &\! = \!&
  \sum_{i=1}^{\nf} \left\{ \Delta f_{q_i^{}}
                         - \Delta f_{{\bar q}_i^{}} \right\}
\eea
that occur in the (so far practically irrelevant) structure functions $g_3^{}$ 
and $g_4^{}$ in polarized charged-current DIS which has been analyzed at NLO 
in Ref.~\cite{SVWgiCC}. The corresponding NNLO corrections may be addressed
in a future publication together with the generalization of 
Refs.~\cite{MRcc,MRVcc} to all~$N$.
It appears safe to assume
$\Delta P_{\rm ns}^{\,-(2)} \,=\, P_{\rm ns}^{\,+(2)}$ as given in Eq.~(3.7) 
of Ref.~\cite{mvvPns}, $\Delta P_{\rm ns}^{\,\rm v (2)}$ is unknown though at
this point.

%
\setcounter{equation}{0}
\section{The NNLO splitting functions in {\bf x}-space}
\label{sec:xres}
%
%
The expressions for the $x$-space splitting functions in Eq.~(\ref{Pexp}) in 
terms of harmonic polylogarithms~\cite{Hpols} can be obtained from their 
$N$-space counterparts in terms of harmonic sums \cite{Hsums} by a completely 
algebraic procedure \cite{Hpols,MV99} based on the fact that latter functions 
occur as coefficients of the Taylor expansion of the former.
Our notation for the harmonic polylogarithms follows Ref.~\cite{Hpols}, 
with the lowest-weight ($w = 1$) functions $H_m(x)$ given by
\beq
\label{hpol1}
  H_0^{}(x)    \equal \ln x \:\: , \quad\quad
  H_{\pm 1}(x) \equal \mp \, \ln (1 \mp x) 
\eeq
and the higher-weight ($w \geq 2$) functions recursively defined as
\beq
\label{hpol2}
  H_{m_1,...,m_w}(x) \equal
    \left\{ \begin{array}{cl}
    \displaystyle{ \frac{1}{w!}\,\ln^w x \:\: ,}
       & \quad {\rm if} \:\:\: m^{}_1,...,m^{}_w = 0,\ldots ,0 \\[2ex]
    \displaystyle{ \int_0^x \! dz\: f_{m_1}(z) \, H_{m_2,...,m_w}(z)
       \:\: , } & \quad {\rm else}
    \end{array} \right.
\eeq
with
\beq
\label{hpolf}
  f_0(x)       \equal \frac{1}{x} \;\; , \quad\quad
  f_{\pm 1}(x) \equal \frac{1}{1 \mp x} \;\; .
\eeq
For chains of indices zero we employ the abbreviated notation
\beq
\label{eq:habbr}
  H_{{\footnotesize \underbrace{0,\ldots ,0}_{\scriptstyle m} },\,
  \pm 1,\, {\footnotesize \underbrace{0,\ldots ,0}_{\scriptstyle n} },
  \, \pm 1,\, \ldots}(x) \equal H_{\pm (m+1),\,\pm (n+1),\, \ldots}(x)
  \:\: .
\eeq
 
Also here we recall, for completeness, the LO and NLO contributions
\bea
\label{dPij0x}
  \Delta P^{\,(0)}_{\,\rm ns}(x) & \! = \! &  
  2 \, \* \colour4colour{\cf} \, \* \Big(
          \dpqq(x)
          + 3/2\, \* \delta(1 - x)
          \Big)
\:\: , \nn \\
  \dPps(0)(x) & \! = \! & 0  
\:\: , \nn \\[0.5mm]
  \dPqg(0)(x) & \! = \! &  
          2 \, \* \colour4colour{\nf} \, \* \left( - 1 + 2\, \*x\right)
\:\: , \nn \\[1mm]
  \dPgq(0)(x) & \! = \! &  
          2 \, \* \colour4colour{\cf} \, \* \left( 2 - x\right)
\:\: , \nn \\[0.5mm]
  \dPgg(0)(x) & \! = \! &
         4 \, \* \colour4colour{\ca}  \, \*  \Big(
            \dpgg(x)
          + 11/12 \, \* \delta(1 - x)
          \Big)
          - 2/3 \, \, \* \colour4colour{\nf} \, \, \* \delta(1 - x)
\:\: ,
\eea
and
\bea
\label{dPns1x}
  \Delta P^{\,+(1)}_{\rm ns}(x) \!\! & \! = \! &
          4 \, \* \colour4colour{\cfs}  \, \*  \Big(
            2 \, \* \dpqq( - x) \, \* (\z2 + 2 \, \* \Hh(-1,0) - \Hh(0,0))
          + 2 \, \* \dpqq(x) \, \* (\Hh(1,0) + \H(2) - 3/4 \, \* \H(0))
\nn\\[-1mm] && \mbox{}
          - 9 \, \* (1 - x)
          - (1 + x) \, \* \Hh(0,0)
          - 1/2 \, \* (7 + 11 \, \* x) \, \* \H(0)
       + \delta(1 - x)  \, \*  (
            3/8
          + 6 \, \* \z3
          - 3 \, \* \z2
          )
          \Big)
\quad \nn\\ && \mbox{\hspn}
        + 4 \, \* \colour4colour{\ca \, \* \cf}  \, \*  \Big(
          - \dpqq( - x) \, \* (\z2 + 2 \, \* \Hh(-1,0) - \Hh(0,0))
          + \dpqq(x) \, \* (\Hh(0,0) + 11/6 \, \* \H(0) - \z2
\nn\\[-1mm] && \mbox{}
            + 67/18 )
          + 26/3 \, \* (1 - x)
          + 2 \, \* (1 + x) \, \* \H(0)
       + \delta(1 - x)  \, \*  (
            17/24
          - 3 \, \* \z3
          + 11/3 \, \* \z2
          )
          \Big)
\nn\\ && \mbox{\hspn}
        + 4/3 \, \* \colour4colour{\cf \, \* \nf}  \, \*  \Big(
          - \dpqq(x) \, \* ( 5/3 + \H(0) )
          - 2 \, \* (1 - x)
       - \delta(1 - x)  \, \*  (
            1/4
          + 2 \, \* \z2
          )
          \Big)
\:\: , \\[1mm]
\label{dPps1x}
  \dPps(1)(x) & \! = \! & 
  4 \, \* \colour4colour{\cf\, \*\nf} \, \* \Big(
          - (1 - 3\, \*x)\, \*\H(0)
          + 1
          - x
          - 2\, \*(1 + x)\, \*\Hh(0,0)
          \Big)
\:\: , \\[1mm]
\label{dPqg1x}
  \dPqg(1)(x) & \! = \! &
  4 \, \* \colour4colour{\ca\, \*\nf} \, \* \Big(
            2\, \*(1 - 2\, \*x)\, \*\Hh(1,1)
          + 4\, \*(1 - x)\, \*\H(1)
          - 2\, \*(1 + 2\, \*x)\, \*(\Hh(-1,0) + \Hh(0,0))
\nn\\[-1mm] && \mbox{}
          + (1 + 8\, \*x)\, \*\H(0)
          - 2\, \*\z2
          + 12
          - 11\, \*x
          \Big)
\nn\\ && \mbox{\hspn}
  + 2 \, \* \colour4colour{\cf\, \*\nf} \, \* \Big(
            4\, \*(1 - 2\, \*x)\, \*(\z2 - 1/2\, \*\Hh(0,0) 
          - \Hh(1,0) - \Hh(1,1) - \H(2))
          - 8\, \*(1 - x)\, \*\H(1) 
\nn\\[-1mm] && \mbox{}
          - 9\, \*\H(0)
          - 22
          + 27\, \*x
          \Big)
\:\: , \\[1mm]
\label{dPgq1x}
  \dPgq(1)(x) & \! = \! &
    4 \, \* \colour4colour{\cf\, \*\ca} \, \* \Big(
            2\, \*(2 - x)\, \*(\Hh(1,0) + \Hh(1,1) + \H(2))
          + 2\, \*(2 + x)\, \*(\Hh(-1,0) + \Hh(0,0))
\nn\\[-1mm] && \mbox{}
          + (4 - 13\, \*x)\, \*\H(0)
          - 1/3\, \*(10 + x)\, \*\H(1)
          + 41/9
          + 35/9\, \*x
          + 2\, \*x\, \*\z2
          \Big)
\nn\\[-0.5mm] && \mbox{\hspn}
  + 2 \, \* \colour4colour{\cfs} \, \* \Big(
            2\, \*(2 - x)\, \*(\Hh(0,0) - 2\, \*\Hh(1,1))
          + 2\, \*(2 + x)\, \*\H(1)
          - (4 - x)\, \*\H(0)
          - 17
          + 8\, \*x
          \Big)
\nn\\[-0.5mm] && \mbox{\hspn}
  + 8/3 \, \* \colour4colour{\cf\, \*\nf} \, \* \Big(
            (2 - x)\, \*\H(1)
          - 4/3
          - 1/3\, \*x
          \Big)
\:\: , \\[1mm]
\label{dPgg1x}
  \dPgg(1)(x) & \! = \! &
   4 \, \* \colour4colour{\cas} \, \* \Big(
            2\, \*\dpgg( - x)\, \*(\z2 + 2\, \*\Hh(-1,0) - \Hh(0,0))
          - 2\, \*\dpgg(x)\, \*(\z2 - \Hh(0,0) - 2\, \*\Hh(1,0) 
\nn\\ && \mbox{}
          - 2\, \*\H(2) - 67/18)
          - 19/2 \, \*(1 - x)
          + 8\, \*(1 + x)\, \*\Hh(0,0)
          + 1/3\, \*(29 - 67\, \*x)\, \*\H(0)
\nn\\[-0.5mm] && \mbox{}
       + \delta(1 - x) \, \* (
            8/3
          + 3\, \*\z3
          )
          \Big)
\nn\\[-0.5mm] && \mbox{\hspn}
  + 8/3\, \*\colour4colour{\ca\, \*\nf} \, \* \Big(
          - 5/3\, \*\dpgg(x)
          - 3\, \*(1 - x)
          - (1 + x)\, \*\H(0)
          - \delta(1 - x)
          \Big)
\nn\\[-0.5mm] && \mbox{\hspn}
  + 2 \, \* \colour4colour{\cf\, \*\nf} \, \* \Big(
          - 10\, \*(1 - x)
          - 4\, \*(1 + x)\, \*\Hh(0,0)
          - 2\, \*(5 - x)\, \*\H(0)
          - \delta(1 - x)
          \Big)
\:\: .
\eea
Here and in Eqs.~(\ref{dPns2x}) -- (\ref{dPgg2x}) we have suppressed the
argument $x$ of the polylogarithms and used
\bea
\label{dpqqgg}
  \Delta p_{\rm{qq}}(x) &\! =\! & 
   2\, (1 - x)^{-1} - 1 - x
\:\: ,\nn \\[0.5mm]
  \Delta p_{\rm{gg}}(x) &\! =\! & 
  (1-x)^{-1} + 1 - 2\, \*x
\:\: .
\eea
Divergences for $x \to 1 $ are to be understood as plus-distributions.

\begin{figure}[p]
\vspace*{-5mm}
\centerline{\hspace*{-1mm}\epsfig{file=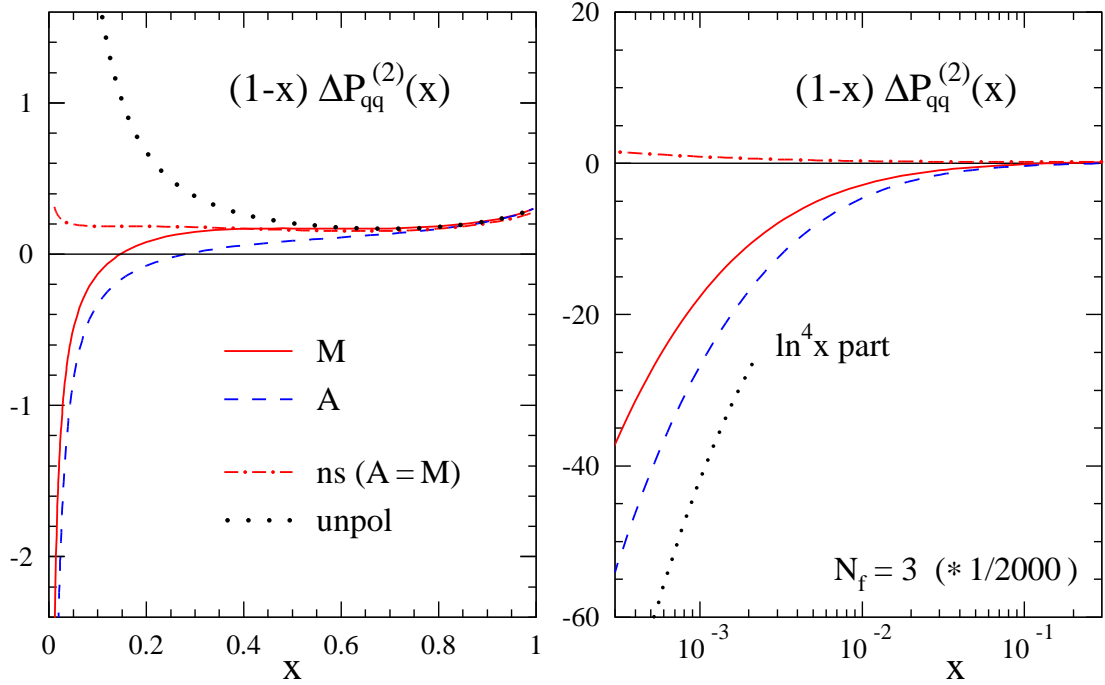,width=15.4cm}}
\vspace{-3mm}
\caption{\label{dPqqFig}
The polarized NNLO quark-quark splitting function in the standard \MSb\ scheme 
(M), as given by the sum of equations (\ref{dPns2x}) and (\ref{dPps2x}) for 
three flavours, multiplied by $\x1$ for display purposes. 
Also shown are the non-singlet contribution, the leading-logarithmic small-$x$ 
part \cite{BVpol}, and the splitting function in the alternative 
scheme (A) with Eq.~(\ref{ZgqA}), see also Appendix~A.
}
\vspace*{-3mm}
\end{figure}

\begin{figure}[p]
\vspace*{-5mm}
\centerline{\hspace*{-1mm}\epsfig{file=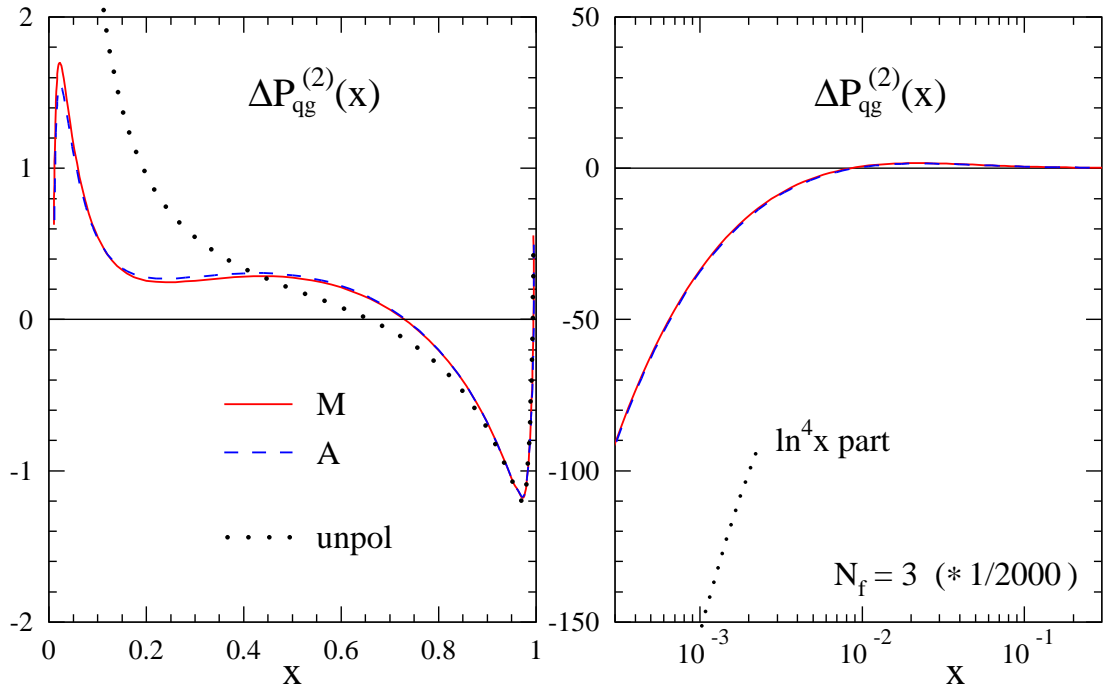,width=15.4cm}}
\vspace{-3mm}
\caption{\label{dPqgFig}
As Figure~\ref{dPqqFig}, but for the gluon-quark splitting function 
(\ref{dPqg2x}) and its A-scheme analogue.
The multiplication with $1/2000 \,\simeq\, 1/(4 \pi)^3$ approximately 
converts the results to a series in $\als$.
}
\vspace*{-4mm}
\end{figure}

\begin{figure}[p]
\vspace*{-5mm}
\centerline{\hspace*{-1mm}\epsfig{file=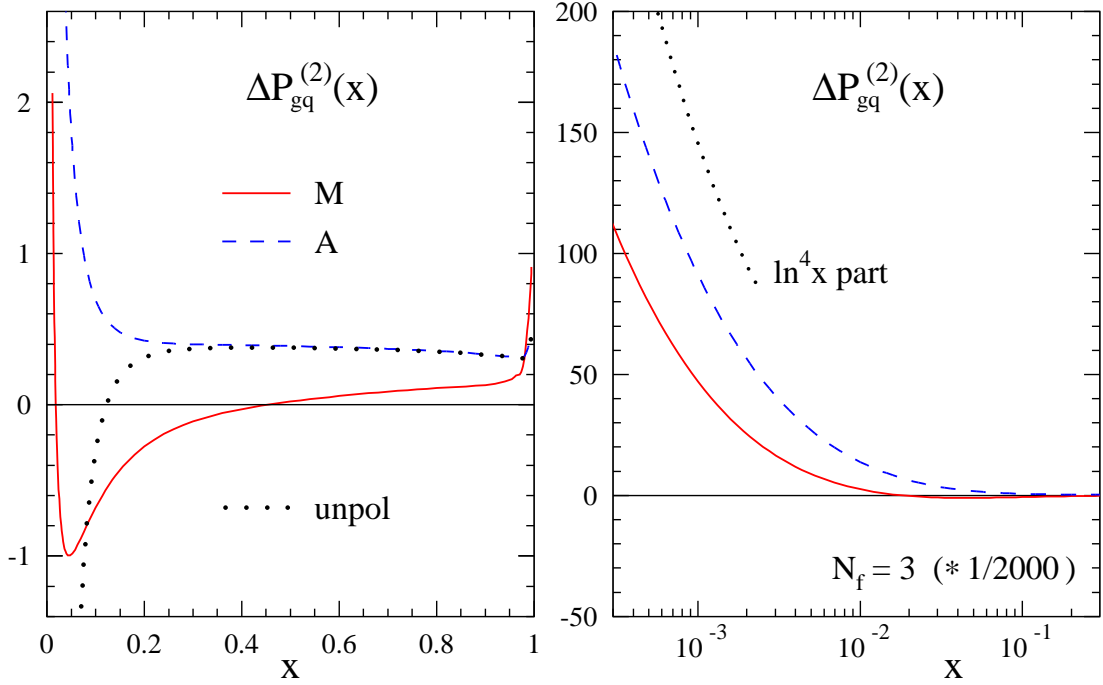,width=15.4cm}}
\vspace{-3mm}
\caption{\label{dPgqFig}
The polarized NNLO quark-gluon splitting function (\ref{dPgq2x}) for the 
standard (M) transformation (\ref{ZikM}) \cite{MSvN98} from the Larin scheme 
and an alternative (A) which also includes Eq.~(\ref{ZgqA}).
As for $\dPqg(2)(x)$ shown in the previous figure, the leading small-$x$
coefficient is different from Ref.~\cite{BVpol}, which provides the 
$\ln^{\,4}x$ terms of the physical kernels $K_{\bar{4}6}$ and $K_{6\bar{4}}$ 
in these cases.
}
\vspace*{-4mm}
\end{figure}

\begin{figure}[p]
\vspace*{-5mm}
\centerline{\hspace*{-1mm}\epsfig{file=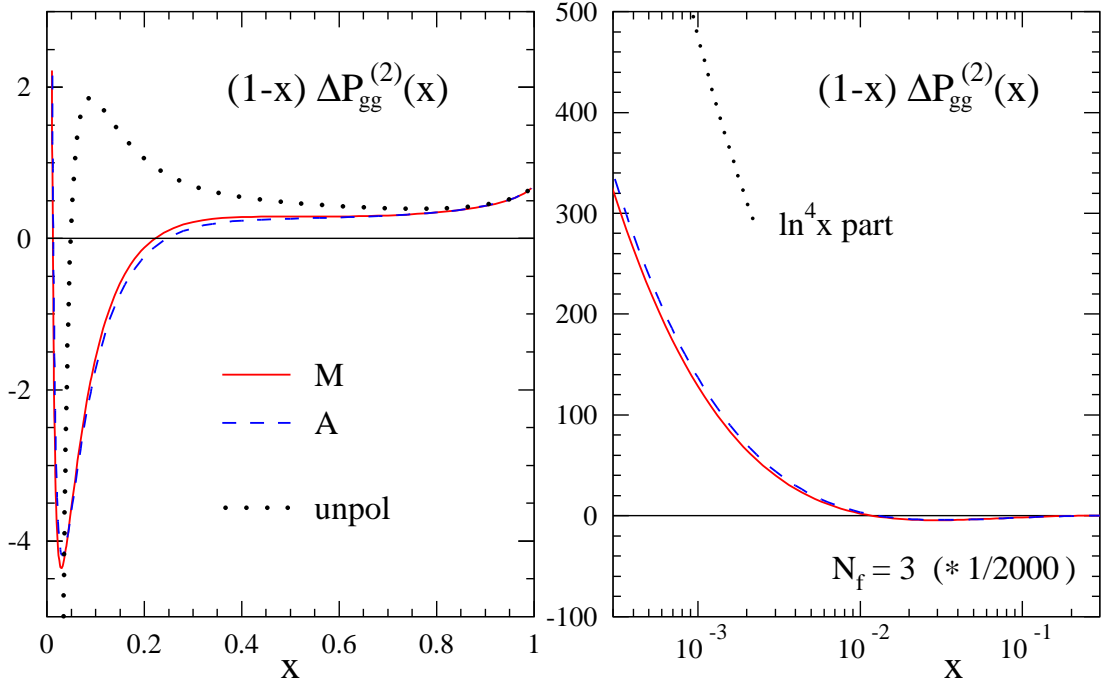,width=15.4cm}}
\vspace{-3mm}
\caption{\label{dPggFig}
As Figure~\ref{dPqqFig}, including the multiplication with $\x1$, but for the 
second diagonal NNLO entry of the splitting-function matrix~(\ref{Sevol}) 
given by Eq.~(\ref{dPgg2x}) in the standard \MSb\ scheme.
}
\vspace*{-4mm}
\end{figure}

The polarized NNLO non-singlet and pure singlet quark-quark splitting 
functions, obtained by Mellin-inverting Eqs.~(\ref{dPns2N}) and (\ref{dPps2N})
are given by
\bea
\label{dPns2x}
  && \hspn \dPnsp(2)(x) \equal
         16 \, \* \colour4colour{\cft}  \, \*  \Big(
         2 \, \* \dpqq( - x)\, \* (
            9/4 \, \* \z3 - 7/4 \, \* \zss
          + 3\, \* \Hh(-3,0) - 16\, \* \H(-2)\, \* \z2 - 4\, \* \Hhh(-2,-1,0) 
\qquad \nn\\[-0.5mm] && \mbox{}
          + 13\, \* \Hhh(-2,0,0) + 14\, \* \Hh(-2,2) 
          - 4\, \* \Hhh(-1,-2,0) + 24\, \* \Hh(-1,-1)\, \* \z2 
          - 20\, \* \Hhhh(-1,-1,0,0) - 24\, \* \Hhh(-1,-1,2) 
\nn\\[0.2mm] && \mbox{}
          - 20\, \* \Hh(-1,0)\, \* \z2 
          + 11\, \* \Hhhh(-1,0,0,0) + 2\, \* \Hhh(-1,2,0) 
          + 16\, \* \Hh(-1,3) + 7\, \* \Hh(0,0)\, \* \z2 
          - 3\, \* \Hhh(0,0,0,0) - 3\, \* \H(2)\, \* \z2 
\nn\\[0.2mm] && \mbox{}
          - \Hh(3,0) - 6\, \* \H(4)
          - 3/2 \, \* \Hh(-2,0) - 3\, \* \H(-1)\, \* \z2 
          - 3/2 \, \* \Hhh(-1,0,0) + 3\, \* \Hh(-1,2) 
          + 3/4 \, \* \H(0)\, \* \z2 
          - 3/2 \, \*\H(3)
\nn\\[0.2mm] && \mbox{}
          + 9/4\, \* \Hhh(0,0,0) - 18 \, \* \H(-1)\, \* \z3
          + 3/4 \, \* \H(0) + 13/2 \, \* \H(0)\, \* \z3
          )
          + 2\, \* \dpqq(x)\, \* (
            9/20\, \* \zss - \Hh(-3,0) 
\nn\\[0.2mm] && \mbox{}
          + 3\, \* \H(-2)\, \* \z2 + 6\, \* \Hhh(-2,-1,0) 
          - 3\, \* \Hhh(-2,0,0) - \Hhhh(0,0,0,0) + 4\, \* \Hhh(1,-2,0) 
          - 2\, \* \Hhhh(1,0,0,0) + 2\, \* \Hhh(1,2,0) 
\nn\\[0.2mm] && \mbox{}
          + 4\, \* \Hh(1,3) + \Hhh(2,0,0)
          + 2\, \* \Hhh(2,1,0) + 2\, \* \Hh(2,2) 
          + 2\, \* \Hh(3,0) + 2\, \* \Hh(3,1) + \H(4)
          - 3/4\, \* \H(0)\, \* \z2 
          - 3\, \* \Hhh(1,0,0) 
\nn\\[0.2mm] && \mbox{}
          - 3/2\, \* \Hh(2,0)
          - 3/32\, \* \H(0)
          + 1/2\, \* \H(0)\, \* \z3
          + 13/16\, \* \Hh(0,0)
          + 6\, \* \H(1)\, \* \z3
          )
          - (11 + 31\, \* x)\, \* \H(3)
\nn\\[0.2mm] && \mbox{}
          + (1 - x)\, \*(
          - 25\, \* \H(1) - 151/8
          - 4\, \* \Hhh(-2,0,0) - \Hh(0,0)\, \* \z2 
          + 3\, \* \Hhhh(0,0,0,0)
          - 6\, \* \H(1)\, \* \z2 - 9\, \* \Hh(1,0)
          )
\nn\\[0.2mm] && \mbox{}
          + (1 + x)\, \*(
            37/10\, \* \zss
          - 18\, \* \H(-1)\, \* \z2 
          + 24\, \* \Hh(-1,2)
          + 14\, \* \Hhh(-1,0,0) 
          + 12\, \* \Hhh(-1,-1,0) 
          - 3\, \* \Hhh(2,0,0) 
\nn\\[0.2mm] && \mbox{}
          - 2\, \* \Hh(3,0) 
          - \H(4)
          - 6\, \* \Hh(-1,0)
          )
          + 1/16 \, \* ( - 307 + 437\, \* x)\, \* \H(0)
          + (1 - 5\, \* x)\, \* \Hh(-2,0)
          + 6\, \* x\, \* \H(0)\, \* \z3
\nn\\[0.2mm] && \mbox{}
          - 2\, \* (1 - 3\, \* x)\, \* \Hh(-3,0)
          - 3\, \* (2 + 5\, \* x)\, \* \Hhh(0,0,0)
          + 3/2\, \* (5 + 11\, \* x)\, \* \z3
          - 1/2\, \* (5 + 13\, \* x)\, \* \Hh(2,0)
\nn\\[0.2mm] && \mbox{}
          + (12 + 31\, \* x)\, \* \H(0)\, \* \z2
          + 3/4\, \* (17 + x)\, \* \z2
          - 3/4\, \* (25 + x)\, \* \H(2)
          - 1/8\, \* (73 - 15\, \* x)\, \* \Hh(0,0)
\nn\\[-0.5mm] && \mbox{}
       + \delta(1 - x) \, \*  (
            29/32
          + 9/8\, \* \z2
          + 17/4\, \* \z3
          + 18/5\, \* \zss
          - 15\, \* \z5
          - 2\, \* \z2\, \* \z3
          )
          \Big)
\nn\\ && \mbox{\hspn}
         + 8 \, \* \colour4colour{\cfs \, \* \ca}  \, \*  \Big(
         2 \, \* \dpqq( - x) \, \* (
          - 31/4 \, \* \z3
          - 1/4 \, \* \zss
          + 67/9 \, \* \z2 + 134/9 \, \* \Hh(-1,0) - 67/9 \, \* \Hh(0,0)
\nn\\[-0.5mm] && \mbox{} 
          - 5\, \* \Hh(-3,0) + 32 \, \* \H(-2) \, \* \z2 
          + 4 \, \* \Hhh(-2,-1,0) - 21 \, \* \Hhh(-2,0,0) 
          - 30 \, \* \Hh(-2,2) + 36 \, \* \Hhhh(-1,-1,0,0) 
\nn\\[0.2mm]  && \mbox{} 
          + 4 \, \* \Hhh(-1,-2,0) - 56 \, \* \Hh(-1,-1) \, \* \z2 
          + 56 \, \* \Hhh(-1,-1,2) + 42 \, \* \Hh(-1,0) \, \* \z2 
          - 17 \, \* \Hhhh(-1,0,0,0) - 2 \, \* \Hhh(-1,2,0) 
\nn\\[0.2mm]  && \mbox{} 
          - 32 \, \* \Hh(-1,3)
          - 13 \, \* \Hh(0,0) \, \* \z2 + 5 \, \* \Hhhh(0,0,0,0) 
          + 7 \, \* \H(2) \, \* \z2 + \Hh(3,0) + 10 \, \* \H(4)
          + 31/6 \, \* \Hh(-2,0) 
\nn\\[0.2mm]  && \mbox{} 
          + 31/3 \, \* \H(-1) \, \* \z2 + 31/6 \, \* \Hhh(-1,0,0) 
          - 31/3 \, \* \Hh(-1,2) - 13/12 \, \* \H(0) \, \* \z2 
          - 89/12 \, \* \Hhh(0,0,0) 
          + 31/6 \, \* \H(3)
\nn\\ && \mbox{} 
          + 42 \, \* \H(-1) \, \* \z3
          - 9/4 \, \* \H(0)
          - 29/2 \, \* \H(0) \, \* \z3
          )
          + 2 \, \* \dpqq(x) \, \* (
            5/6 \, \* \z3
          - 69/20 \, \* \zss
          - \Hh(-3,0) 
\nn\\[0.2mm] && \mbox{} 
          - 3 \, \* \H(-2) \, \* \z2 - 14 \, \* \Hhh(-2,-1,0) 
          + 5 \, \* \Hhh(-2,0,0) - 4 \, \* \Hh(-2,2) 
          - 4 \, \* \Hh(0,0) \, \* \z2 + 5 \, \* \Hhhh(0,0,0,0) 
          - 16 \, \* \Hhh(1,-2,0) 
\nn\\[0.2mm] && \mbox{} 
          - 2 \, \* \Hh(1,0) \, \* \z2 
          + 11 \, \* \Hhhh(1,0,0,0) + 8 \, \* \Hhhh(1,1,0,0) 
          - 8 \, \* \Hh(1,3) - 2 \, \* \H(2) \, \* \z2 
          + 5 \, \* \Hhh(2,0,0) + \Hh(3,0) + \H(4)
\nn\\[0.2mm] && \mbox{} 
          + 3 \, \* \Hh(-2,0) + 41/12 \, \* \H(0) \, \* \z2 
          - 23/12 \, \* \Hhh(0,0,0) + 31/3 \, \* \Hhh(1,0,0) 
          + 11/3 \, \* \Hh(2,0) + 2/3 \, \* \H(3)
\nn\\[0.2mm] && \mbox{} 
          - 13/4 \, \* \Hh(0,0) + 67/9 \, \* \Hh(1,0) + 67/9 \, \* \H(2)
          - 151/48 \, \* \H(0)
          - 17/2 \, \* \H(0) \, \* \z3
          - 24 \, \* \H(1) \, \* \z3
          )
\nn\\[0.2mm] && \mbox{} 
          + 4 \, \* (1 - 2 \, \* x) \, \* \Hh(-3,0)
          + 2 \, \* (1 - x) \, \* (
            379/12
          - \H(-2) \, \* \z2 - 2 \, \* \Hhh(-2,-1,0) + 3 \, \* \Hhh(-2,0,0)
          + 7 \, \* \H(1) \, \* \z2 
\nn\\[0.2mm] && \mbox{} 
          + 4 \, \* \Hhh(1,0,0)
          + 26/3 \, \* \Hh(1,0)
          + 251/6 \, \* \H(1)
          )
          + 2 \, \* (1 + x) \, \* ( 25 \, \* \H(-1) \, \* \z2 
            - 14 \, \* \Hhh(-1,-1,0) 
            - 32 \, \* \Hh(-1,2) 
\nn\\[0.2mm] && \mbox{} 
            - 13 \, \* \Hhh(-1,0,0) 
            + 2 \, \* \Hh(2,0)
          + \H(2) \, \* \z2 + 2 \, \* \Hhh(2,0,0) - 3 \, \* \H(4)
          + 19/3 \, \* \Hh(-1,0)
          )
          - (6 + 7 \, \* x) \, \* \zss
\nn\\[0.2mm] && \mbox{} 
          + 2 \, \* (2 - 3 \, \* x) \, \* \H(0) \, \* \z3
          - 5 \, \* (3 - 7 \, \* x) \, \* \Hh(-2,0)
          + 2 \, \* (5 + 3 \, \* x) \, \* \Hh(0,0) \, \* \z2
          + 2 \, \* (9 + 31 \, \* x) \, \* \H(3)
\nn\\[0.2mm] && \mbox{} 
          - (33 + 62 \, \* x) \, \* \H(0) \, \* \z2
          + 1/18 \, \* (157 - 557 \, \* x) \, \* \Hh(0,0)
          - (39 + 17 \, \* x) \, \* \z2
          - 1/2 \, \* (97 + 39 \, \* x) \, \* \z3
\nn\\[0.2mm] && \mbox{} 
          + 1/2 \, \* (35 + 13 \, \* x) \, \* \Hhh(0,0,0)
          + 1/72 \, \* (2627 - 3869 \, \* x) \, \* \H(0)
          + (155/3 + 17 \, \* x) \, \* \H(2)
          - 8 \, \* \Hhhh(0,0,0,0)
\nn\\[-0.5mm] && \mbox{} 
       + \delta(1 - x)  \, \*  (
           151/32
          - 205/12 \, \* \z2
         + 211/6 \, \* \z3 
         - 247/30 \, \* \zss
         + 15 \, \* \z5
         + 2 \, \* \z2 \, \* \z3
          )
          \Big)
\nn\\ && \mbox{\hspn}
         + 8 \, \* \colour4colour{\cf \, \* \cas}  \, \*  \Big(
            2 \, \* \dpqq( - x) \, \* ( 
          11/4 \, \* \z3
          + \zss
          - 67/18 \, \* \z2 - 67/9 \, \* \Hh(-1,0) + 67/18 \, \* \Hh(0,0)
          + \Hh(-3,0) 
\nn\\[-0.5mm] && \mbox{}
          - 8 \, \* \H(-2) \, \* \z2 + 4 \, \* \Hhh(-2,0,0) 
          + 8 \, \* \Hh(-2,2) + 16 \, \* \Hh(-1,-1) \, \* \z2 
          - 8 \, \* \Hhhh(-1,-1,0,0) - 16 \, \* \Hhh(-1,-1,2) 
\nn\\[0.2mm] && \mbox{}
          - 11 \, \* \Hh(-1,0) \, \* \z2 
          + 3 \, \* \Hhhh(-1,0,0,0) + 8 \, \* \Hh(-1,3) 
          + 3 \, \* \Hh(0,0) \, \* \z2 - \Hhhh(0,0,0,0) 
          - 2 \, \* \H(2) \, \* \z2 - 2 \, \* \H(4)
\nn\\[0.2mm] && \mbox{}
          - 11/6 \, \* \Hh(-2,0) - 11/3 \, \* \H(-1) \, \* \z2 
          - 11/6 \, \* \Hhh(-1,0,0) + 11/3 \, \* \Hh(-1,2)
          + 1/6 \, \* \H(0) \, \* \z2 
          + 31/12 \, \* \Hhh(0,0,0) 
\nn\\[0.2mm] && \mbox{}
          - 11/6 \, \* \H(3)
          - 12 \, \* \H(-1) \, \* \z3
          + 3/4 \, \* \H(0)
          + 4 \, \* \H(0) \, \* \z3)
          + 2 \, \* \dpqq(x) \, \* (245/48 
          + 1/2 \, \* \z3
          + 12/5 \, \* \zss
\nn\\[0.2mm] && \mbox{}
          - 67/18 \, \* \z2 + 389/72 \, \* \Hh(0,0)
          + \Hh(-3,0) 
          + 4 \, \* \Hhh(-2,-1,0) - \Hhh(-2,0,0) + 2 \, \* \Hh(-2,2) 
          - \Hhhh(0,0,0,0) 
\nn\\[0.2mm] && \mbox{}
          + 6 \, \* \Hhh(1,-2,0) 
          - \Hh(1,0) \, \* \z2 - 3 \, \* \Hhhh(1,0,0,0)
          - 4 \, \* \Hhhh(1,1,0,0) 
          + 4 \, \* \Hh(1,3) - 2 \, \* \Hhh(2,0,0) + \H(4)
          - 3/2 \, \* \Hh(-2,0) 
\nn\\[0.2mm] && \mbox{}
          - 31/12 \, \* \H(0) \, \* \z2 + 31/12 \, \* \Hhh(0,0,0) 
          - 11/4 \, \* \Hhh(1,0,0) + 11/12 \, \* \H(3)
          + 1043/216 \, \* \H(0)
          + 4 \, \* \H(0) \, \* \z3
\nn\\[0.2mm] && \mbox{}
          + 9 \, \* \H(1) \, \* \z3)
          - (1 - x) \, \*  (74/3 \, \* \H(1)
          - 391/27 
          + \Hh(-3,0) - \H(-2) \, \* \z2 
          - 2 \, \* \Hhh(-2,-1,0) + \Hhh(-2,0,0)
\nn\\[0.2mm] && \mbox{}
          + 4 \, \* \H(1) \, \* \z2 + 4 \, \* \Hhh(1,0,0))
          - (1 + x) \, \* (16 \, \* \H(-1) \, \* \z2 
          - 8 \, \* \Hhh(-1,-1,0) - 6 \, \* \Hhh(-1,0,0) 
          - 20 \, \* \Hh(-1,2)
\nn\\[0.2mm] && \mbox{}
          + 10/3 \, \* \Hh(-1,0) + 28/3 \, \* \H(2)
          + \H(2) \, \* \z2 + 1/2 \, \* \Hhh(2,0,0) - 3/2 \, \* \H(4))
          + 1/4 \, \* (3 + 5 \, \* x) \, \* \zss
          - 2 \, \* \H(0) \, \* \z3
\nn\\[0.2mm] && \mbox{}
          + 9 \, \* (1 + 2 \, \* x) \, \* \H(0) \, \* \z2
          - 2 \, \* (1 + 9 \, \* x) \, \* \H(3)
          + 2/3 \, \* (3 + 10 \, \* x) \, \* \Hh(0,0)
          - 1/2 \, \* (5 + 3 \, \* x) \, \* \Hh(0,0) \, \* \z2
\nn\\[0.2mm] && \mbox{}
          + (7 - 15 \, \* x) \, \* \Hh(-2,0)
          + 2/3 \, \* (9 + 14 \, \* x) \, \* \z2
          + 1/9 \, \* (43 - 21 \, \* x) \, \* \H(0)
          + 1/2 \, \* (41 + 3 \, \* x) \, \* \z3
\nn\\[-0.5mm] && \mbox{}
          - 7 \, \* \Hhh(0,0,0)
          + \Hhhh(0,0,0,0)
       - \delta(1 - x)  \, \*  (
            1657/288
          - 5 \, \* \z5
          + 194/9 \, \* \z3
          - 562/27 \, \* \z2
          + 1/4 \, \* \zss
          )
          \Big)
\nn\\ && \mbox{\hspn}
       + 8/3 \, \* \colour4colour{\cfs \, \* \nf}  \, \*  \Big(
       4 \, \* \dpqq( - x) \, \* (
            3/2 \, \* \z3
          - 5/3 \, \* \z2 - 10/3 \, \* \Hh(-1,0) + 5/3 \, \* \Hh(0,0)
          - \Hh(-2,0) - 2 \, \* \H(-1) \, \* \z2 
\nn\\[-0.5mm] && \mbox{}
          - \Hhh(-1,0,0) + 2 \, \* \Hh(-1,2) 
          + 1/2 \, \* \H(0) \, \* \z2 + \Hhh(0,0,0) - \H(3)
          )
       + 2 \, \* \dpqq(x) \, \* (          
          - 55/16 
          + 5  \, \* \z3 + \H(0) \, \* \z2 
\nn\\[0.2mm] && \mbox{}
          - \Hhh(0,0,0) 
          - 4 \, \* \Hhh(1,0,0) - 2 \, \* \Hh(2,0) - 2 \, \* \H(3)
          + 3/2 \, \* \Hh(0,0) - 10/3 \, \* \Hh(1,0) - 10/3 \, \* \H(2)
          + 5/8  \, \* \H(0)
          )
\nn\\[0.2mm] && \mbox{}
          + (1 - x) \, \* (34 - 8 \, \* \H(1) - 4 \, \* \Hh(1,0))
          - (1 + x) \, \* (8 \, \* \Hh(-1,0) - 3 \, \* \Hhh(0,0,0))
          + 1/3 \, \* (31 + 55 \, \* x) \, \* \Hh(0,0)
\nn\\[-0.5mm] && \mbox{}
          + 1/12 \, \* (269 + 253 \, \* x) \, \* \H(0)
          - 8 \, \* \H(2)
       - \delta(1 - x)  \, \*  (
            69/8
          - 5/2 \, \* \z2
          + 17 \, \* \z3
          - 29/5 \, \* \zss
          )
          \Big)
\nn\\ && \mbox{\hspn}
       + 8/3 \, \* \colour4colour{\ca \, \* \cf \, \* \nf}  \, \*  \Big(
       2 \, \* \dpqq( - x) \, \* (
          - 3/2 \, \* \z3
          + 5/3 \, \* \z2 + 10/3 \, \* \Hh(-1,0) - 5/3 \, \* \Hh(0,0)
          + \Hh(-2,0) 
\nn\\[-0.5mm] && \mbox{}
          + 2 \, \* \H(-1) \, \* \z2 
          + \Hhh(-1,0,0) - 2 \, \* \Hh(-1,2) 
          - 1/2 \, \* \H(0) \, \* \z2 - \Hhh(0,0,0) + \H(3)
          )
       + 2 \, \* \dpqq(x) \, \* (
          - 209/72 
\nn\\[0.2mm] && \mbox{}
          - 9/2  \, \* \z3
          + 5/3 \, \* \z2 
          - 7/2 \, \* \Hh(0,0)
          + \H(0) \, \* \z2 - \Hhh(0,0,0) + 3/2 \, \* \Hhh(1,0,0) - 1/2 \, \* \H(3)
          - 167/36 \, \* \H(0))
\nn\\[0.2mm] && \mbox{}
          - (1 - x) \, \* (440/9 - 2 \, \* \H(1))
          + (1 + x) \, \* (4 \, \* \Hh(-1,0) + \H(2))
          + (3 - x) \, \* \z2
          - (6 + 5 \, \* x) \, \* \Hh(0,0)
\nn\\[-0.5mm] && \mbox{}
          - 2/3 \, \* (33 - x) \, \* \H(0)
       + \delta(1 - x)  \, \*  (
            15/2
          - 167/9 \, \* \z2
          + 25/3 \, \* \z3
          + 3/10 \, \* \zss
          )
          \Big)
\nn\\ && \mbox{\hspn}
       + 8/9  \, \*  \colour4colour{\cf \, \* \nfs}  \, \*  \Big(
          \dpqq(x) \, \* (- 1/3 + 5/3 \, \* \H(0) + \Hh(0,0))
          + (1 - x) \, \* (13/3 + 2 \, \* \H(0))
\nn\\[-1mm] && \mbox{}
       - \delta(1 - x) \, \*  (
            17/8
          - 10/3 \, \* \z2
          + 2 \, \* \z3
          )
          \Big)
\eea
and
\bea
\label{dPps2x}
  && \hspn \dPps(2)(x) \equal
         4 \, \* \colour4colour{\ca \, \* \cf \, \* \nf}  \, \*  \Big(
           4 \* (1 - x) \, \* (5/2 \, \* \H(1) \, \* \z2 
          - 33/4 \, \* \Hhh(1,0,0) + 5 \, \* \Hhh(1,1,0) + 5 \, \* \Hhh(1,1,1)
          - 4439/54
\qquad \nn\\[-0.5mm] && \mbox{}
          - \H(-2) \, \* \z2 
          - 2 \, \* \Hhh(-2,-1,0) - 3 \, \* \Hhh(-2,0,0) 
          - 1/2 \, \* \Hh(0,0) \, \* \z2
          + 17/2 \, \* \H(1,0) + 65/12 \, \* \Hh(1,1)
          + 266/9 \, \* \H(1))
\nn\\[0.2mm] && \mbox{}
          - 2 \, \* (1 + x) \, \* (\H(-1) \, \* \z2 
          + 10 \, \* \Hhh(-1,-1,0) + 17 \, \* \Hhh(-1,0,0) + 4 \, \* \Hh(-1,2)
          - 2 \, \* \H(2) \, \* \z2 
          + 7 \, \* \Hhh(2,0,0) - 4 \, \* \Hhh(2,1,0) 
\nn\\[0.2mm] && \mbox{}
          - 4 \, \* \Hhh(2,1,1) - 2 \, \* \Hh(3,1) + \H(4)
          - 37 \, \* \H(-1,0))
          + 1/5 \, \* (117 + 107 \, \* x) \, \* \zss
          - 1/9 \, \* (427 - 1151 \, \* x) \, \* \Hh(0,0)
\nn\\[0.2mm] && \mbox{}
          - 1/27 \, \* (2257 + 8899 \, \* x) \, \* \H(0)
          - 4 \, \* (1 - 5 \, \* x) \, \* \Hh(-3,0)
          - 4 \, \* (3 - 4 \, \* x) \, \* \Hhhh(0,0,0,0)
          + 2 \, \* (6 + x) \, \* \Hh(2,0)
\nn\\[0.2mm] && \mbox{}
          + 2 \, \* (9 - 19 \, \* x) \, \* \Hh(-2,0)
          + 4 \, \* (9 + 13 \, \* x) \, \* \H(0) \, \* \z3
          + 2/3 \, \* (19 - 11 \, \* x) \, \* \Hh(2,1)
          + 14/3 \, \* (25 - 26 \, \* x) \, \* \z3
\nn\\[0.2mm] && \mbox{}
          - 4/3 \, \* (19 + 37 \, \* x) \, \* \Hhh(0,0,0)
          - 1/3 \, \* (29 + 47 \, \* x) \, \* \H(0) \, \* \z2
          + 1/3 \, \* (83 + 47 \, \* x) \, \* \H(3)
\nn\\[-0.5mm] && \mbox{}
          + 1/9 \, \* (91 - 134 \, \* x) \, \* \z2
          + 1/9 \, \* (575 + 134 \, \* x) \, \* \H(2)
          \Big)
\nn\\ && \mbox{\hspn}
       + 4 \, \* \colour4colour{\cfs \, \* \nf}  \, \*  \Big(
            10 \* (1 - x) \* (\Hhh(1,0,0) - 2 \, \*  \Hhh(1,1,0) 
          - 2\, \* \Hhh(1,1,1) - 6/5 - 6/5 \, \* \Hh(1,0) 
          - 13/10 \, \* \Hh(1,1) 
\nn\\[-0.5mm] && \mbox{}
          - 25/2 \, \* \H(1) )
          - 4 \, \* (1 + x) \, \* ( 37/10 \, \* \zss 
            + 7 \, \* \H(0) \, \* \z3 - 6 \, \* \Hh(0,0) \, \* \z2 
            + 4 \, \* \Hhhh(0,0,0,0) - \Hhh(2,0,0) + 2 \, \* \Hhh(2,1,0) 
\nn\\[0.2mm] && \mbox{}
            + 2 \, \* \Hhh(2,1,1) + 4 \, \* \Hh(3,0) 
            + \Hh(3,1) + 6 \, \* \H(4) )
          - 4 \, \* (2 - 3 \, \* x) \, \* \Hh(2,1)
          + 20 \, \* (2 - x) \, \* (\H(0) \, \* \z2 - \H(3))
\nn\\[0.2mm] && \mbox{}
          - 4 \, \* (4 - 7 \, \* x) \, \* \Hh(2,0)
          - 4 \, \* (5 - 6 \, \* x) \, \* \Hhh(0,0,0)
          - 4 \, \* (11 - 21 \, \* x) \, \* \z3
          - (25 - 114 \, \* x) \, \* \H(0)
\nn\\[-0.5mm] && \mbox{}
          - (32 + 25 \, \* x) \, \* \Hh(0,0)
          + (64 + 27 \, \* x) \, \* (\z2 - \H(2))
          \Big)
\nn\\ && \mbox{\hspn}
       + 2/9 \, \* \colour4colour{\cf \, \* \nfs}  \, \*  \Big(
       4 \* (1 - x) \, \* (
            86/3 
          + 2 \, \* \H(1)
          + 15 \, \* \Hh(1,1) )
          + 8 \, \* (5 - 4 \, \* x) \* (\z2 - \H(2))
          + 4 \, \* (23 + 17 \, \* x) \* \Hh(0,0)
\nn\\[-1mm] && \mbox{}
          + 24 \, \* (1 + x) \, \* (\z3 + 2 \, \* \H(0) \, \* \z2 
            + \Hhh(0,0,0) + \Hh(2,1) - 2 \, \* \H(3) )
          + 4/3 \, \* (65 - 43 \, \* x) \, \* \H(0)
          \Big)
\:\: .
\eea
Eqs.~(\ref{dPqg2N}) and (\ref{dPgq2N}) result in the third-order gluon-quark 
and quark-gluon splitting functions 
\bea
\label{dPqg2x}
  && \hspn \dPqg(2)(x) \equal
       8 \, \* \colour4colour{\cas \, \* \nf}  \, \*  \Big(
            (1 - 2 \, \* x) \, \* (31\, \* \H(1) \, \* \z3 
              + 6 \, \* \Hhh(1,-2,0) + 10 \, \* \Hh(1,0) \, \* \z2 
              - 5 \, \* \Hhhh(1,0,0,0) + 2 \, \* \Hh(1,1) \, \* \z2 
\nn\\[-0.5mm] && \mbox{}
	      + 2 \, \* \Hhhh(1,1,0,0) + 8 \, \* \Hhhh(1,1,1,0) 
              - 4 \, \* \Hhhh(1,1,1,1) + 4 \, \* \Hhh(1,1,2) 
              + 4 \, \* \Hhh(1,2,0) - 6 \, \* \Hh(1,3)
          - 11/6 \, \* \Hhh(1,1,1) )
\nn\\[0.2mm] && \mbox{}
          - 1/36 \, \* (16099 - 16346 \, \* x)
          + 1/18 \, \* (733 + 12 \, \* x + 54 \, \* x^2) \, \* \z2
          + 1/6 \, \* (273 - 4 \, \* x) \, \* \H(2)
\nn\\[0.2mm] && \mbox{}
          - 1/18 \, \* (675 - 2356 \, \* x + 54 \, \* x^2) \, \* \Hh(0,0)
          - (1 - 18 \, \* x) \, \* \Hh(-3,0)
          + (1 + x) \, \* (
            8 \, \* \Hhh(2,1,0) + 4 \, \* \Hhh(2,1,1)
\nn\\[0.2mm] && \mbox{}
          - 13 \, \* \H(-1) \, \* \z2 - 42 \, \* \Hhh(-1,-1,0) 
          - 8 \, \* \Hh(-1,2))
          - (1 - 14 \, \* x) \, \* \H(-2) \, \* \z2
          + 1/20 \, \* (495 + 538 \, \* x) \, \* \zss
\nn\\[0.2mm] && \mbox{}
          - 4 \, \* (1 + 2 \, \* x) \, \* (\Hh(-2,2) 
            + 5/4 \, \* \H(-1) \, \* \z3 + 1/2 \, \* \Hhh(-1,-2,0) 
            - 1/2 \, \* \Hh(-1,-1) \, \* \z2 - 3 \, \* \Hhhh(-1,-1,-1,0) 
\nn\\[0.2mm] && \mbox{}
	    - \Hhhh(-1,-1,0,0) - \Hhh(-1,-1,2) + 3/4 \, \* \Hhhh(-1,0,0,0) 
            - \Hhh(-1,2,1) + \Hh(-1,3) )
          + 4 \, \* (2 + 3 \, \* x) \, \* \Hh(3,1)
\nn\\[0.2mm] && \mbox{}
          - 1/9 \, \* (2 + 65 \, \* x) \, \* \Hh(1,1)
          - 3/2 \, \* (3 - 2 \, \* x) \, \* \H(4)
          + 4 \, \* (3 - x) \, \* \Hh(2,0)
          + 12 \, \* (3 + 4 \, \* x) \, \* \H(0) \, \* \z3
\nn\\[0.2mm] && \mbox{}
          - 2 \, \* (5 - 6 \, \* x) \, \* \Hhh(-2,-1,0)
          + (5 + 6 \, \* x) \, \* \H(2) \, \* \z2
          + 1/2 \, \* (7 - 31 \, \* x) \, \* \H(0) \, \* \z2
          + 2 \, \* (7 - 10 \, \* x) \, \* \Hh(2,1)
\nn\\[0.2mm] && \mbox{}
          + 3 \, \* (7 - 9 \, \* x) \, \* \Hh(-2,0)
          + 1/2 \, \* (7 - 6 \, \* x) \, \* \Hh(0,0) \, \* \z2
          - (13 - 6 \, \* x) \, \* \Hhh(-2,0,0)
          - (15 - 16 \, \* x) \, \* \Hhhh(0,0,0,0)
\nn\\[0.2mm] && \mbox{}
          - 1/2 \, \* (25 + 42 \, \* x) \, \* \Hhh(2,0,0)
          + 1/6 \, \* (35 - 46 \, \* x) \, \* \Hh(1,2)
          + 1/2 \, \* (35 + 31 \, \* x) \, \* \H(3)
          + (36 - 35 \, \* x) \, \* \Hh(1,0)
\nn\\[0.2mm] && \mbox{}
          - 2/3 \, \* (41 + 40 \, \* x) \, \* \Hhh(0,0,0)
          + 1/6 \, \* (91 - 80 \, \* x) \, \* \H(1) \, \* \z2
          - 1/3 \, \* (104 + 115 \, \* x) \, \* \Hhh(-1,0,0)
\nn\\[0.2mm] && \mbox{}
          + 1/6 \, \* (157 - 146 \, \* x) \, \* \Hhh(1,1,0)
          - 1/6 \, \* (212 - 223 \, \* x) \, \* \Hhh(1,0,0)
          - 1/108 \, \* (11468 + 40643 \, \* x) \, \* \H(0)
\nn\\[-0.5mm] && \mbox{}
          + 1/2 \, \* (315 - 268 \, \* x) \, \* \z3
          + 5/108 \, \* (1006 - 911 \, \* x) \, \* \H(1)
          + 1/9 \, \* (776 + 709 \, \* x + 27 \, \* x^2) \, \* \Hh(-1,0)
          \Big)
\nn\\ && \mbox{\hspn}
       + 8 \, \* \colour4colour{\ca \, \* \cf \, \* \nf}  \, \*  \Big(
          - 2 \, \* (1 - 2 \, \* x) \, \* (39/2 \, \* \H(1) \, \* \z3 
            + 3 \, \* \Hhh(1,-2,0) + 9 \, \* \Hh(1,0) \, \* \z2 
            - 3/2 \, \* \Hhhh(1,0,0,0)
\nn\\[-0.5mm] && \mbox{}
            + 6 \, \* \Hh(1,1) \, \* \z2 + \Hhhh(1,1,0,0) 
              + 3 \, \* \Hhhh(1,1,1,0) - 4 \, \* \Hhhh(1,1,1,1) 
              - 3 \, \* \Hhh(1,1,2) - 6 \, \* \Hhh(1,2,1) - 6 \, \* \Hh(1,3) )
\nn\\[0.2mm] && \mbox{}
          - (59/2 - 31 \, \* x + 3 \, \* x^2) \, \* \H(1) \, \* \z2
          - 4 \, \* (1 - x) \, \* \H(2) \, \* \z2
          - 3 \, \* (5 + 6 \, \* x + 2 \, \* x^2) \, \* \Hhh(-1,0,0)
\nn\\[0.2mm] && \mbox{}
          - (89/6 - 65/3 \, \* x - 6 \, \* x^2) \, \* \Hhh(0,0,0)
          - (701/72 + 1357/36 \, \* x - 9 \, \* x^2) \, \* \Hh(0,0)
          - (11 - 35 \, \* x 
\nn\\[0.2mm] && \mbox{}
            - 6 \, \* x^2) \, \* \H(3)
          + (1 - 14 \, \* x) \, \* \Hhh(2,1,1)
          - 2 \, \* (5 + 8 \, \* x + 3 \, \* x^2) \, \* \Hh(-1,2)
          - (8 + 17 \, \* x + 9 \, \* x^2) \, \* \Hh(-1,0)
\nn\\[0.2mm] && \mbox{}
          - 2 \, \* (1 + 16 \, \* x + 3 \, \* x^2) \, \* \Hh(-2,0)
          - 2 \, \* (1 - 6 \, \* x) \, \* \Hh(-3,0)
          + 1/288 \, \* (69407 - 68990 \, \* x)
\nn\\[0.2mm] && \mbox{}
          + 8 \, \* (1 + x) \, \* \Hhh(-2,-1,0)
          - 1/3 \, \* (370 - 293 \, \* x + 45 \, \* x^2) \, \* \z3
          - (101/9 + 85/18 \, \* x + 9 \, \* x^2) \, \* \z2
\nn\\[0.2mm] && \mbox{}
          - 1/10 \, \* (101 + 146 \, \* x) \, \* \zss
          + 2 \, \* (1 + 2 \, \* x) \, \* (17/2 \, \* \H(-1) \, \* \z3 
            + 3 \, \* \Hhh(-1,-2,0) - 9 \, \* \Hh(-1,-1) \, \* \z2 
\nn\\[0.2mm] && \mbox{}
            - 6 \, \* \Hhhh(-1,-1,-1,0) + 6 \, \* \Hhhh(-1,-1,0,0) 
	    + 6 \, \* \Hhh(-1,-1,2) + 9 \, \* \Hh(-1,0) \, \* \z2 
            - 11/2 \, \* \Hhhh(-1,0,0,0) 
\nn\\[0.2mm] && \mbox{}
            - 2 \, \* \Hhh(-1,2,0) 
            - 2 \, \* \Hhh(-1,2,1) - 6 \, \* \Hh(-1,3) 
            - 2 \, \* \Hhhh(0,0,0,0) )
          - 8 \, \* (1 + 3 \, \* x) \, \* \H(4)
          + 6 \, \* (1 + 4 \, \* x) \, \* \Hh(0,0) \, \* \z2
\nn\\[0.2mm] && \mbox{}
          - 4 \, \* (3 + 4 \, \* x) \, \* \Hh(3,0)
          + (1 + 22 \, \* x) \, \* \Hhh(2,0,0)
          - 5 \, \* (2 - 7 \, \* x) \, \* \Hh(2,0)
          + 2 \, \* (13 + 16 \, \* x + 3 \, \* x^2) \, \* \Hhh(-1,-1,0)
\nn\\[0.2mm] && \mbox{}
          - 4 \, \* (1 + 6 \, \* x) \, \* \Hh(-2,2)
          - 2 \, \* (3 + 8 \, \* x) \, \* (\Hhh(-2,0,0) + \Hh(3,1))
          - 3/2 \, \* (11 - 10 \, \* x) \, \* (\Hhh(1,1,0) - \Hh(1,2))
\nn\\[0.2mm] && \mbox{}
          + (9 - 35 \, \* x - 12 \, \* x^2) \, \* \H(0) \, \* \z2
          - (17/9 - 46/9 \, \* x) \, \* \Hh(1,0)
          + (37/3 - 47/3 \, \* x) \, \* \Hhh(1,1,1)
\nn\\[0.2mm] && \mbox{}
          + (317/9 - 313/9 \, \* x) \, \* \Hh(1,1)
          + (29/9 + 85/18 \, \* x) \, \* \H(2)
          + (61/3 - 59/3 \, \* x) \, \* \Hhh(1,0,0)
\nn\\[0.2mm] && \mbox{}
          + 4 \, \* (2 + 7 \, \* x) \, \* \H(-2) \, \* \z2
          - 12 \, \* \Hhh(2,1,0)
          + (23 + 32 \, \* x + 9 \, \* x^2) \, \* \H(-1) \, \* \z2
          - (41 + 22 \, \* x) \, \* \H(0) \, \* \z3
\nn\\[-0.5mm] && \mbox{}
          + (41/6 + 46/3 \, \* x) \, \* \Hh(2,1)
          + 1/27 \, \* (1195 - 1433 \, \* x) \, \* \H(1)
          + 1/216 \, \* (15259 + 25645 \, \* x) \, \* \H(0)
          \Big)
\nn\\ && \mbox{\hspn}
       + 8 \, \* \colour4colour{\cfs \, \* \nf}  \, \*  \Big(
            2\, \* (1 - 2 \, \* x) \, \* (7/2 \, \* \Hh(0,0) \, \* \z2 
            + 7 \, \* \H(1) \, \* \z3 + 2 \, \* \Hhh(1,-2,0) 
            + 7 \, \* \Hh(1,0) \, \* \z2 
            - 3 \, \* \Hhhh(1,0,0,0) 
\nn\\[-0.5mm] && \mbox{}
            + 5 \, \* \Hh(1,1) \, \* \z2 
	    - 4 \, \* \Hhhh(1,1,0,0) - \Hhhh(1,1,1,0) - 2 \, \* \Hhhh(1,1,1,1) 
            - 5 \, \* \Hhh(1,1,2) - 6 \, \* \Hhh(1,2,0) - 6 \, \* \Hhh(1,2,1) 
            - 7 \, \* \Hh(1,3) 
\nn\\[0.2mm] && \mbox{}
            - 2 \, \* \Hhh(2,1,0) - 5/2 \, \* \Hhh(2,1,1) - 4 \, \* \Hh(2,2) 
            - 4 \, \* \Hh(3,0) - 5 \, \* \Hh(3,1) - 7/2 \, \* \H(4) 
            - 5 \, \* \Hh(-2,0) )
          + 681/16 \, \* x
\nn\\[0.2mm] && \mbox{}
          + 2 \, \* (1 - x) \, \* (13 \, \* \H(1) \, \* \z2 
            - 11 \, \* \Hhh(1,0,0) - 5 \, \* \Hhh(1,1,0) - 11 \, \* \Hh(1,2) 
          + 4 \, \* \H(2) \, \* \z2)
          - 4 \, \* (3/5 - 2 \, \* x) \, \* \zss
\nn\\[0.2mm] && \mbox{}
          - 1357/32 
          - 2 \, \* (10 + 9 \, \* x - 3 \, \* x^2) \, \* \Hh(-1,0)
          - 4 \, \* (1 + x) \, \* (\H(-1) \, \* \z2 
          + 2 \, \* \Hhh(-1,-1,0) + 2 \, \* \Hhh(-1,0,0))
\nn\\[0.2mm] && \mbox{}
          + (59/2 - 18 \, \* x + 6 \, \* x^2) \, \* \z2
            + 4 \, \* (1 + 2 \, \* x) \, \* (\Hhh(-1,-2,0) 
            + 2 \, \* \Hhhh(-1,-1,0,0) - 1/2 \, \* \Hhhh(-1,0,0,0))
\nn\\[0.2mm] && \mbox{}
          - 4 \, \* (1 + 4 \, \* x) \, \* \Hhh(-2,0,0)
          - 2 \, \* (3 - 2 \, \* x) \, \* \Hhh(2,0,0)
          - 7/2 \, \* (5 - 2 \, \* x) \, \* \Hh(2,1)
          - 3/2 \, \* (7 - 8 \, \* x) \, \* \Hhh(1,1,1)
\nn\\[0.2mm] && \mbox{}
          - 2 \, \* (10 - x) \, \* \Hh(2,0)
          - 9/2 \, \* (11 - 4 \, \* x) \, \* \H(2)
          + (13 - 14 \, \* x) \, \* \H(0) \, \* \z3
          + 1/2 \, \* (15 - 4 \, \* x) \, \* \Hhh(0,0,0)
\nn\\[0.2mm] && \mbox{}
          - 2 \, \* (17 - 22 \, \* x) \, \* \Hh(1,0)
          + (23/2 - 2 \, \* x) \, \* \H(0) \, \* \z2
          + (25 - 11 \, \* x) \, \* \z3
          - (29 - 36 \, \* x) \, \* \Hh(1,1)
\nn\\[0.2mm] && \mbox{}
          - (43/2 - 2 \, \* x) \, \* \H(3)
          - 1/8 \, \* (77 - 397 \, \* x) \, \* \H(0)
          + 1/8 \, \* (59 + 458 \, \* x - 48 \, \* x^2) \, \* \Hh(0,0)
\nn\\[-0.5mm] && \mbox{}          
          - (78 - 329/4 \, \* x) \, \* \H(1)
          - 4 \, \* x \, \* (4 \, \* \Hh(-3,0) 
            - \Hhhh(0,0,0,0) - 2 \, \* \H(-2) \, \* \z2 
          - 4 \, \* \Hhh(-2,-1,0))
          \Big)
\nn\\ && \mbox{\hspn}
       + 2/9 \, \* \colour4colour{\ca \, \* \nfs}  \, \*  \Big(
            12 \, \* (1 - 2 \, \* x) \, \* (\H(1) \, \* \z2 + \Hhh(1,0,0) + \Hhh(1,1,0) + \Hhh(1,1,1) - \Hh(1,2))
          + 24 \, \* (1 + 2 \, \* x) \, \* \Hhh(-1,0,0)
\nn\\[-0.5mm] && \mbox{}
          + 48 \, \* (1 - x) \, \* \Hhh(0,0,0)
          - 8 \, \* (2 - 7 \, \* x) \, \* \Hh(1,1)
          + 8 \, \* (2 + 7 \, \* x) \, \* \Hh(-1,0)
          + 4 \, \* (4 + 3 \, \* x) \, \* \z2
          + 472 - 527 \, \* x
\nn\\[-0.5mm] && \mbox{}
          - 4/3 \, \* (23 - 4 \, \* x) \, \* \H(1)
          + 2 \, \* (99 + 68 \, \* x) \, \* \Hh(0,0)
          - 36 \, \* \z3 
	  + 1067/3 \, \* \H(0) 
	  + 200/3 \, \* x \, \* \H(0)
	  - 12 \, \* x \, \* \H(2)
          \Big)
\nn\\ && \mbox{\hspn}
       + 2/9 \, \* \colour4colour{\cf \, \* \nfs}  \, \*  \Big(
            12 \, \* (1 - 2 \, \* x) \, \* (\z3 + 2 \, \* \Hhh(1,0,0) - \Hhh(1,1,1) - \Hh(2,1) + 12 \, \* \Hhhh(0,0,0,0))
          + 4 \, \* (11 - 16 \, \* x) \, \* \Hh(1,0)
\qquad \nn\\[-0.5mm] && \mbox{}
          + 1/4 \, \* (4193 - 4226 \, \* x)
          - 8 \, \* (2 - 7 \, \* x) \, \* (\z2 - \Hh(1,1) - \H(2))
          + 2/3 \, \* (28 + 19 \, \* x) \, \* \H(1)
\nn\\[-0.5mm] && \mbox{}
          + 12 \, \* (43 + 10 \, \* x) \, \* \Hhh(0,0,0)
          + 17 \, \* (53 + 14 \, \* x) \, \* \Hh(0,0)
          + 1/3 \, \* (3217 - 59 \, \* x) \, \* \H(0)
          \Big)
\eea
and
\bea
\label{dPgq2x}
  && \hspn \dPgq(2)(x) \equal
         8 \, \* \colour4colour{\cas \, \* \cf}  \, \*  \Big(
            4 \, \* (1 - 2 \, \* x) \, \* \Hhh(2,1,1)
          + 1/12 \, \* (3718 - 3349 \, \* x)
          - 1/20 \, \* (366 + 193 \, \* x) \, \* \zss
\nn\\[-0.5mm] && \mbox{}
          + 16 \, \* (1 + x) \, \* \Hh(-2,2)
          + 2 \, \* (2 - 11 \, \* x) \, \* \Hhh(-2,-1,0)
          + (2 - 9 \, \* x) \, \* \Hh(-3,0)
          - (106/3 + 3 \, \* x^{-1} 
\nn\\[0.2mm] && \mbox{}
           + 173/3 \, \* x) \, \* \Hh(-1,0)
          - 1/54 \, \* (1442 - 403 \, \* x) \, \* \H(1)
          - (46/3 + 3 \, \* x^{-1} - 121/6 \, \* x) \, \* \H(1) \, \* \z2
\nn\\[0.2mm] && \mbox{}
          + (2 - x) \, \* (7 \, \* \H(1) \, \* \z3 - 2 \, \* \Hhh(1,-2,0) 
             - 4 \, \* \Hh(1,0) \, \* \z2 + 3 \, \* \Hhhh(1,0,0,0) 
             - 2 \, \* \Hh(1,1) \, \* \z2 + 4 \, \* \Hhhh(1,1,0,0) 
             + 8 \, \* \Hhh(1,1,2) 
\nn\\[0.2mm] && \mbox{}
	     + 8 \, \* \Hhhh(1,1,1,0) + 4 \, \* \Hhhh(1,1,1,1) 
             + 12 \, \* \Hhh(1,2,0) 
             + 8 \, \* \Hhh(1,2,1) + 8 \, \* \Hh(1,3) + 6 \, \* \Hhh(2,1,0) 
             + 6 \, \* \Hh(2,2) 
             - 55/6 \, \* \Hhh(1,1,1) ) 
\nn\\[0.2mm] && \mbox{}
          - 4 \, \* (2 + x) \, \* (23/4 \, \* \H(-1) \, \* \z3 
             + 5/2 \, \* \Hhh(-1,-2,0) - 13/2 \, \* \Hh(-1,-1) \, \* \z2 
             - 3 \, \* \Hhhh(-1,-1,-1,0) - 4 \, \* \Hh(-1,3)
\nn\\[0.2mm] && \mbox{}
             + 6 \, \* \Hhhh(-1,-1,0,0) + 5 \, \* \Hhh(-1,-1,2) 
             + 5 \, \* \Hh(-1,0) \, \* \z2 - 11/4 \, \* \Hhhh(-1,0,0,0) 
             - \Hhh(-1,2,0) - \Hhh(-1,2,1) )
\nn\\[0.2mm] && \mbox{}
          - 7/2 \, \* (2 + 5 \, \* x) \, \* \Hh(2,1)
          - 1/9 \, \* (5 - 148 \, \* x) \, \* \Hh(1,0)
          + 4 \, \* (6 - x) \, \* \Hh(3,1)
          + 4 \, \* (8 + x) \, \* \Hh(3,0)
\nn\\[0.2mm] && \mbox{}
          - (3 - 122/3 \, \* x) \, \* \Hh(-2,0)
          - 2 \, \* (10 + 7 \, \* x) \, \* \H(0) \, \* \z3
          - (14 + 5 \, \* x) \, \* \H(2) \, \* \z2
          - (14 + 27 \, \* x) \, \* \H(-2) \, \* \z2
\nn\\[0.2mm] && \mbox{}
          - 1/6 \, \* (14 + 41 \, \* x) \, \* (\Hhh(1,1,0) + \Hh(1,2))
          + 2 \, \* (15 - 4 \, \* x) \, \* \Hhhh(0,0,0,0)
          + 1/36 \, \* (50 + 581 \, \* x) \, \* \Hh(1,1)
\nn\\[0.2mm] && \mbox{}
          + (13/3 - 9 / x - 4/3 \, \* x) \, \* \H(-1) \, \* \z2
          + 1/3 \, \* (38 - 139 \, \* x) \, \* \Hh(2,0)
          + (38 + 11 \, \* x) \, \* \Hhh(-2,0,0)
\nn\\[0.2mm] && \mbox{}
          - (23 + 13/2 \, \* x) \, \* \Hh(0,0) \, \* \z2
          - 1/6 \, \* (47 - 419 \, \* x) \, \* \H(0) \, \* \z2
          + 1/6 \, \* (245 - 223 \, \* x) \, \* \Hhh(1,0,0)
\nn\\[0.2mm] && \mbox{}
          + (25 + 13/2 \, \* x) \, \* \H(4)
          + 2/9 \, \* (49 + 73 \, \* x) \, \* \H(2)
          + (21 + 13/2 \, \* x) \, \* \Hhh(2,0,0)
          + (47 - 5/3 \, \* x) \, \* \Hhh(0,0,0)
\nn\\[0.2mm] && \mbox{}
          + (40/3 + 6 \, \* x^{-1} + 44/3 \, \* x) \, \* \Hh(-1,2)
          - 1/3 \, \* (161 - 194 \, \* x) \, \* \z3
          - 2/9 \, \* (208 + 73 \, \* x) \, \* \z2
\nn\\[0.2mm] && \mbox{}
          + (133/3 + 6 \, \* x^{-1} + 137/3 \, \* x) \, \* \Hhh(-1,0,0)
          + (106/3 - 6 \, \* x^{-1} + 80/3 \, \* x) \, \* \Hhh(-1,-1,0)
\nn\\[-0.5mm] && \mbox{}
          + 1/6 \, \* (29 - 419 \, \* x) \, \* \H(3)
          + 1/18 \, \* (1444 - 2351 \, \* x) \, \* \Hh(0,0)
          + 1/108 \, \* (11998 + 18649 \, \* x) \, \* \H(0)
          \Big)
\nn\\&& \mbox{\hspn}
       + 8 \, \* \colour4colour{\ca \, \* \cfs}  \, \*  \Big(
            8/3 \, \* (1 + x) \, \* \H(3)
          + (2 - 7 \, \* x) \, \* (2 \, \* \Hh(-3,0) - \Hhh(2,1,1))
          - 1/216 \, \* (13037 - 4423 \, \* x) \, \* \H(0)
\nn\\[-0.5mm] && \mbox{}
          - 1/5 \, \* (46 + 49 \, \* x) \, \* \zss
          - (6 - 9 \, \* x^{-1} + x) \, \* \Hh(-1,0)
          - 1/48 \, \* (2911 - 11273/6 \, \* x)
\nn\\[0.2mm] && \mbox{}
          + (2 - x) \, \* (3 \, \* \H(1) \, \* \z3 
            + 10 \, \* \Hhh(1,-2,0) + 2 \, \* \Hh(1,0) \, \* \z2 
            + 5 \, \* \Hhhh(1,0,0,0) - 4 \, \* \Hh(1,1) \, \* \z2 
            + 10 \, \* \Hhhh(1,1,0,0) 
\nn\\[0.2mm] && \mbox{}
            - 6 \, \* \Hhhh(1,1,1,0) 
            - 8 \, \* \Hhhh(1,1,1,1) - 2 \, \* \Hhh(1,1,2) 
            - 4 \, \* \Hhh(1,2,1) + 2 \, \* \Hhh(2,1,0) + 4 \, \* \Hh(2,2) 
	    + 4 \, \* \Hh(3,0))
\nn\\[0.2mm] && \mbox{}
          + (2 + x) \, \* (11 \, \* \H(-1) \, \* \z3 + 6 \, \* \Hhh(-1,-2,0) 
              - 10 \, \* \Hh(-1,-1) \, \* \z2 - 12 \, \* \Hhhh(-1,-1,-1,0) 
              + 8 \, \* \Hhhh(-1,-1,0,0) 
\nn\\[0.2mm] && \mbox{}
              + 4 \, \* \Hhh(-1,-1,2) 
              - 2 \, \* \Hh(-1,0) \, \* \z2 + 3 \, \* \Hhhh(-1,0,0,0) 
              - 4 \, \* \Hhh(-1,2,1) + 6 \, \* \Hhh(-1,0,0) )
           - 3 \, \* (2 - 3 \, \* x) \, \* \Hh(-2,0)
\nn\\[0.2mm] && \mbox{}
          - (2 + 7 \, \* x) \, \* \Hh(3,1)
          - 2 \, \* (4 - 5 \, \* x) \, \* \H(2) \, \* \z2
          + 4 \, \* (4 - x) \, \* \H(0) \, \* \z3
          + 2 \, \* (4 + x) \, \* (\Hh(-1,2) + 2 \, \* \Hhh(0,0,0,0))
\nn\\[0.2mm] && \mbox{}
          - 2/3 \, \* (5 - x) \, \* \Hh(2,0)
          - 4 \, \* (7 + 4 \, \* x) \, \* \Hhh(-1,-1,0)
          - 1/3 \, \* (8 - 43 \, \* x) \, \* \Hh(2,1)
          + 2 \, \* (8 - 7 \, \* x) \, \* \Hhh(2,0,0)
\nn\\[0.2mm] && \mbox{}
          + 5/3 \, \* (8 - x) \, \* \Hh(1,0)
          - (10 - 3 \, \* x) \, \* \Hhh(1,0,0)
          + 1/6 \, \* (10 + 13 \, \* x) \, \* \Hh(1,2)
          - 2 \, \* (11 + 5 \, \* x) \, \* \H(-1) \, \* \z2
\nn\\[0.2mm] && \mbox{}
          - 2/3 \, \* (13 + 4 \, \* x) \, \* \H(0) \, \* \z2
          - 1/18 \, \* (31 + 274 \, \* x) \, \* \Hh(1,1)
          + 2/9 \, \* (32 - 73 \, \* x) \, \* \H(2)
\nn\\[0.2mm] && \mbox{}
          + 7/6 \, \* (32 - 25 \, \* x) \, \* \z3
          + 1/3 \, \* (89 - 88 \, \* x) \, \* \Hhh(0,0,0)
          - 1/6 \, \* (70 - 71 \, \* x) \, \* \Hhh(1,1,0)
          + 8 \, \* \H(4)
\nn\\[0.2mm] && \mbox{}
          + 1/3 \, \* (73 - 23 \, \* x) \, \* \Hhh(1,1,1)
          + 1/6 \, \* (74 - 61 \, \* x) \, \* \H(1) \, \* \z2
          - 6 \, \* x \, \* (\Hhh(-2,0,0) 
            - \H(-2) \, \* \z2 - 2 \, \* \Hhh(-2,-1,0) ) 
\nn\\[-0.5mm] && \mbox{}
          - 2/9 \, \* (59 - 73 \, \* x) \, \* \z2
          + 1/108 \, \* (586 + 613 \, \* x) \, \* \H(1)
          + 1/72 \, \* (730 - 821 \, \* x) \, \* \Hh(0,0)
          - 4 \, \* \Hh(0,0) \, \* \z2
          \Big)
\nn\\ && \mbox{\hspn}
       + 8\, \* \colour4colour{\cft}  \, \*  \Big(
            8 \, \* (1 + x) \, \* (\H(-1) \, \* \z2 + 2 \, \* \Hhh(-1,-1,0) 
            + 1/2 \, \* \Hhh(-1,0,0)+\Hhh(-2,0,0) )
          + 2331/16 - 4547/32 \, \* x
\qquad \nn\\[-0.5mm] && \mbox{}
          - 12 \, \* (1 - x) \, \* \H(0) \, \* \z3
          - 2 \, \* (13 + 3 \, \* x^{-1} + 12 \, \* x) \, \* \Hh(-1,0)
          + (3 + 93/10 \, \* x) \, \* \zss
          - (2 - 5 \, \* x) \, \* \Hhh(2,0,0)
\nn\\[0.2mm] && \mbox{}
          + 2 \, \* (2 - 3 \, \* x) \, \* \H(2) \, \* \z2
          - 2 \, \* (2 - x) \, \* (3/2 \, \* \Hh(0,0) \, \* \z2 
            + 8 \, \* \H(1) \, \* \z3 + 6 \, \* \Hhh(1,-2,0) 
            + 2 \, \* \Hh(1,0) \, \* \z2 
            - 3 \, \* \Hh(1,1) \, \* \z2 
\nn\\[0.2mm] && \mbox{} 
            + 3 \, \* \Hhhh(1,1,0,0) + \Hhhh(1,1,1,0) 
	    - 2 \, \* \Hhhh(1,1,1,1) + 3 \, \* \Hhh(1,1,2) 
            + 2 \, \* \Hhh(1,2,0) + 2 \, \* \Hhh(1,2,1)
	    + 1/2 \, \* \Hhh(2,1,1) + \Hh(2,2) 
\nn\\[0.2mm] && \mbox{} 
            + 3/2 \, \* \Hh(3,1) - 3/2 \, \*\H(4) + \Hh(-2,0))
          - 4 \, \* (2 + x) \, \* (\Hhh(-1,-2,0) + 2 \, \* \Hhhh(-1,-1,0,0) 
             - \Hh(-1,0) \, \* \z2 
\nn\\[0.2mm] && \mbox{} 
             - 1/2 \, \* \Hhhh(-1,0,0,0) )
          - (4 - 7 \, \* x) \, \* \Hh(1,2)
          - 4 \, \* (5 - 2 \, \* x) \, \* \Hh(1,0)
          - 2 \, \* (4 + x) \, \* (\Hhhh(0,0,0,0) + 3/4 \, \* \Hhh(1,1,1))
\nn\\[0.2mm] && \mbox{} 
          - (4 - x) \, \* (\H(1) \, \* \z2 - 3/2 \, \* \Hhh(1,0,0) 
            - 2 \, \* \Hhh(1,1,0) )
          - 1/2 \, \* (8 - x) \, \* \H(3)
          - 1/4 \, \* (8 - 7 \, \* x) \, \* \Hh(1,1)
\nn\\[0.2mm] && \mbox{} 
          + 1/2 \, \* (8 + 7 \, \* x) \, \* \Hh(2,1)
          - (23 + 3 \, \* x) \, \* \Hhh(0,0,0)
          - 1/2 \, \* (25 + 23 \, \* x) \, \* \z2
          - 3/8 \, \* (26 + 31 \, \* x) \, \* \Hh(0,0)
\nn\\[0.2mm] && \mbox{} 
          - 1/2 \, \* (27 - 23 \, \* x) \, \* \H(2)
          + 1/4 \, \* (32 - 23 \, \* x) \, \* \H(1)
          - (52 - 21 \, \* x) \, \* \z3
          + 1/8 \, \* (389 + 721 \, \* x) \, \* \H(0)
\nn\\[-0.5mm] && \mbox{} 
          - x \, \* (8 \, \* \Hhh(-2,-1,0) + 4 \, \* \H(-2) \, \* \z2 
            + 1/2 \, \* \H(0) \, \* \z2 - 2 \, \* \Hh(2,0) 
            - 16 \, \* \Hh(-3,0))
          \Big)
\nn\\[0.2mm] && \mbox{\hspn}
       + 8/3 \, \* \colour4colour{\ca \, \* \cf \, \* \nf}  \, \*  \Big(
            369/4 - 643/6 \, \* x
          - 2 \, \* (10 - 3 \, \* x^{-1} + 8 \, \* x) \, \* \Hh(-1,0)
          - 10/3 \, \* (1 + x) \, \* \Hh(1,0)
\nn\\[-0.5mm] && \mbox{}
          + 3 \, \* (2 - 3 \, \* x) \, \* \Hh(2,1)
          - (2 - x) \, \* (\H(1) \, \* \z2 + 2 \, \* \Hhh(1,0,0) 
            - 5 \, \* \Hhh(1,1,0) - 5 \, \* \Hhh(1,1,1) - 5 \, \* \Hh(1,2) 
            - 2 \, \* \Hh(2,0) ) 
\nn\\[0.2mm] && \mbox{}
          - 4 \, \* (2 + x) \, \* (2 \, \* \H(-1) \, \* \z2 
            + 2 \, \* \Hhh(-1,-1,0) 
            - \Hhh(-1,0,0) - \Hh(-1,2) - \H(3))
          - 1/6 \, \* (4 + 91 \, \* x) \, \* \H(2)
\nn\\[0.2mm] && \mbox{}
          - 4 \, \* (5 + x) \, \* \H(0) \, \* \z2
          - 2 \, \* (6 - 5 \, \* x) \, \* \Hh(-2,0)
          - 2 \, \* (23 - 14 \, \* x) \, \* \z3
          - 1/3 \, \* (41 + 74 \, \* x) \, \* \Hh(0,0)
\nn\\[0.2mm] && \mbox{}
          + (194/9 + 3143/36 \, \* x) \, \* \H(0)
          - 1/3 \, \* (80 - 37 \, \* x) \, \* \Hh(1,1)
          + 1/9 \, \* (91 + 118 \, \* x) \, \* \H(1)
          + 2 \, \* x \, \* \Hhh(0,0,0)
\nn\\[-0.5mm] && \mbox{}
          - (58/3 - 91/6 \, \* x) \, \* \z2
          \Big)
\nn\\ && \mbox{\hspn}
       + 8/3 \, \* \colour4colour{\cfs \, \* \nf}  \, \*  \Big(
            2 \, \* (2 - x) \, \* (31/2 \, \* \z3 + 6 \, \* \Hh(-2,0) 
            - \H(1) \, \* \z2 + 3/2 \, \* \Hhh(1,0,0) + 2 \, \* \Hhh(1,1,0) 
            - 5/2 \, \* \Hhh(1,1,1) 
\nn\\[-0.5mm] && \mbox{}
            + \Hh(1,2) 
            + \Hh(2,0) + 1/2 \, \* \Hh(2,1) + \H(3) 
          - 6 \, \* \Hhhh(0,0,0,0) )
          - 1/36 \, \* (3155 + 3893 \, \* x) \, \* \H(0)
          - 1733/8 
\nn\\[0.2mm] && \mbox{}
          + 10043/48 \, \* x
          - 31/9 \, \* (1 + x) \, \* \H(1)
          - (4 + x) \, \* (\Hh(1,0) + 4/3 \, \* \H(2))
          + 2 \, \* (10 + x) \, \* \H(0) \, \* \z2
\nn\\[0.2mm] && \mbox{}
          + 4/3 \, \* (22 + x) \, \* \z2
          + 1/3 \, \* (41 - 4 \, \* x) \, \* \Hh(1,1)
          - (50 + 29 \, \* x) \, \* \Hhh(0,0,0)
          + 6 \, \* (4 - x^{-1} + 5 \, \* x) \, \* \Hh(-1,0)
\nn\\[-0.5mm] && \mbox{}
          - 1/12 \, \* (674 - 457 \, \* x) \, \* \Hh(0,0)
          \Big)
\nn\\[-0.5mm] && \mbox{\hspn}
       + 8/3 \, \* \colour4colour{\cf \, \* \nfs}  \, \*  \Big(
            (2 - x) \, \* \Hh(1,1)
          - 1/3 \, \* (4 + x) \, \* \H(1)
          - 2 + 5/3 \, \* x
          \Big)
\:\: .
\eea
Finally the $x$-space expression corresponding to Eq.~(\ref{dPgg2N}) for the 
polarized NNLO gluon-gluon splitting function reads
\bea
\label{dPgg2x}
 && \hspn \dPgg(2)(x) \equal
        16 \, \* \colour4colour{\cat}  \, \*  \Big(
          4 \, \* \dpgg( - x) \, \* (- 11/8 \, \* \zss
            + \Hh(-3,0) - 4 \, \* \H(-2) \, \* \z2 - 2 \, \* \Hhh(-2,-1,0) 
            + 3 \, \* \Hh(-2,2) 
\qquad \nn\\[-0.5mm] && \mbox{}
            + 9/2 \, \* \Hhh(-2,0,0) - 3 \, \* \H(-1) \, \* \z3 
            - 2 \, \* \Hhh(-1,-2,0) + 4 \, \* \Hh(-1,-1) \, \* \z2 
            - 6 \, \* \Hhhh(-1,-1,0,0) - 4 \, \* \Hhh(-1,-1,2) 
\nn\\[0.2mm] && \mbox{} 
            - 9/2 \, \* \Hh(-1,0) \, \* \z2 + 4 \, \* \Hhhh(-1,0,0,0) 
            + \Hhh(-1,2,0) + 4 \, \* \Hh(-1,3) + 5/4 \, \* \H(0) \, \* \z3 
            + 2 \, \* \Hh(0,0) \, \* \z2 - \Hhhh(0,0,0,0) 
\nn\\[0.2mm] && \mbox{} 
	       - 1/2 \, \* \H(2) \, \* \z2 - 1/2 \, \* \Hh(3,0) 
            - 2 \, \* \H(4) + 11/24 \, \* \H(0) \, \* \z2  
            + 67/36 \, \* ( \z2 + 2 \, \* \Hh(-1,0) - \Hh(0,0) ) )
\nn\\[0.2mm] && \mbox{} 
          + 4 \, \* \dpgg(x) \, \* (
              245/96 - 3/40  \, \* \zss
            - \Hh(-3,0) + 3/2 \, \* \H(-2) \, \* \z2 + \Hhh(-2,-1,0) 
            - \Hhh(-2,0,0) - \Hh(-2,2) 
\nn\\[0.2mm] && \mbox{} 
            - 7/4 \, \* \H(0) \, \* \z3 
            - 2 \, \* \Hh(0,0) \, \* \z2 + \Hhhh(0,0,0,0) 
            - 3/2 \, \* \H(1) \, \* \z3 - \Hhh(1,-2,0) 
            - 3/2 \, \* \Hh(1,0) \, \* \z2 + 2 \, \* \Hhhh(1,0,0,0) 
\nn\\[0.2mm] && \mbox{} 
            + 2 \, \* \Hhhh(1,1,0,0) + 2 \, \* \Hhh(1,2,0) 
	       + 2 \, \* \Hh(1,3) - \H(2) \, \* \z2 
            + 5/2 \, \* \Hhh(2,0,0) + 2 \, \* \Hhh(2,1,0) + 2 \, \* \Hh(2,2) 
            + 5/2 \, \* \Hh(3,0) 
\nn\\[0.2mm] && \mbox{} 
            + 2 \, \* \Hh(3,1) + 2 \, \* \H(4)
            + 11/12  \, \* \z3 + 11/12  \, \* \Hh(-2,0) 
            + 11/24 \, \* \Hhh(1,0,0) + 11/24 \, \* \H(3)
\nn\\[0.2mm] && \mbox{} 
            - 67/36 \, \* ( \z2 - \Hh(0,0) -  2\, \* \Hh(1,0) 
            - 2 \, \* \H(2) ) + 1/24 \, \* \H(0) )
          - 1/3 \, \* (72 - 185 \, \* x - 22 \, \* x^2) \, \* \H(0) \, \* \z2
\nn\\[0.2mm] && \mbox{} 
          - 1/3 \, \* (32 - 161 \, \* x - 11 \, \* x^2) \, \* \Hh(-2,0)
          + 4 \, \* (1 - 5 \, \* x) \, \* \Hh(-3,0)
          - 1/6 \, \* (312 - 393 \, \* x - 55 \, \* x^2) \, \* \z3
\nn\\[0.2mm] && \mbox{} 
          + (1 - x) \, \* ( 5579/18 
            + 4 \, \* \H(-2) \, \* \z2 + 8 \, \* \Hhh(-2,-1,0) 
            + 12 \, \* \Hhh(-2,0,0)
          - 21/2 \, \* \H(1) \, \* \z2 + 37 \, \* \Hhh(1,0,0)
\nn\\[0.2mm] && \mbox{} 
            + 1/18 \, \* \H(1) - 19/2 \, \* \Hh(1,0) )
          - 1/5 \, \* (43 + 33 \, \* x) \, \* \zss
          - 8 \* (1 + 3 \, \* x) \, \* \H(0) \, \* \z3
          - 2 \, \* (11 + 13 \, \* x) \, \* \Hh(0,0) \, \* \z2
\nn\\[0.2mm] && \mbox{} 
          + (1 + x) \, \* (
            21 \, \* \Hhh(-1,-1,0) - 25/2 \, \* \H(-1) \, \* \z2 
            + 65 \, \* \Hhh(-1,0,0) + 23 \, \* \Hh(-1,2)
            - 4 \, \* \H(2) \, \* \z2 + 10 \, \* \Hhh(2,0,0) 
\nn\\[0.2mm] && \mbox{} 
            + 16 \, \* \Hh(3,0) + 26 \, \* \H(4)
          - 215/3 \, \* \Hh(-1,0) )
          - 1/9 \, \* (74 - 97 \, \* x) \, \* \H(2)
          + 1/3 \, \* (77 - 115 \, \* x) \, \* \Hh(2,0)
\nn\\[0.2mm] && \mbox{} 
          + 1/3 \, \* (40 - 185 \, \* x - 11 \, \* x^2) \, \* \H(3)
          - 1/9 \, \* (571 + 97 \, \* x) \, \* \z2
          + 1/3 \, \* (158 - 87 \, \* x - 11 \, \* x^2) \, \* \Hhh(0,0,0)
\nn\\[0.2mm] && \mbox{} 
          + 1/12 \, \* (1019 - 1489 \, \* x) \, \* \Hh(0,0)
          + 1/216 \, \* (24625 + 40069 \, \* x) \, \* \H(0)
          - 11/6 \, \* (x^{-1} - x^2) \, \* \H(1) \, \* \z2
\nn\\[0.2mm] && \mbox{} 
          + 28 \, \* \Hhhh(0,0,0,0)
          - 11/2 \, \* (x^{-1} + x^2) \, \* (\H(-1) \, \* \z2 
          + 2/3 \, \* \Hhh(-1,-1,0) - 2/3 \, \* \Hhh(-1,0,0) 
          - 2/3 \, \* \Hh(-1,2))
\nn\\[-0.5mm] && \mbox{} 
	  + \delta(1 - x) \, \*  (
            79/32
          - 5 \, \* \z5
          + 67/6 \, \* \z3
          + 1/6 \, \* \z2
          - \z2 \, \* \z3
          + 11/24 \, \* \zss
          )
          \Big)
\nn\\ && \mbox{\hspn}
       + 8 \, \* \colour4colour{\cas \, \* \nf}  \, \*  \Big(
            2/3 \, \* \dpgg(x) \, \* ( 10/3  \, \* \z2 
            - 10/3 \, \* \Hh(0,0) - 20/3 \, \* \Hh(1,0) 
            - 20/3 \, \* \H(2) - 209/36 - 8  \, \* \z3 
\nn\\[-0.5mm] && \mbox{}
            - 2 \, \* \Hh(-2,0) - \Hhh(1,0,0) - \H(3) - 1/2 \, \* \H(0) )
          + 2/9 \, \* \dpgg( - x) \, \* ( 10 \, \* \Hh(0,0) 
              - 10 \, \* \z2 - 20 \, \* \Hh(-1,0) 
\nn\\[0.2mm] && \mbox{}
              - 3 \, \* \H(0) \, \* \z2 )
          - 1/6 \, \* (51 - 61 \, \* x - 16 \, \* x^2) \, \* \H(0) \, \* \z2
          - 1/18 \, \* (146 + 227 \, \* x + 36 \, \* x^2) \, \* \Hh(0,0)
\nn\\[0.2mm] && \mbox{}
          - 1/3 \, \* (23 + 43 \, \* x - 4 \, \* x^2) \, \* \Hh(-2,0)
          - 1/3 \, \* (1 - 12 \, \* x + 4 \, \* x^2) \, \* \Hhh(0,0,0)
          - 2 \, \* (1 - 5 \, \* x) \, \* \Hh(-3,0)
\nn\\[0.2mm] && \mbox{}
          + 2 \, \* (1 - x) \, \* (512/9 
          + 3 \, \* \H(-2) \, \* \z2 + 6 \, \* \Hhh(-2,-1,0) 
          - 3 \, \* \Hhh(-2,0,0)
          - 11/2 \, \* \H(1) \, \* \z2 + 11/4 \, \* \Hhh(1,0,0)
\nn\\[0.2mm] && \mbox{}
          + 1087/72 \, \* \H(1)
          - 2 \, \* \Hh(1,0) )
          + (1 + x) \, \* (
            7 \, \* \H(-1) \, \* \z2 + 22 \, \* \Hhh(-1,-1,0) 
            - 9 \, \* \Hhh(-1,0,0) + 4 \, \* \Hh(-1,2) 
\nn\\[0.2mm] && \mbox{}
            - 4/3 \, \* \Hh(2,0)
            - 6 \, \* \H(2) \, \* \z2 + 3 \, \* \Hhh(2,0,0) + 3 \, \* \H(4)
            - 19 \, \* \Hh(-1,0) )
          - 2/39 \, \* (507 - 195 \, \* x - 65 \, \* x^2) \, \* \z3
\nn\\[0.2mm] && \mbox{}
          - 1/18 \, \* (499 + 301 \, \* x - 36 \, \* x^2) \, \* \z2
          + 3/10 \, \* (13 + 23 \, \* x) \, \* \zss
          + 1/6 \, \* (5 - 61 \, \* x - 8 \, \* x^2) \, \* \H(3)
\nn\\[0.2mm] && \mbox{}
          - (5 + 3 \, \* x) \, \* \Hh(0,0) \, \* \z2
          + 1/18 \, \* (157 + 301 \, \* x) \, \* \H(2)
          + 1/108 \, \* (2422 + 7609 \, \* x) \, \* \H(0)
          - 12 \, \* \H(0) \, \* \z3
\nn\\[0.2mm] && \mbox{}
          - 2/3 \, \* (x^{\, -1} - x^2) \, \* \H(1) \, \* \z2
          - 2 \, \* (x^{\, -1} + x^2) \, \* (\H(-1) \, \* \z2 
          + 2/3 \, \* \Hhh(-1,-1,0) 
          - 2/3 \, \* \Hhh(-1,0,0) 
\nn\\[-0.5mm] && \mbox{}
          - 2/3 \, \* \Hh(-1,2) - \Hh(-1,0))
          + 2 \, \* \Hhhh(0,0,0,0)
	  - 1/3 \, \* \delta(1 - x)  \, \*  (
            233/48
          + 10 \, \* \z3
          + \z2
          + 1/2 \, \* \zss
          )
          \Big)
\nn\\ && \mbox{\hspn}
       + 8/3 \, \* \colour4colour{\ca \, \* \cf \, \* \nf}  \, \*  \Big(
            4 \, \* \dpgg(x) \, \* (3 \, \* \z3 - 55/16 )
          + 3 \, \* (1 - x) \, \* (
              8 \, \* \Hhh(-2,0,0) - 7507/27 
            - 16 \, \* \H(-2) \, \* \z2 
\nn\\[-0.5mm] && \mbox{}
            - 32 \, \* \Hhh(-2,-1,0) + 30 \, \* \H(1) \, \* \z2 
            - 29 \, \* \Hhh(1,0,0)
            - 10 \, \* \Hhh(1,1,0) - 10 \, \* \Hhh(1,1,1)
            - 26/3 \, \* \H(1,0) - 65/6 \, \* \Hh(1,1)
\nn\\[0.2mm] && \mbox{}
            - 1127/18 \, \* \H(1) )
          + 6 \, \* (1 + x) \, \* (
              61/6 \, \* \Hh(-1,0)
            - 11 \, \* \H(-1) \, \* \z2 - 30 \, \* \Hhh(-1,-1,0) 
            + 3 \, \* \Hhh(-1,0,0) 
\nn\\[0.2mm] && \mbox{}
            - 4 \, \* \Hh(-1,2)
            + 6 \, \* \Hh(0,0) \, \* \z2 + 8 \, \* \H(2) \, \* \z2 
            - 7 \, \* \Hhh(2,0,0) - 2 \, \* \Hhh(2,1,0) 
            - 2 \, \* \Hhh(2,1,1) - 4 \, \* \Hh(3,0) - \Hh(3,1) 
            - 6 \, \* \H(4) )
\nn\\[0.2mm] && \mbox{}
          + (125 + 38 \, \* x - 20 \, \* x^2) \, \* \z3
          + 1/6 \, \* (848 + 341 \, \* x - 108 \, \* x^2) \, \* \z2
          - 1/18 \, \* (8363 + 3362 \, \* x) \, \* \H(0)
\nn\\[0.2mm] && \mbox{}
          - (181 + 88 \, \* x - 8 \, \* x^2) \, \* \Hhh(0,0,0)
          - 1/6 \, \* (1723 - 692 \, \* x - 108 \, \* x^2) \, \* \Hh(0,0)
          - 3/5 \, \* (43 + 83 \, \* x) \, \* \zss
\nn\\[0.2mm] && \mbox{}
          - (32 - 43 \, \* x - 8 \, \* x^2) \, \* \H(3)
          - 24 \, \* (3 - 2 \, \* x) \, \* \Hhhh(0,0,0,0)
          + 6 \, \* (9 - x) \, \* \H(0) \, \* \z3
          - (19 - 11 \, \* x) \, \* \Hh(2,1)
\nn\\[0.2mm] && \mbox{}
          + 8 \, \* (3 + 12 \, \* x - x^2) \, \* \Hh(-2,0)
          + (56 - 43 \, \* x - 16 \, \* x^2) \, \* \H(0) \, \* \z2
          - 1/6 \, \* (482 + 341 \, \* x) \, \* \H(2)
\nn\\[0.2mm] && \mbox{}
          - (38 - 37 \, \* x) \, \* \Hh(2,0)
          + 4 \, \* (x^{\,-1} - x^2) \, \* \H(1) \, \* \z2
          + 4 \, \* (x^{\,-1} + x^2) \, \* (3 \, \* \H(-1) \, \* \z2 
            + 2 \, \* \Hhh(-1,-1,0) 
\nn\\[-0.5mm] && \mbox{}
            - 2 \, \* \Hhh(-1,0,0) 
          - 2 \, \* \Hh(-1,2)
          - 9/2 \, \* \Hh(-1,0))
          - 48 \, \* x \, \* \H(-3,0)
	  - 241/48 \, \* \delta(1 - x)
          \Big)
\nn\\ && \mbox{\hspn}
       + 8 \, \* \colour4colour{\cfs \, \* \nf}  \, \*  \Big(
            8 \, \* (1 - x) \, \* ( \H(-2) \, \* \z2
            + 1 + 2 \, \* \Hhh(-2,-1,0) - \Hhh(-2,0,0) 
            - 2\, \* \H(1) \, \* \z2 
            + 11/8 \, \* \Hhh(1,0,0) 
\nn\\[-0.5mm] && \mbox{}
            + 5/4 \, \* ( \Hhh(1,1,0) + \Hhh(1,1,1) )
            - 7/8 \, \* \Hh(1,0) 
            + 13/16 \, \* \Hh(1,1) + 41/16 \, \* \H(1) )
          + 4 \, \* (1 + x) \, \* (4 \, \* \H(-1) \, \* \z2 
\nn\\[0.2mm] && \mbox{}
            + 8 \, \* \Hhh(-1,-1,0) - 4 \, \* \Hhh(-1,0,0)
            + \Hh(0,0) \, \* \z2 - \Hhhh(0,0,0,0) 
            - 2 \, \* \H(2) \, \* \z2 + 3/2 \, \* \Hhh(2,0,0) 
            + \Hhh(2,1,0) + \Hhh(2,1,1) 
\nn\\[0.2mm] && \mbox{}
            + 1/2 \, \* \Hh(3,1) 
            - \H(4) + 5/2 \, \* \Hh(-1,0) )
          + (8 - 19/2 \, \* x + 4 \, \* x^2) \, \* \z2
          - (23 + 3/2 \, \* x + 4 \, \* x^2) \, \* \Hh(0,0)
\nn\\[0.2mm] && \mbox{}
          + (9 + 13 \, \* x) \, \* \zss
          - 2 \, \* (1 - 7 \, \* x) \, \* \H(0) \, \* \z3
          + 2 \, \* (2 - 3 \, \* x) \, \* \Hh(2,1)
          + 2 \, \* (4 - x) \, \* (\H(0) \, \* \z2 - \H(3))
\nn\\[0.2mm] && \mbox{}
          - 2 \, \* (3 + 4 \, \* x) \, \* \Hh(2,0)
          + (2 + 19/2 \, \* x) \, \* \H(2)
          - 5/2 \, \* (5 - 2 \, \* x) \, \* \H(0)
          - 2 \, \* (7 - 3 \, \* x) \, \* \Hhh(0,0,0)
\nn\\[-0.5mm] && \mbox{}
          - 2 \, \* (5 + 21 \, \* x) \, \* \z3
          + 4 \, \* (x^{-1} + x^2) \, \* \Hh(-1,0)
          - 16 \, \* x \, \* (2 \, \* \Hh(-2,0) - \Hh(-3,0))
	  + 1/8 \, \* \delta(1 - x)
          \Big)
\nn\\ && \mbox{\hspn}
       + 2/27 \, \* \colour4colour{\ca \, \* \nfs}  \, \*  \Big(
          - 8 \, \* \dpgg(x)
          + 48 \, \* (1 + x) \, \* (\z2 - 1/2 \, \* \Hh(0,0) - \H(2))
          - 3 \, \* (1 - x) \, \* ( 33 + 41 \, \* \H(1) )
\nn\\[-1mm] && \mbox{}
          - (56 - 67 \, \* x) \, \* \H(0) 
          + 87/4 \, \* \delta(1 - x)
          \Big)
\nn\\&& \mbox{\hspn}
       + 2/27 \, \* \colour4colour{\cf \, \* \nfs}  \, \*  \Big(
          - 4 \, \* (1 - x) \, \* ( 146 
            + 90 \, \* \Hh(1,0) + 45 \, \* \Hh(1,1)
            + 78 \, \* \H(1) )
          - 72 \, \* (1 + x) \, \* (\z3 - 2 \, \* \H(0) \, \* \z2 
\nn\\[-0.5mm] && \mbox{}
            + \Hhh(0,0,0) + 2 \, \* \Hh(2,0) + \Hh(2,1) + 2 \, \* \H(3))
          + 24 \, \* (13 - 8 \, \* x) \, \* (\z2 - \H(2))
          - 12 \, \* (7 - 23 \, \* x) \, \* \Hh(0,0)
\nn\\[-0.5mm] && \mbox{}
          - 52 \, \* (5 - x) \, \* \H(0)
	     + 33/2 \, \* \delta(1 - x)
          \Big)
\:\: .
\eea
The functions (\ref{dPns2x}) -- (\ref{dPgg2x}) are shown in Figs.~\ref{dPqqFig}
-- \ref{dPggFig} for $\nf = 3$ effectively massless quark flavours. For the 
numerical evaluation of the harmonic polylogarithms we have made use of
Ref.~\cite{HPLnum}.

Except for the case of $\Delta P_{\rm gq}$, the respective first two terms in 
the expansion of the entries of the matrix (\ref{Sevol}) powers of $\x1$ are
identical to their unpolarized counterparts, i.e., Eq.~(\ref{dPij1xto1}) holds 
also for the differences $\,\delta_{ik}^{\,(2)}(x)$ defined in 
Eq.~(\ref{delta}). The NNLO counterpart to Eq.~(\ref{dPgq1xto1}) is
\bea
  && \hspn \hspn \delta_{\,\rm gq}^{\,(2)}(x) \equal
          \Lnt(3) \: \: \* 8\, \* \cf\, \* (\ca - \cf)^2
\nn \\[1mm] && \mbox{\hspn}
       +  \Lnt(2)\, \* \Bigg[ \,
            { 2 \over 3 }\: \* \cf\, \* (\ca - \cf)\, \* 
              (77\, \* \ca - 45\, \* \cf)
          - { 28 \over 3 }\: \* \cf\, \* (\ca - \cf)\, \* \nf
         \Bigg]
\nn \\[1mm] && \mbox{\hspn}
       +  \LntO\, \* \Bigg[ \,
            { 1870 \over 9 }\: \* \cf\, \* \cas
          - { 2260 \over 9 }\: \* \cfs\, \* \ca
          + 54\, \* \cft
\nn \\[-1mm] && \mbox{}
          - 8\, \* \z2\, \* \cf\, \* (\ca - \cf)\, \* (5\, \* \ca - 2\, \* \cf)
          - { 424 \over 9 }\: \* \cf\, \* \ca\, \* \nf
          + { 304 \over 9 }\: \* \cfs\, \* \nf
          + { 8 \over 3 }\: \* \cf\, \* \nfs
         \Bigg]
\nn \\ && \mbox{\hspn}
       +  \cf\, \* \cas\, \* \bigg(\, { 2068 \over 9 } 
            - { 154 \over 3 }\: \* \z2 \bigg)
          - \cfs\, \* \ca\, \* \bigg( \, { 466 \over 3 } - 30\, \* \z2 \bigg)
          + 24\, \* \cft
          + { 52 \over 3 }\: \* \cfs\, \* \nf
          + { 40 \over 9 }\: \* \cf\, \* \nfs
\nn \\[1mm] && \mbox{}
          + 8\, \* \z3\, \* \cf\, \* (\ca - \cf)\, \* (5\, \* \ca + 4\, \* \cf)
          - \cf\, \* \ca\, \* \nf\, \* \bigg( \, { 632 \over 9 } 
            - { 28 \over 3 }\: \* \z2 \bigg)
\nn \\ && \mbox{\hspn}
       +  (1 - x)\, \* \Lnt(3) \: \: \* 8\, \* \cf\, \* (\ca - \cf)^2
\nn \\[1mm] && \mbox{\hspn}
       +  (1 - x)\, \* \Lnt(2) \, \* \Bigg[ \,
            { 2 \over 3 }\: \* \cf\, \* (\ca - \cf)\, \* 
              (41\, \* \ca + 15\, \* \cf \bigg)
          - { 28 \over 3 }\: \* \cf\, \* \nf\, \* (\ca - \cf)
         \Bigg]
\nn \\[1mm] && \mbox{\hspn}
       +  (1 - x)\, \* \LntO \, \* \Bigg[ \,
            { 1690 \over 9 }\: \* \cf\, \* \cas
          - { 1504 \over 9 }\: \* \cfs\, \* \ca
          + 22\, \* \cft
          + { 16 \over 9 }\: \* \cfs\, \* \nf
          + { 8 \over 3 }\: \* \cf\, \* \nfs
\nn \\[-1mm] && \mbox{}
          - 8\, \* \z2\, \* \cf\, \* (\ca - \cf)\, \* (5\, \* \ca - 2\, \* \cf)
          - { 208 \over 9 }\: \* \cf\, \* \ca\, \* \nf
         \Bigg]
\\ && \mbox{\hspn}
       +  (1 - x)\, \* \Bigg[
            \cf\, \* \cas\, \* \bigg( \, { 104 \over 3 } 
               - { 34 \over 3 }\: \* \z2 \bigg)
          + \cfs\, \* \ca\, \* \bigg( \, { 574 \over 9 } - 42\, \* \z2 \bigg)
          - \cft\, \* (16 - 32\, \* \z2)
\nn \\ && \mbox{}
          + 8\, \* \z3\, \* \cf\, \* (\ca - \cf)\, \* (5\, \* \ca + 4\, \* \cf)
          - \cf\, \* \ca\, \* \nf\, \* \bigg( \, { 280 \over 9 } 
              - { 28 \over 3 }\, \* \z2 \bigg)
          - { 4 \over 9 }\, \* \cfs\, \* \nf
          - { 8 \over 9 }\, \* \cf\, \* \nfs
         \Bigg]
\nn \\ && \mbox{\hspn} + {\cal O} \left( (1-x)^2 \right)
\nn \:\: .
\eea
All terms shown in this equations are removed by including the additional 
contribution (\ref{P2trf}) to the transformation (\ref{ZgqA}) from the Larin 
scheme.

At small-$x$ the polarized splitting functions are double-logarithmically 
enhanced, i.e., terms up to $\ln^{\, 2n} x$ occur at N$^n$LO. Using the 
notation
\beq
\label{dP2xto0}
  \Delta P_{ik}^{\,(2)}(x) \equal 
       D_{ik}^{\,(0)} \ln^{\,4} x \:
  + \: D_{ik}^{\,(1)} \ln^{\,3} x \: 
  + \: D_{ik}^{\,(2)} \ln^{\,2} x \:
  + \: D_{ik}^{\,(3)} \ln x \: 
  + \: {\cal O}(1) 
\eeq
for the leading logarithmic (LL), next-to-leading logarithmic (NLL) 
contributions etc at NNLO, the small-$x$ terms of the non-singlet and 
pure-singlet splitting function are given by
\bea
\label{eq:Dns-exact}
  D_{\rm ns}^{\,(0)} &\! =\! & \mbox{}
  - \cf \, \* \cas + 4 \, \* \cfs \, \* \ca - {10 \over 3} \, \* \cft
\nn\\[1mm]
&\!\cong\! & 1.43210
 \:\: ,
\nn\\[2mm]
  D_{\rm ns}^{\,(1)} &\! =\! & 
  {40 \over 9} \, \* \cf \, \* \cas 
              - {14 \over 9} \, \* \cfs \, \* \ca - 4 \, \* \cft 
              + {20 \over 9} \, \* \cfs \, \* \nf 
              - {16 \over 9} \, \* \cf \, \* \ca \, \* \nf 
\nn\\[2mm]
&\!\cong\! & 35.5556 - 3.16049\, \* \nf
 \:\: ,
\nn\\[2mm]
  D_{\rm ns}^{\,(2)} &\! =\! &  
  \bigg( 81 + 14\, \* \z2 \bigg) \, \* \cf \, \* \cas
              - \bigg(\, {152 \over 3} 
              + 96\, \* \z2 \bigg) \, \* \cfs \, \* \ca
              - ( 60 - 104 \, \* \z2 ) \, \* \cft 
\nn\\[0.5mm] && \mbox{}
              - {196 \over 9} \, \* \cf \, \* \ca \, \* \nf
              + {80 \over 3} \, \* \cfs \, \* \nf 
              + {4 \over 9} \, \* \cf \, \* \nfs 
\nn\\[2mm]
&\!\cong\! & 399.205 - 39.7037\, \* \nf + 0.592592\, \* \nfs
 \:\: ,
\nn\\[2mm]
  D_{\rm ns}^{\,(3)} &\! =\! &  
  \bigg(\, {3442 \over 27} + {100 \over 3}\, \* \z2 
                      + 112\, \* \z3 \bigg) \, \* \cf \, \* \cas
             + \bigg(\, {1850 \over 9} - {680 \over 3}\, \* \z2 
                      - 336\, \* \z3 \bigg) \, \* \cfs \, \* \ca
\nn\\[0.5mm] && \mbox{}
             - ( 286 - 192\, \* \z2 - 224\, \* \z3 ) \, \* \cft
             - \bigg(\, {2252 \over 27} 
               - {8 \over 3}\, \* \z2 \bigg) \, \* \cf \, \* \ca \, \* \nf 
             + \bigg(\, {568 \over 9} 
               + {32 \over 3}\, \* \z2 \bigg) \, \* \cfs \, \* \nf 
\nn\\ && \mbox{}
             + {88 \over 27} \, \* \cf \, \* \nfs 
\nn\\[2mm]
&\!\cong\! & 1465.93 - 172.693\, \* \nf + 4.34568\, \* \nfs
\eea
and
\bea
\label{eq:Dps-exact}
  D_{\rm ps}^{\,(0)} &\! =\! & 
          - 2 \, \* \nf \, \* \cf \, \* \ca
          - {8 \over 3} \, \* \nf \, \* \cfs
\nn\\[2mm]
&\!\cong\! & \mbox{} - 12.7407\, \*\nf
\:\: ,
\nn\\[2mm]
  D_{\rm ps}^{\,(1)} &\! =\! & 
          - {152 \over 9} \, \* \nf \, \* \cf \, \* \ca
          - {40 \over 3} \, \* \nf \, \* \cfs
          + {8 \over 9} \, \* \nfs \, \* \cf
\nn\\[2mm]
&\!\cong\! & \mbox{} - 91.2593\, \* \nf + 1.18519\, \* \nfs
\, ,
\nn\\[2mm]
  D_{\rm ps}^{\,(2)} &\! =\! &  
          - \bigg(\, {854 \over 9} 
             + 4 \, \* \z2 \bigg)\, \* \nf \, \* \cf \, \* \ca
          - (64 - 48 \, \* \z2) \, \* \nf \, \* \cfs
          + {92 \over 9} \, \* \nfs \, \* \cf
\nn\\[2mm]
&\!\cong\! & \mbox{} - 379.285\, \* \nf + 13.6296\, \* \nfs
\:\: ,
\nn\\[2mm]
  D_{\rm ps}^{\,(3)} &\! =\! &  
          - \bigg(\,{9028 \over 27} + {116 \over 3} \, \* \z2 
            - 144 \, \* \z3\bigg)\, \* \nf \, \* \cf \, \* \ca
          - (100 - 160 \, \* \z2 + 112 \, \* \z3) \, \* \nf \, \* \cfs
\nn\\[1mm] &&
          + \bigg(\, {520 \over 27} 
            + {32 \over 3} \, \* \z2 \bigg)\, \* \nfs \, \* \cf
\nn\\[2mm]
&\!\cong\! & \mbox{} - 848.741\, \* \nf + 49.0736\, \* \nfs
\eea
where the respective last lines provide the QCD values rounded to six 
significant figures. \\
The corresponding coefficients for $\dPqg(2)$ and $\dPgq(2)$ read
\bea
\label{eq:Dqg-exact}
  D_{\rm qg}^{\,(0)} &\! =\! & \mbox{}
          - 5 \, \* \nf \, \* \cas
          - {4 \over 3} \, \* \nf \, \* \cf \, \* \ca
          + {4 \over 3} \, \* \nfs \, \* \cf
\nn\\[1mm]
&\!\cong\! & \mbox{} - 50.3333\, \* \nf + 1.77778\, \* \nfs
\, ,
\nn\\[2mm]
  D_{\rm qg}^{\,(1)} &\! =\! & \mbox{}
          - {328 \over 9} \, \* \nf \, \* \cas
          - {178 \over 9} \, \* \nf \, \* \cf \, \* \ca
          + 10 \, \* \nf \, \* \cfs
          + {16 \over 9} \, \* \nfs \, \* \ca
          + {172 \over 9} \, \* \nfs \, \* \cf
\nn\\[1mm]
&\!\cong\! & \mbox{} - 389.334\, \* \nf + 30.8148\, \* \nfs
\, ,
\nn\\[2mm]
  D_{\rm qg}^{\,(2)} &\! =\! &  \mbox{}
          - (150 - 14 \, \* \z2) \, \* \nf \, \* \cas
          - \bigg(\,{701 \over 18} 
            - 24 \, \* \z2 \bigg)\, \* \nf \, \* \cf \, \* \ca
          + \bigg(\, {59 \over 2} + 28 \, \* \z2\bigg) \, \* \nf \, \* \cfs
\nn\\ && \mbox{}
          + 22 \, \* \nfs \, \* \ca
          + {901 \over 9} \, \* \nfs \, \* \cf
\nn\\[1mm]
&\cong & \mbox{} - 1006.28\, \* \nf + 199.481\, \* \nfs
\, ,
\nn\\[2mm]
  D_{\rm qg}^{\,(3)} &\! =\! &  \mbox{}
          - \bigg(\, {22936 \over 27} - 28 \, \* \z2 
            - 288 \, \* \z3\bigg) \, \* \nf \, \* \cas
          + \bigg(\, {15259 \over 27} + 72 \, \* \z2 
            - 328 \, \* \z3\bigg) \, \* \nf \, \* \cf \, \* \ca
\nn\\[1mm] && \mbox{}
          - (77 - 92 \, \* \z2 - 104 \, \* \z3)\, \* \nf \, \* \cfs
          + {2134 \over 27} \, \* \nfs \, \* \ca
          + {6434 \over 27} \, \* \nfs \, \* \cf
\nn\\[1mm]
&\!\cong\! & \mbox{} - 2603.45\, \* \nf + 554.840\, \* \nfs
\eea
and
\bea
\label{eq:Dgq-exact}
  D_{\rm gq}^{\,(0)} &\! =\! & 
            10 \, \* \cf \, \* \cas
          + {16 \over 3} \, \* \cfs \, \* \ca
          - {8 \over 3} \, \* \cft
          - {8 \over 3} \, \* \nf \, \* \cfs
\nn\\[1mm]
&\!\cong\! & 142.123
          - 4.74074\, \* \nf
\:\: ,
\nn\\[2mm]
  D_{\rm gq}^{\,(1)} &\! =\! & 
            {188 \over 3} \, \* \cf \, \* \cas
          + {356 \over 9} \, \* \cfs \, \* \ca
          - {92 \over 3} \, \* \cft
          - {200 \over 9} \, \* \nf \, \* \cfs
\nn\\[2mm]
&\!\cong\! & 890.272
          - 39.5062\, \* \nf
\:\: ,
\nn\\[2mm]
  D_{\rm gq}^{\,(2)} &\! =\! &  
            \bigg(\, {3104 \over 9} - 92 \, \* \z2\bigg) \, \* \cf \, \* \cas
          + \bigg(\, {365 \over 9} - 16 \, \* \z2\bigg) \, \* \cfs \, \* \ca
          - (39 + 24 \, \* \z2) \, \* \cft
\nn\\[0.5mm] &&
          - {164 \over 9} \, \* \nf \, \* \cf \, \* \ca
          - {674 \over 9} \, \* \nf \, \* \cfs
\nn\\[2mm]
&\!\cong\! & 2212.57
         - 206.025\, \* \nf
\:\: ,
\nn\\[2mm]
  D_{\rm gq}^{\,(3)} &\! =\! &  
          \bigg(\, {22052 \over 27} - {188 \over 3} \, \* \z2 
            - 160 \, \* \z3\bigg) \, \* \cf \, \* \cas
          - \bigg(\, {11093 \over 27} + {208 \over 3} \, \* \z2 
            - 128 \, \* \z3\bigg) \, \* \cfs \, \* \ca
\nn\\[1mm] && \mbox{}
          + (341 - 96 \, \* \z3)\, \* \cft
          + \bigg(\, {1984 \over 27} 
            - {160 \over 3} \, \* \z2\bigg)\, \* \nf \, \* \cf \, \* \ca
          - \bigg(\, {6742 \over 27} 
              - {160 \over 3} \, \* \z2\bigg)\, \* \nf \, \* \cfs
\nn\\[2mm]
&\!\cong\! & 4811.85
          - 344.947\, \* \nf
 \:\: .
\eea
Finally the small-$x$ coefficients (\ref{dP2xto0}) of the polarized NNLO
gluon-gluon splitting function~are
\bea
\label{eq:Dgg-exact}
  D_{\rm gg}^{\,(0)} &\! =\! & 
            {56 \over 3} \, \* \cat
          + {2 \over 3} \, \* \nf \, \* \cas
          - 8 \, \* \nf \, \* \cf \, \* \ca
          - {4 \over 3} \, \* \nf \, \* \cfs
\nn\\[1mm]
&\!\cong\! & 504 - 28.3704\, \* \nf
\:\: ,
\nn\\[2mm]
  D_{\rm gg}^{\,(1)} &\! =\! & 
            {1264 \over 9} \, \* \cat
          - {4 \over 9} \, \* \nf \, \* \cas
          - {724 \over 9} \, \* \nf \, \* \cf \, \* \ca
          - {56 \over 3} \, \* \nf \, \* \cfs
          - {8 \over 9} \, \* \nfs \, \* \cf
\nn\\[1mm]
&\!\cong\! & 3792 - 358.963\, \* \nf - 1.18519\, \* \nfs
\:\: ,
\nn\\[2mm]
  D_{\rm gg}^{\,(2)} &\! =\! &  
            \bigg(\, {2126 \over 3} - 176 \, \* \z2\bigg) \, \* \cat
          - \bigg(\, {244 \over 9} + 20 \, \* \z2\bigg) \, \* \nf \, \* \cas
          - \bigg(\, {3542 \over 9} 
            - 48 \, \* \z2\bigg) \, \* \nf \, \* \cf \, \* \ca
\nn\\[0.5mm] &&
          - (92 - 16 \, \* \z2) \, \* \nf \, \* \cfs
          - {8 \over 9} \, \* \nfs \, \* \ca
          - {28 \over 9} \, \* \nfs \, \* \cf
\nn\\[1mm]
&\!\cong\! & 11317.3
          - 1915.25\, \* \nf
          - 6.81481\, \* \nfs
\:\: ,
\nn\\[2mm]
  D_{\rm gg}^{\,(3)} &\! =\! &  
          \bigg(\, {47810 \over 27} - {976 \over 3} \, \* \z2 
            - 192 \, \* \z3\bigg) \, \* \cat
          + \bigg(\, {4844 \over 27} - {236 \over 3} \, \* \z2 
            - 96 \, \* \z3\bigg) \, \* \nf \, \* \cas
\nn\\[1mm] &&
          - \bigg(\, {34172 \over 27} - {448 \over 3} \, \* \z2 
            - 144 \, \* \z3\bigg) \, \* \nf \, \* \cf \, \* \ca
          - (68 - 64 \, \* \z2 + 16 \, \* \z3) \, \* \nf \, \* \cfs
\nn\\[1mm] &&
          - {112 \over 27} \, \* \nfs \, \* \ca
          - \bigg(\, {520 \over 27} 
            - {32 \over 3} \, \* \z2\bigg) \, \* \nfs \, \* \cf
\nn\\[2mm]
&\!\cong\! & 27129.4 - 3944.01\, \* \nf - 14.7288\, \* \nfs
 \:\: .
\eea
The coefficients $D_{\rm ns}^{\,(0)}$, $D_{\rm ps}^{\,(0)}$ and 
$D_{\rm gg}^{\,(0)}$, which are identical to the coefficients of the 
corresponding physical kernels, agree directly with Refs.~\cite{BVns,BVpol}, 
for $D_{\rm qg}^{\,(0)}$ and $D_{\rm gq}^{\,(0)}$ agreement with
Ref.~\cite{BVpol} is obtained after taking into account Eq.~(\ref{PhysK}).


The small-$x$ behaviour in the right parts of Figs.~\ref{dPqqFig} -- 
\ref{dPggFig} is due to the above contributions, which exhibit the usual 
pattern of alternating LL, NLL etc terms with coefficients strongly 
increasing towards lower logarithms. 
Consequently the leading logarithms alone do not provide a good approximation 
for any practically relevant values of $x$ as illustrated in the figures. 
Yet~it is also clear, from the scale of the ordinates
in those right panels and Eq.~(\ref{eq:Dps-exact}) -- (\ref{eq:Dgg-exact}),
that these logarithms lead to a huge small-$x$ enhancement that can 
potentially spoil the stability of the expansion in $\als$ at $x$-values that 
would be accessible to an electron-proton collider with polarized beams.

Given the length and complexity of the exact expressions (\ref{dPns2x}) -- 
(\ref{dPgg2x}), it may be useful to also have at one's disposal compact and
accurate approximate expressions for the case of QCD, i.e., $C_A \,=\, 3$
and $C_F \,=\, 4/3$. 
Such approximations can be build up, besides powers of $x$, from the 
non-logarithmic plus distribution and end-point logarithms
\beq
\label{D0etc}
  \DD_{\,0} \equal 1/\x1_+ \; ,
  \quad L_1 \equal \ln \x1 \; ,
  \quad L_0 \equal \ln x 
\:\: .
\eeq
Due to $\dPnsp(2) = \Pnsm(2)$, the result (4.23) of Ref.~\cite{mvvPns} can be
used also here; it is given by
\bea
\label{dPns2A}
 && \hspn\hspn \dPnsp(2)(x) \:\:\cong\:\:
       1174.898\: \* \DD_0 + 1295.470\: \* \delta (1-x) + 714.1\: \* L_1
     + 1860.2 - 3505\: \* x 
\nn \\ & & \mbox{}
     + 297.0\: \* x^2
     - 433.2\: \* x^3 + L_0 \* L_1 \* (684 + 251.2\: \* L_0)+ 1465.2\: \* L_0
     + 399.2\: \* L_0^2 
\nn \\ & & \mbox{}
     + 320/9\: \* L_0^3 + 116/81\: \* L_0^4
\nn \\ && \hspn + \nf \: \* \left(
     - 183.187\: \* \DD_0 - 173.933\: \* \delta (1-x)
     - 5120/81\: \* L_1 - 216.62 + 406.5\: \* x + 77.89\: \* x^2 
\right. \nn \\ & & \left. \mbox{}
     + 34.76\: \* x^3 - 1.136\: \* x\*L_0^3 - 65.43\: \* L_0 \* L_1
     - 172.69\: \* L_0
     - 3216/81\: \* L_0^2 - 256/81\: \* L_0^3 \right)
\nn \\ &&  \hspn + \nfs \: \*\left(
     - \DD_0 - (51/16 + 3\*\z3 - 5\*\z2) \: \* \delta (1-x)
     + x\*(1-x)^{-1}\* L_0\* (3/2\: \* L_0 + 5) + 1
\right. \nn \\ & & \left.
     + (1-x)\* (6 + 11/2\: \* L_0 + 3/4\: \* L_0^2) \right) \: \* 64/81
  \, .
\eea
The polarized pure-singlet NNLO splitting function (\ref{dPps2x}) can be
parametrized as
\bea
\label{dPps2A}
  && \hspn\hspn\hspn \dPps(2)(x) \:\:\cong\:\:
  \nf \: \* (1-x) \* \left(
    - 344/27\: \*  L_0^4 - (90.9198 + 81.50\: \* x) \* L_0^3
\right. \nn \\ & & \left. \mbox{}
    - (368.6 - 349.9\: \* x) \* L_0^2
    - (739.0 - 232.57\: \* L_1) \* L_0
    - 1362.6 + 1617.4\: \* x 
\right. \nn \\ & & \left. \mbox{}
    - 674.8\: \* x^2 + 167.41\: \* x^3
    - 204.76\: \* L_1 - 12.61\: \* L_1^2 - 6.541\: \* L_1^3
   \right)
\nn \\ && \hspn \mbox{} + \nfs \: \* (1-x) \* \left(
  (1.1741 - 0.8253\: \* x) \* L_0^3  + (13.287 + 10.657\: \* x) \* L_0^2
  + 45.482\: \* L_0 
\right. \nn \\ & & \left. \mbox{}
  + 49.13 - 30.77\: \* x - 4.307\: \* x^2
  - 0.5094\: \* x^3 + 9.517\: \* L_1 + 1.7805\: \* L_1^2 \right)
  \:\: .
\eea
Sufficiently accurate parametrizations of the corresponding off-diagonal
quantities in Eqs.~(\ref{dPqg2x}) and (\ref{dPgq2x}) are given by
\bea
\label{dPqg2A}
 && \hspn\hspn\hspn \dPqg(2)(x) \:\:\cong\:\:
  \nf \: \* \left(
    - 151/3\: \* L_0^4 - (385.64 + 73.30\: \* x) \* L_0^3
    - (894.8 - 1145.3\: \* x) \* L_0^2
\right. \nn \\ & & \left.
    - (1461.2 - 825.4\: \* L_1) \* L_0
    - 2972.4 + 4672\: \* x - 1221.6\: \* x^2 - 18.0\: \* x^3
\right. \nn \\ & & \left.
    + 278.32\: \* L_1
    - 90.26\: \* L_1^2 - 5.30\: \* L_1^3 + 3.784\: \* L_1^4
   \right)
\nn \\ &&  \mbox{} \hspn + \nfs \: \* \left(
  16/9\: \* L_0^4 + (30.739  + 10.186\: \* x) \* L_0^3
  + (196.96 + 179.1\: \* x) \* L_0^2
\right. \nn \\ & & \left.
  + (526.3  - 47.30\: \* L_1) \* L_0
  + 499.65 - 432.18\: \* x - 141.63\: \* x^2 - 11.34\: \* x^3
\right. \nn \\ & & \left.
  - 6.256\: \* L_1 + 7.32\: \* L_1^2 + 0.7374\: \* L_1^3
   \right)
\eea
and
\bea
\label{dPgq2A}
 && \hspn\hspn\hspn \dPgq(2)(x) \:\:\cong\:\:
  11512/81\: \* L_0^4 + (888.003  + 175.1\: \* x) \* L_0^3
  + (2140 - 850.7\: \* x) \* L_0^2
\nn \\ & &  \mbox{}
  + (4046.6 - 1424.8\: \* L_1) \* L_0
  + 6159 - 3825.9\: \* x + 1942\: \* x^2 - 742.1\: \* x^3
\nn \\ & &  \mbox{}
  + 1843.7\: \* L_1
  + 451.55\: \* L_1^2 + 59.3\: \* L_1^3 + 5.143\: \* L_1^4
\nn \\ & & \hspn +  \nf \: \* \left(
  - 128/27\: \* L_0^4 - (39.3872 + 30.023\: \* x) \* L_0^3
  - (202.46 + 126.53\: \* x) \* L_0^2
\right. \nn \\ & & \left. \mbox{}
  - (308.98 + 16.18\: \* L_1) \* L_0
  - 301.07 - 296.0\: \* x + 406.13\: \* x^2 - 101.62\: \* x^3
\right. \nn \\ & & \left.
  - 171.78\: \* L_1
  - 47.86\: \* L_1^2 - 4.963\: \* L_1^3
   \right)
\nn \\ & & \hspn + \nfs \: \* \left(
     16/27\: \* ( - 12 + 10\: \* x + ( 8 + 2\: \* x) \* L_1
         + (6 - 3\: \* x) \* L_1^2 )
   \right)
\:\: .
\eea
Finally the gluon-gluon splitting function (\ref{dPgg2x}) can be approximately
represented by
\bea
\label{dPgg2A}
 && \hspn\hspn \dPgg(2)(x) \:\:\cong\:\:
  2643.521\: \* \DD_0 + 4427.762\: \* \delta (1-x)
  + 504 \* L_0^4 + (3777.5  + 1167\: \* x) \* L_0^3
\nn \\ & & \mbox{}
  + (10902 - 863\: \* x) \* L_0^2 + (23091 - 12292\: \* L_1) \* L_0
  + 30988 - 39925\: \* x + 13447\: \* x^2
\nn \\ & & \mbox{}
  - 4576\: \* x^3 - 13247\: \* (1-x)\*L_1 + 3801\: \* L_1
\nn \\ & & \hspn + \nf \: \* \left(
  - 412.172\: \* \DD_0 - 528.536\: \* \delta (1-x)
  - 766/27\: \* L_0^4 - (357.798 - 131\: \* x)\* L_0^3
\right. \nn \\ & & \left. \mbox{}
  - (1877.2 - 613.1\: \* x) \* L_0^2 - (3524 + 7932\: \* L_1) \* L_0
  - 1173.5 + 2648.6\: \* x - 2160.8\: \* x^2
\right. \nn \\ & & \left. \mbox{}
  + 1251.7\: \* x^3 - 6746\: \* (1-x)\*L_1 - 295.7\: \* L_1
   \right)
\nn \\ & & \hspn + \nfs \: \* \left(
  - 16/9\: \* \DD_0 + 6.4607\: \* \delta (1-x)
  - 1.1809\: \* L_0^3 - (6.679 - 15.764\: \* x) \* L_0^2
\right. \nn \\ & & \left. \mbox{}
  - (13.29 + 16.944\: \* L_1) \* L_0 - 16.606 + 32.905\: \* x
  - 18.30\: \* x^2 + 2.637\: \* x^3 - 0.210\: \* L_1
   \right)
\:\: . \quad
\eea
These expressions can be readily transformed to Mellin space for any 
$N \neq -n$, $\,n = 0,\,1,\,2,\,\ldots\,$; the~most complex objects needed there
are the logarithmic derivatives of Euler's $\Gamma$-function.
 
The $\nfs$ contributions in Eqs.~(\ref{dPns2A}) and (\ref{dPgq2A}) are exact.
The same holds for all coefficients of $\ln^{\,4}x$ and, up to the truncation
of irrational numbers, those of $1/\x1_+$ in Eqs.~(\ref{dPns2A}) and~%
(\ref{dPgg2A}). The other terms at $x <\! 1$ have fitted to the exact results,
evaluated by the {\sc Fortran} code of Ref.~\cite{HPLnum}, at $10^{-6} \leq x 
\leq 1 - 10^{-6}$ using the {\sc Minuit} package \cite{Minuit1,Minuit2}.
Except for $x$ values very close to zeros of the splitting functions, the above
parametrizations deviate from the exact results by less than one part in 
thousand, which should be sufficient for any foreseeable phenomenological 
application.
As in the unpolarized case \cite{mvvPns,mvvPsg}, the coefficients of 
$\delta (1-x)$
have been adjusted in Eq.~(\ref{dPgg2A}) using low integer moments in order to
achieve a maximal accuracy of the parametrization and its convolutions with
the polarized gluon distribution. For a brief discussion of this slightly 
subtle point the reader is referred to Ref.~\cite{mvvPsg} (penultimate 
paragraph of section 4).

\begin{figure}[p]
\vspace*{-1mm}
\centerline{\hspace*{-1mm}\epsfig{file=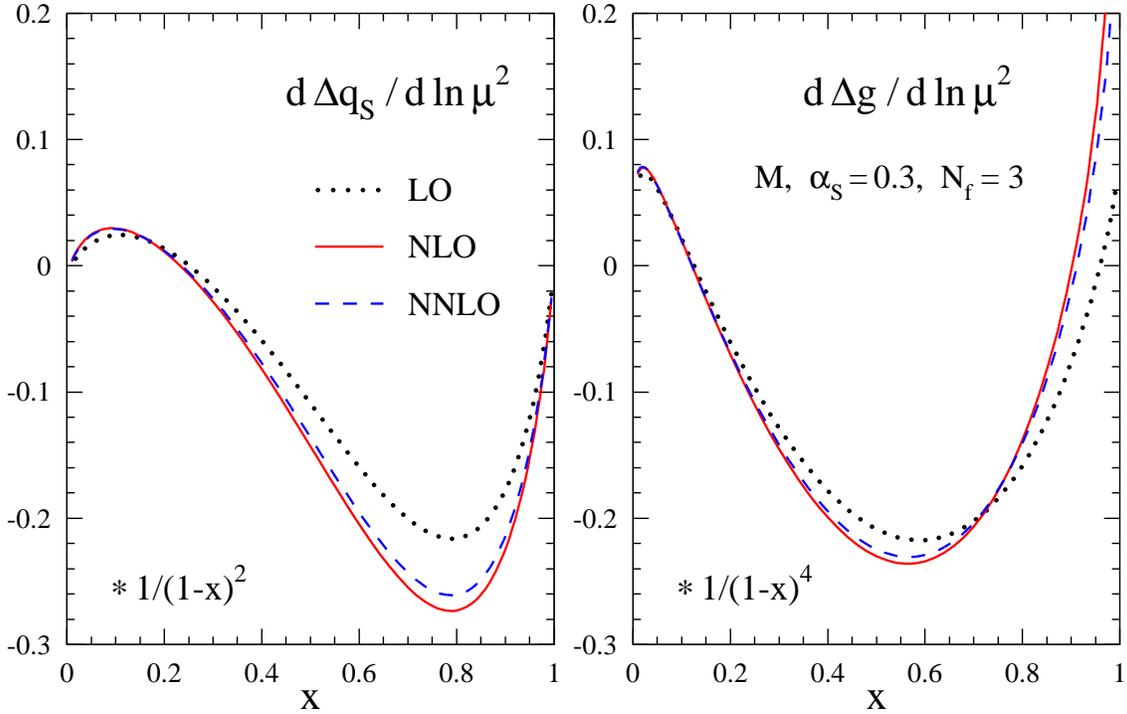,width=15.4cm}}
\vspace{-4mm}
\caption{\label{evolQ}
The perturbative expansion of the scale derivatives of the polarized
singlet-quark and gluon distributions in the standard \MSb\ scheme (M)
\cite{MSvN98}, for the low-scale input distributions in Eq.~(\ref{QGinp}) 
and a rather large value of the strong coupling $\als$. The results have 
been multiplied by powers of $\x1$ suitable to clearly display the NLO and 
NNLO effects up to rather large~$x$.
}
\end{figure}
\begin{figure}[p]
\vspace*{-1mm}
\centerline{\hspace*{-1mm}\epsfig{file=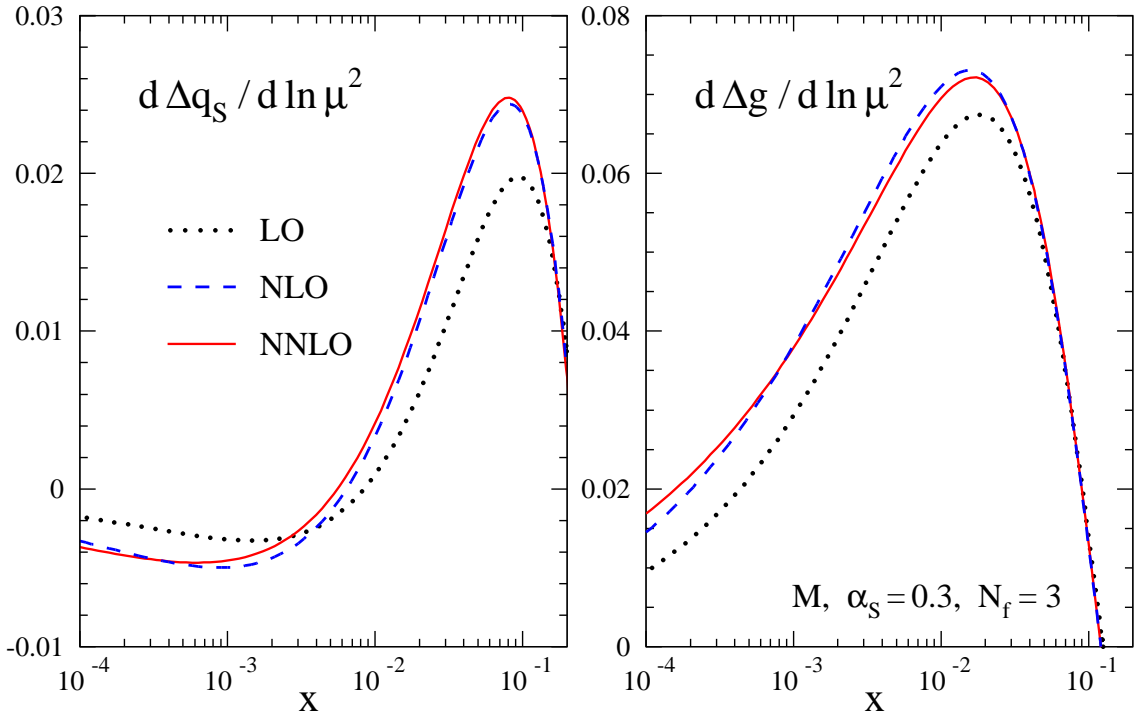,width=15.4cm}}
\vspace{-4mm}
\caption{\label{evolG}
As Figure~\ref{evolG}, but using a logarithmic scale in $x$ to show the 
results down to small $x$.
}
\vspace*{-1mm}
\end{figure}

The effect of the new results (\ref{dPps2x}) -- (\ref{dPgg2x}) on the evolution
of polarized parton densities is briefly illustrated in Figs.~\ref{evolQ} and
\ref{evolG}, where the respective first and second lines of Eq.~(\ref{Sevol})
have been evaluated for the schematic, but sufficiently realistic low-scale 
distributions
\bea
\label{QGinp}
  \Delta f_{\rm q}^{}(x,\mu_0^{\:2}) &\! = \! &
           0.8 \, x^{\,0.7} \, \x1^3 \, (1 + 3\, x + 2.5\, x^{\,2} )
    \:-\: 0.25 \, x^{\,0.7} \, \x1^7
\:\: ,\nn \\[0.5mm]
  \Delta f_{\rm g}^{}(x,\mu_0^{\:2}) &\! = \! &
           1.5 \, x^{\,0.5} \, \x1^5
\eea
used for the evolution benchmarks in Refs.~\cite{Pegasus,GSbnch2}, for 
$\als(\mu_0^{\:2}) = 0.3$ and $\nf = 3$. After the convolution with the 
distributions (\ref{QGinp}), the NNLO corrections are fairly small down to 
small $x$.

%
\section{Summary}
\label{sec:summ}

We have extended the determination of the helicity-difference (polarized)
splitting functions $\Delta P_{ik\,}$, which were only known at the first 
\cite{Altarelli:1977zs,Sasaki:1975hk,Ahmed:1976ee} and second 
\cite{MvN95,WVdP1a,WVdP1b} order in the strong coupling constant $\als$
so~far, to the third order (next-to-next-to-leading order, NNLO) in massless 
perturbative~QCD. 
These corrections are relevant to the structure function $g_1^{}$ in polarized 
deep-inelastic scattering (DIS), for which we also confirm the results of 
Ref.~\cite{ZvNpol} for the NNLO coefficient functions, and all other 
observables that are sensitive to the polarized quark and gluon distributions 
$\Delta f_{q_i^{}} + \Delta f_{{\bar q}_i^{}}$ and~$\Delta f_g$.
The so far practically irrelevant polarized quark-antiquark differences 
have not been addressed here; the corresponding splitting functions can be 
calculated, e.g., by extending the analysis of weak-interaction structure
functions in Ref.~\cite{SVWgiCC} to NNLO accuracy.

The calculation of the upper row of the matrix of NNLO flavour-singlet
splitting functions, i.e., of $\dPqq(2)(x)$ and $\dPqg(2)(x)$, was carried out 
via the structure function $g_1^{}$ as a direct extension of our previous
calculations of the helicity-averaged (unpolarized) case \cite{mvvPns,mvvPsg},
for an earlier brief account see Ref.~\cite{mvvLL08}. 
The corresponding lower-row quantities $\dPgq(2)(x)$ and $\dPgg(2)(x)$ have 
been determined in a different manner from graviton-exchange DIS, see 
Ref.~\cite{LamLi}, which includes structure functions sensitive to the 
polarized gluon distribution at the Born level.

We have first calculated the relevant structure function at fixed odd moments
to $N = 25$, using a large-$N$ optimized version \cite{jvLL14} of the
{\sc Mincer} program \cite{MINCER1,MINCER2} in {\sc (T)Form} 
\cite{TFORM,FORM4}. 
Exploiting in particular the close relation between the polarized and
unpolarized splitting functions for the highest-weight harmonic sums
\cite{Hsums} and for the threshold limit, cf.~Ref.~\cite{BBSsuppr}
-- which includes the so-called supersymmetric relation, see Refs.~\cite
{Furmanski:1980cm,Antoniadis:1981zv}, as far as it can be addressed in \MSb\ --
we have then been able to determine the all-$N$ expressions of $\dPgq(2)$ and 
$\dPgg(2)$.
It was crucial for this step that the coefficients of the harmonic sums are
integer, up to low powers of 2 and 3 that can be removed by a suitable
normalization, which allows the use of advanced tools \cite{axbWeb,LLL,axbAlg} 
for systems of Diophantine equations; this was observed and exploited before 
in a comparable but somewhat simpler situation in Ref.~\cite{Veliz}.
Finally the results have been validated by comparing the next two moments
of all-$N$ expressions to additional results calculated using {\sc Mincer} 
up to $N=29$. 

Our results have been presented above in $N$-space and $x$-space, using the 
transformation of Ref.~\cite{MSvN98} from the so-called Larin scheme for
$\gamma_5^{}$ \cite{g5L1,g5L2} in dimensional regularization to \MSb.
This scheme shows an unphysical feature in the threshold limit of
the quark-gluon splitting function $\Delta P_{\rm gq}$ already at NLO, which
can be removed to NNLO by simple additional terms in the scheme 
transformation. Yet this situation does not appear to necessitate a change of 
the factorization scheme in practical calculations after almost two decades of
NLO analyses in QCD spin physics.

The new functions $\Delta P_{ik\,}^{\,(2)}(x)$ are consistent with all
known limits and partial results, e.g., for the leading large-$\nf$ terms
\cite{BGracey}, and expectations. In particular, the first moment of 
$\dPgg(2)(x)$, which is not directly accessible in graviton-exchange DIS
\cite{LamLi} but can be determined from the $x$-space results in terms of
harmonic polylogarithms \cite{Hpols}, is identical to the NNLO coefficient of
the beta function of QCD \cite{beta2a,beta2b} as theoretically required.
We have checked our calculations of graviton-exchange DIS also by 
re-calculating, and obtaining full agreement for, $\dPqq(2)$ and $\dPqg(2)$ 
to fairly high values odd of $N$ and all unpolarized flavour-singlet NNLO 
splitting functions at even $N \leq 10$. 
As those results, the present polarized splitting functions lead to fairly
small NNLO corrections, down to low values of $x$, after the convolution with 
realistic polarized quark and gluon distributions, despite a double-logarithmic
small-$x$ enhancement that dwarfs that of the non-singlet cases.

Our results allow NNLO analyses of spin-dependent hard-scattering observables,
provided that the corresponding coefficient functions are known to this 
accuracy as for the structure function $g_1^{}$ in DIS \cite{ZvNpol}, for a 
fixed number of effectively massless flavours $\nf$. 
The extension to analyses in the so-called
variable flavour-number scheme, where effective theories for different
values of $\nf$ are used together, requires non-trivial matching coefficients
for the strong coupling \cite{Chetyrkin:1997un} and the parton densities at 
this order. The latter coefficients have been calculated in 
Ref.~\cite{Buza:1996wv} for the unpolarized case. 
As far as we know, the corresponding results for the helicity-difference
parton distributions are not yet available in the literature though.

{\sc Form} and {\sc Fortran} files of our main analytical results in $N$-space
and $x$-space, and compact high-accuracy parametrizations of the functions 
$\Delta P_{ik\,}^{\,(2)}(x)$, can be obtained by downloading the source of 
this article from {\tt http://arxiv.org/} or from the authors upon request.

%
\vspace*{5mm}
\subsection*{Acknowledgments}

We would like to thank John Gracey for useful discussions.
This work has been supported by
the \mbox{German}
{\it Bundesministerium f\"ur Bildung und Forschung} through contract 05H12GU8,
the {\it European Research Council}$\,$ (ERC) Advanced Grant no.~320651,
{\it HEPGAME} and the UK {\it Science \& Technology Facilities Council}$\,$
(STFC) grant ST/G00062X/1.
We are particularly grateful for the opportunity to use a substantial part of
the {\tt ulgqcd} computer cluster in Liverpool which was funded by STFC under
grant number ST/H008837/1.

%
\vspace*{5mm}
\section*{Appendix A~~~~Transformation to the \MSb\ scheme}
\label{sec:AppA}
\renewcommand{\theequation}{A.\arabic{equation}}
\setcounter{equation}{0}

Here we collect, for completeness, the functions entering the transformation 
of the splitting and coefficient functions from the Larin scheme to \MSb\ as 
discussed in Section 2, Eqs.~(\ref{g5Larin}) -- (\ref{Ptrf}) and 
Eqs.~(\ref{ZikM}) -- (\ref{P2trf}), and Section 3, see Eq.~(\ref{ZgqA}).

The NLO and NNLO quark-quark elements (\ref{ZikM}) of the transformation 
matrix $Z(x,\muS)$ read
\bea
   z_{\,\rm ns}^{\,(1)}(x) &\! =\! &
        - 8\, \* \cf\, \* ( 1 - x )
%
\:\: ,\\[2mm] 
   z_{\,\rm ns}^{\,(2)}(x) &\! =\! &
         8\, \* \cfs\, \* \Big( 
           (1 - x) \* ( 
              5 
            - 2\, \* \Hh(1,0) 
            - 2\, \* \H(2)
           )
           - 2\, \* (1 + x)\, \* ( 
              2\, \*\Hh(-1,0)
            - \Hh(0,0)
            + \z2
           )
           + (1 + 2\,\*x)\, \* \H(0)
         \Big)
\nn \\[-0.5mm] & & \hspn \mbox{}
       + 4\, \* \cf \* \ca\, \* \Big(
           4\, \* (1 + x)\, \* \H(-1,0)
           - 4\, \* ( \Hh(0,0) - \z2 )
           - ( 29 + 7\,\* x)/3\: \* \H(0)
           - 211/9\: \* (1 - x)
         \Big)
\quad \nn \\[-0.5mm] & & \hspn \mbox{}
       + 8/9\: \* \cf  \* \nf \, \* (1 - x) \, \* \Big(
           3\, \* \H(0) + 5
         \Big)
%
\:\: ,\\[2mm]
   z_{\,\rm ps}^{\,(2)}(x) &\! =\! &
         4\, \* \cf \* \nf\, \* \Big(
           (2 + x)\, \* \Hh(0,0)
         + (3 - x)\, \* \H(0)
         + 2\,\* (1 - x) 
         \Big)
\eea
for the standard transformation, above denoted by `M' where required for 
clarity, of Ref.~\cite{MSvN98} where the critical last line has been 
calculated.

In the alternative (`A') form of the transformation, which restores the
$\x1^2$ suppression for $x \ra 1$ of the difference or the unpolarized and
polarized splitting functions for $\dPgq(1)(x)$ and $\dPgq(2)(x)$, 
there are additional quark-gluon entries (\ref{ZgqA}) given by
\bea
  z_{\,{\rm gq},\,A}^{\,(1)}(x) &\! =\! & 
    -\, 2\, \* \cf \,\* ( 2 - x )
\:\: , \\[2mm]
  z_{\,{\rm gq},\,A}^{\,(2)}(x) &\! =\! & - \frct{1}{2}\: \dPgq(1)(x) 
  - \, 8\,\*\cfs \, \* \Big( 3\, \* (1-x) + (2+x)\, \* \H(0) \Big)
\:\: ,
\eea
where the last line has been expressed in term of the NLO splitting
function (\ref{dPqg1x}) for brevity. 
Furthermore Eq.~(A.3) is replaced by
\beq
  z_{\,{\rm ps},\,A}^{\,(2)}(x) \equal
  z_{\,\rm ps}^{\,(2)}(x) \:+\: 12\,\*\cf\*\nf\,\* (1-x)
\eeq
which ensures that Eq.~(\ref{dPpsnN1}) holds also in the A-scheme.

%
\section*{B~~~~NLO coefficient functions in graviton-exchange DIS}
\label{sec:AppB}
\renewcommand{\theequation}{B.\arabic{equation}}
\setcounter{equation}{0}

\def\pqq(#1){p_{\rm{qq}}(#1)}
\def\pqg(#1){p_{\rm{qg}}(#1)}
\def\pgq(#1){p_{\rm{gq}}(#1)}
\def\pgg(#1){p_{\rm{gg}}(#1)}

The (un-)$\,$polarized graviton DIS structure function $H_{i}$ of 
Ref.~\cite{LamLi} have been introduced briefly in Section~\ref{sec:2loop}.
We have defined combinations of those $H_{i\,}$ which, at Born level, 
are either given by the flavour-singlet (un-)polarized quark distribution 
$(\Delta) f_{\rm q}^{}$ or by the gluon density $(\Delta) f_{\rm g}^{}$, 
cf.~Eqs.~(\ref{fiUnp}), (\ref{fiPol}) and (\ref{qS}).
Their quark and gluon coefficient functions $C_{\,i,\rm q}$ and 
$C_{\,i,\rm g}$ can be expanded in powers of $\ars\,$, see Eq.~(\ref{aDef}).

In the unpolarized case, using the definitions 
$\,H_{\bar{1}}^{} = H_1 - H_3\,$ and  
$\,H_{\bar{2}}^{} = H_2 - 4\, H_3\,$, cf.~Eq.~(\ref{Hunp}), 
the leading-order results for the corresponding non-vanishing coefficient 
functions are
\beq
  \label{unpolCgLO}
  c_{\,\bar{1},\rm q}^{\,(0)}(x) \equal c_{\,\bar{2},\rm q}^{\,(0)}(x) 
  \equal c_{\,3,\rm g}^{\,(0)}(x) \equal \delta(1-x)
\:\: .
\eeq
The normalization of the structure functions is chosen such that all dependence
on $D = 4 - 2 \ep$ is removed from the structure functions $H_{i}$ at
Born level, i.e., the results in Eq.~(\ref{unpolCgLO}) are exact.

The NLO results for the unpolarized graviton-exchange coefficient functions 
read, at $\Qs = \muS$,
\bea
\label{unpolCgNLO}
\label{c1nlo}
  c_{\,\bar{1},\rm q}^{\,(1)}(x) &\! =\! & 
  2\,\*\colour4colour{\cf} \* \bigg(
          - \pqq(x)\*(3/4+\H(0)+\H(1))
          + 1/4\*(25 - x)
       - \delta(1 - x)\* (
            13/2
          + 2\*\z2
          )
          \bigg)
\:\: , \nn \\[0.5mm]
  c_{\,\bar{1},\rm g}^{\,(1)}(x) &\! =\! &
  2/3\:\*\colour4colour{\ca} \* (7\*\pqg(x) + 9)
  - 2\,\*\colour4colour{\nf} \* \bigg(
            \pqg(x)\*(29/6+\H(0)+\H(1))
          - 5/2
          \bigg)
\:\: , \\[2mm]
\label{c2nlo}
  c_{\,\bar{2},\rm q}^{\,(1)}(x) &\! =\! & c_{\,\bar{1},\rm q}^{\,(1)}(x) 
  - 9\,\* x\,\*  \colour4colour{\cf}
\:\: , \nn \\[0.5mm]
  c_{\,\bar{2},\rm g}^{\,(1)}(x) &\! =\! & c_{\,\bar{1},\rm g}^{\,(1)}(x)
  - ( 2\,\* \colour4colour{\ca} \,+\, \colour4colour{\nf} ) 
    \: \* 6\, \* x\,\* (1-x)
\:\: , \\[2mm]
\label{c3nlo}
  c_{\,3,\rm q}^{\,(1)}(x) &\! =\! &
  2\,\*\colour4colour{\cf} \* \bigg(
          - \pgq(x)\*(3/4 +\H(0)+\H(1))
          + 1/4\, \*(6 + x)
          \bigg)
\:\: , \nn \\[0.5mm]
  c_{\,3,\rm g}^{\,(1)}(x) &\! =\! & 
  4\,\*\colour4colour{\ca} \* \bigg(
          - \pgg(x)\*(11/12+\H(0)+\H(1))
          + 11/12\, \*(2 - x + x^2)
\\[-0.5mm]& & \mbox{}
       - \delta(1 - x) \* (
           34/9
          + \z2
          )
          \bigg)
  \:+\: 2/3\:\*\colour4colour{\nf} \* \bigg(
          \pgg(x)
          - 2 + x - x^2
          + 25/6\: \*\delta(1 - x)
          \bigg)
\nn \:\: ,
\eea
where we have used the abbreviations
\bea
  p_{\rm{qq}}(x) &\! =\! & 2\, (1 - x)^{-1} - 1 - x \, ,\nn \\[0.5mm]
  p_{\rm{qg}}(x) &\! =\! & 1 - 2x + 2x^{\,2} \, ,\nn \\[0.5mm]
  p_{\rm{gq}}(x) &\! =\! & 2x^{\,-1} -2 + x \, ,\nn \\[0.5mm]
  p_{\rm{gg}}(x) &\! =\! & (1-x)^{-1} + x^{\,-1} - 2 + x - x^{\,2} 
  \:\: .
\eea
The NLO QCD corrections for unpolarized gravition-exchange DIS at NLO have been
presented before in Ref.~\cite{SVgDIS} in terms of the bare structure functions
$H_1$, $H_2$ and $H_3$ as a Laurent series in $\ep$, i.e., before mass 
factorization.
The results for the coefficient functions in Eq.~(\ref{unpolCgNLO}) can be used
to construct the corresponding expressions to be compared with 
Ref.~\cite{SVgDIS}.  
Accounting, of course, for the different normalization we find agreement except
for the result of the coefficient function $c_{\,3,\rm q}^{\,(1)}$ as given in 
Eq.~(3.3) of Ref.~\cite{SVgDIS}.

In the polarized case we similarly use 
$\,H_{\bar{4}}^{} = 2 ( H_4 - H_6 )\,$ and $\,H_6\,$, recall Eq.~(\ref{Hpol})
with
\beq
  \label{unpoldCgLO}
  c_{\,\bar{4},\rm i}^{\,(0)}(x) \:=\: \delta_{\rm iq}\, \delta(1-x)
\:\:, \quad
  c_{\,6,\rm i}^{\,(0)}(x) \:=\: \delta_{\rm ig}\, \delta(1-x)
\:\: .
\eeq
Again the structure functions are normalized such that there is no 
dependence in $\ep$ at this order.
The NLO results for the polarized graviton DIS coefficient functions 
in the standard \MSb\ scheme, i.e., with the transformation (\ref{ZikM}), 
are given by
\bea
\label{c4nlo}
  c_{\,\bar{4},\rm q}^{\,(1)}(x) &\! =\! & 
  2\,\*\colour4colour{\cf} \* \bigg(
          - \dpqq(x)\*(
            3/4 + \H(0) + \H(1))
          - 1/4\:\*(11 - 17\*x)
       - \delta(1 - x) \* (
            13/2
          + 2\*\z2
          )
          \bigg)
\:\: , \nn \\[0.5mm]
  c_{\,\bar{4},\rm g}^{\,(1)}(x) &\! =\! & 
    32/3\: \*\colour4colour{\ca} \* (2\*x - 1) 
  - 2\,\*\colour4colour{\nf} \* \bigg(
            (2\*x - 1)\*(\H(0) + \H(1))
          - 1/3\*(13 - 20\*x)
          \bigg)
\:\: , \\[2mm]
\label{c6nlo}
  c_{\,6,\rm q}^{\,(1)}(x) &\! =\! &
  \colour4colour{\cf} \* \bigg(
          - 2\*(2 - x)\*(\H(0) + \H(1))
          - (10 - 3/x - 7\*x)
          \bigg)
\:\: , \nn \\[0.5mm]
  c_{\,6,\rm g}^{\,(1)}(x) &\! =\! & 
  4\,\*\colour4colour{\ca} \* \bigg(
          - \dpgg(x) \* (11/12
          + \H(0)
          + \H(1))
          - 1/12\*(35 - 11/x - 46\*x)
\\[-0.5mm]& & \mbox{}
       - \delta(1 - x) \* (
            34/9
          + \z2
          )
          \bigg)
       +  2/3\: \* \colour4colour{\nf} \* \bigg(
            \dpgg(x)
          + 1 - 1/x - 2\*x
       + 25/6\: \* \delta(1 - x)
          \bigg)
\nn \:\: , \quad
\eea
in terms of $\dpqq(x)$ and $\dpgg(x)$ defined in Eq.~(\ref{dpqqgg}).

Analogous to our discussion of relations between the unpolarized and polarized 
splitting functions in $N$-space in Section 3, is may be interesting to note 
that all $H_0$ and $H_1$ contributions to Eqs.~(\ref{c4nlo}) and (\ref{c6nlo})
are related to those in Eqs.~(\ref{c1nlo}) -- (\ref{c3nlo}) by replacing
$p_{\rm ik}^{}(x)$ by their polarized counterparts $\Delta p_{\rm ik}^{}(x)$ 
with, cf.~Eq.~(\ref{dpij0}),
$\Delta p_{\rm qg}(x) \,=\, 2x -1$ and $\Delta p_{\rm gq}(x) \,=\, 2 - x$.

%
\section*{C~~~~Calculation of graviton-exchange DIS}
\label{sec:AppC}
\renewcommand{\theequation}{C.\arabic{equation}}
\setcounter{equation}{0}

Here we present some core ingredients of our diagram calculations, starting
with the Feynman rules as used for graviton-exchange DIS.
They have been taken from various sources~\cite{Han:1998sg,Mathews:2004xp}.
We assume all momenta of the gluons and the graviton to be outgoing, 
while the momenta of the quarks and ghosts follow the arrows on the lines. 
The color indices in the fundamental representation are $i$ and $j$; 
color indices in the adjoint representation are represented by the letters 
a,b,c,d,e; the Lorentz indices of the graviton are $\alpha$ and $\beta$ and 
those of the gluons are $\mu, \nu, \rho, \sigma$. We also use 
a gauge parameter which is indicated by $\xi\,$.

For completeness we start with the QCD propagators and vertices:

\vspace*{1mm}
\noindent
\begin{minipage}{6cm}
\begin{center}
          \SetWidth{1}
	\begin{picture}(120,30)(0,0)
		\Gluon(35,10)(85,10){4}{6}
		\Text(60,22)[]{Q}
		\Text(30,10)[r]{a,$\mu$}
		\Text(90,10)[l]{b,$\nu$}
	\end{picture}
\end{center}
\end{minipage}
\begin{minipage}{10cm}
	\begin{eqnarray}
		- i \,\delta_{ab}\, \bigg( 
            \delta_{\mu\nu}-\xi\:\frac{Q_\mu Q_\nu}{Q\cdot Q} \bigg) /Q\cdot Q
	\end{eqnarray}
\end{minipage}

\noindent
\begin{minipage}{6cm}
\begin{center}
          \SetWidth{1}
	\begin{picture}(120,30)(0,0)
		\ArrowLine(35,10)(85,10)
		\Text(60,20)[]{P}
		\Text(30,10)[r]{j}
		\Text(90,10)[l]{i}
	\end{picture}
\end{center}
\end{minipage}
\begin{minipage}{10cm}
	\begin{eqnarray}
		i \,\delta_{ij}(\gamma_\mu P^{\,\mu})/P\cdot P
	\end{eqnarray}
\end{minipage}

\noindent
\begin{minipage}{6cm}
\begin{center}
          \SetWidth{1}
	\begin{picture}(120,30)(0,0)
		\DashArrowLine(35,10)(85,10){4}
		\Text(60,20)[]{Q}
		\Text(30,10)[r]{a}
		\Text(90,10)[l]{b}
	\end{picture}
\end{center}
\end{minipage}
\begin{minipage}{10cm}
	\begin{eqnarray}
		i \,\delta_{ab}/Q\cdot Q
	\end{eqnarray}
\end{minipage}  \vspace{2mm}


\noindent
\begin{minipage}{6cm}
\begin{center}
          \SetWidth{1}
	\begin{picture}(130,90)(0,0)
		\ArrowLine(60,45)(20,80)
		\ArrowLine(20,10)(60,45)
		\Gluon(60,45)(110,45){4}{5}
		\Vertex(60,45){1.5}
		\Text(15,80)[r]{i}
		\Text(45,68)[lb]{$p_2^{}$}
		\Text(45,25)[lt]{$p_1^{}$}
		\Text(15,10)[r]{j}
		\Text(115,45)[l]{a,$\mu$}
	\end{picture}
\end{center}
\end{minipage}
\begin{minipage}{10cm}
	\begin{eqnarray}
		i g \,T_{ij}^{\,a} \,\gamma_\mu
	\end{eqnarray}
\end{minipage}

\vspace*{2mm}
\noindent
\begin{minipage}{6cm}
\begin{center}
          \SetWidth{1}
	\begin{picture}(120,90)(0,0)
		\Gluon(20,80)(60,45){-4}{5}
		\Gluon(20,10)(60,45){4}{5}
		\Gluon(60,45)(110,45){4}{5}
		\Text(15,80)[r]{a,$\mu$}
		\Text(15,10)[r]{b,$\nu$}
		\Text(115,45)[l]{c,$\rho$}
		\Text(45,68)[lb]{$p_1^{}$}
		\Text(45,25)[lt]{$p_2^{}$}
		\Text(85,53)[b]{$p_3^{}$}
		\Vertex(60,45){1.5}
	\end{picture}
\end{center}
\end{minipage}
\begin{minipage}{10cm}
	\begin{eqnarray} && \mbox{}
		- g f^{abc} (\delta_{\mu\nu}(p_1^{}-p_2^{})_\rho
				\nonumber \\[0.5mm] && \mbox{}
			+\delta_{\nu\rho}(p_2^{}-p_3^{})_\mu
			+\delta_{\rho\mu}(p_3^{}-p_1^{})_\nu)
	\end{eqnarray}
\end{minipage}

\vspace*{2mm}
\noindent
\begin{minipage}{6cm}
\begin{center}
          \SetWidth{1}
	\begin{picture}(140,110)(0,0)
		\Gluon(20,55)(70,55){4}{5}
		\Gluon(70,15)(70,55){4}{4}
		\Gluon(70,55)(70,95){4}{4}
		\Gluon(70,55)(120,55){4}{5}
		\Vertex(70,55){1.5}
		\Text(70,100)[b]{a,$\mu$}
		\Text(15,55)[r]{b,$\nu$}
		\Text(70,10)[t]{c,$\rho$}
		\Text(125,55)[l]{d,$\sigma$}
	\end{picture}
\end{center}
\end{minipage}
\begin{minipage}{10cm}
	\begin{eqnarray}
		- i g^2 \!\! & ( \!\! & \mbox{}
		+ f^{abe} f^{cde} ( \delta_{\mu\rho} \delta_{\nu\sigma}
				- \delta_{\mu\sigma} \delta_{\nu\rho}) 
\nonumber \\[0.5mm] && \mbox{}
		+ f^{ace} f^{dbe} ( \delta_{\mu\sigma} \delta_{\rho\nu}
				- \delta_{\mu\nu} \delta_{\rho\sigma}) 
\nonumber \\[0.5mm] && \mbox{}
		+ f^{ade} f^{bce} ( \delta_{\mu\nu} \delta_{\sigma\rho}
				- \delta_{\mu\rho} \delta_{\sigma\nu})\ \ )
	\end{eqnarray}
\end{minipage}

\noindent
\begin{minipage}{6cm}
\begin{center}
          \SetWidth{1}
	\begin{picture}(120,90)(0,0)
		\DashArrowLine(60,45)(20,80){4}
		\DashArrowLine(20,10)(60,45){4}
		\Gluon(60,45)(110,45){4}{5}
		\Vertex(60,45){1.5}
		\Text(15,80)[r]{b}
		\Text(45,68)[lb]{$p_2^{}$}
		\Text(45,25)[lt]{$p_1^{}$}
		\Text(15,10)[r]{a}
		\Text(115,45)[l]{c,$\mu$}
	\end{picture}
\end{center}
\end{minipage}
\begin{minipage}{10cm}
	\begin{eqnarray}
		- g f^{abc} {p_2^{}}_\mu
	\end{eqnarray}
\end{minipage}


\noindent
The additional vertices involving the gravition are given by

\noindent
\begin{minipage}{6cm}
\begin{center}
          \SetWidth{1}
	\begin{picture}(130,90)(0,0)
		\ArrowLine(60,45)(20,80)
		\ArrowLine(20,10)(60,45)
		\Photon(62,45)(110,45){4}{5}
		\Photon(62,45)(110,45){-4}{5}
		\Line(60,45)(62,45)
		\Line(110,45)(112,45)
		\Vertex(60,45){1.5}
		\Text(15,80)[r]{i}
		\Text(45,68)[lb]{$p_2^{}$}
		\Text(45,25)[lt]{$p_1^{}$}
		\Text(15,10)[r]{j}
		\Text(117,45)[l]{$\alpha,\beta$}
	\end{picture}
\end{center}
\end{minipage}
\begin{minipage}{10cm}
	\begin{eqnarray}
	- i \:\frac{\kappa}{8}\: \delta_{ij} \!\! & ( \!\! & 
		\gamma_\alpha\, (p_1^{}+p_2^{})_\beta
		+\gamma_\beta\, (p_1^{}+p_2^{})_\alpha \nonumber \\ && \mbox{}
	     - 2 \,\delta_{\alpha\beta}\, \gamma_\mu (p_1^{}+p_2^{})^\mu\ \ )
	\end{eqnarray}
\end{minipage}

\noindent
\begin{minipage}{6cm}
\begin{center}
          \SetWidth{1}
	\begin{picture}(140,110)(0,0)
		\ArrowLine(20,55)(70,55)
		\ArrowLine(70,55)(70,95)
		\Gluon(70,55)(70,15){-4}{4}
		\Photon(72,55)(120,55){4}{5}
		\Photon(72,55)(120,55){-4}{5}
		\Line(70,55)(72,55)
		\Line(120,55)(122,55)
		\Vertex(70,55){1.5}
		\Text(70,100)[b]{i}
		\Text(15,55)[r]{j}
		\Text(70,10)[t]{a,$\mu$}
		\Text(117,45)[l]{$\alpha,\beta$}
	\end{picture}
\end{center}
\end{minipage}
\begin{minipage}{10cm}
	\begin{eqnarray}
		i g \:\frac{\kappa}{4}\: T_{ij}^a \!\! & ( \!\! &
			\delta_{\alpha\mu}\gamma_\beta
			+\delta_{\beta\mu}\gamma_\alpha
			-2\delta_{\alpha\beta}\gamma_\mu\ \ )
	\end{eqnarray}
\end{minipage}

\noindent
\begin{minipage}{6cm}
\begin{center}
          \SetWidth{1}
	\begin{picture}(130,90)(0,0)
		\Gluon(20,80)(60,45){-4}{5}
		\Gluon(20,10)(60,45){4}{5}
		\Photon(62,45)(110,45){4}{5}
		\Photon(62,45)(110,45){-4}{5}
		\Line(60,45)(62,45)
		\Line(110,45)(112,45)
		\Vertex(60,45){1.5}
		\Text(15,80)[r]{a,$\mu$}
		\Text(45,68)[lb]{$p_1^{}$}
		\Text(45,25)[lt]{$p_2^{}$}
		\Text(15,10)[r]{b,$\nu$}
		\Text(117,45)[l]{$\alpha,\beta$}
	\end{picture}
\end{center}
\end{minipage}
\begin{minipage}{10cm}
	\begin{eqnarray} 
	- i \:\frac{\kappa}{2}\: \delta^{ab} \!\! & ( \!\! &
		p_1^{}\cdot p_2^{}\ C_{\alpha\beta,\mu\nu}
		+ D_{\alpha\beta,\mu\nu}(p_1^{},p_2^{})
		\nonumber \\ &&
		+ \,\frac{1}{1-\xi}\ E_{\alpha\beta,\mu\nu}(p_1^{},p_2^{})\ \ )
	\end{eqnarray}
\end{minipage}

\noindent
\begin{minipage}{6cm}
\begin{center}
          \SetWidth{1}
	\begin{picture}(140,110)(0,0)
		\Gluon(20,55)(70,55){4}{5}
		\Gluon(70,15)(70,55){4}{4}
		\Gluon(70,55)(70,95){4}{4}
		\Photon(72,55)(120,55){4}{5}
		\Photon(72,55)(120,55){-4}{5}
		\Line(70,55)(72,55)
		\Line(120,55)(122,55)
		\Vertex(70,55){1.5}
		\Text(70,100)[b]{a,$\mu$}
		\Text(15,55)[r]{b,$\nu$}
		\Text(70,10)[t]{c,$\rho$}
		\Text(77,75)[l]{$p_1$}
		\Text(45,62)[b]{$p_2$}
		\Text(77,35)[l]{$p_3$}
		\Text(127,55)[l]{$\alpha,\beta$}
	\end{picture}
\end{center}
\end{minipage}
\begin{minipage}{10cm}
	\begin{eqnarray}
		- g \:\frac{\kappa}{2}\: f^{abc} \!\! & ( \!\! & \mbox{}
				+C_{\alpha\beta,\mu\nu}(p_1^{}-p_2^{})_\rho
			\nonumber \\[-0.5mm] && \mbox{}
				+C_{\alpha\beta,\mu\rho}(p_3^{}-p_1^{})_\nu
			\nonumber \\[0.5mm] && \mbox{}
				+C_{\alpha\beta,\nu\rho}(p_2^{}-p_3^{})_\mu
			\nonumber \\[0.5mm] && \mbox{}
				+F_{\alpha\beta,\mu\nu\rho}(p_1^{},p_2^{},p_3^{})\ \ )
	\end{eqnarray}
\end{minipage}

\noindent
\begin{minipage}{6cm}
\begin{center}
          \SetWidth{1}
	\begin{picture}(130,110)(0,0)
		\Gluon(15,85)(60,55){-4}{5}
		\Gluon(15,25)(60,55){4}{5}
		\Gluon(60,55)(70,95){-4}{4}
		\Gluon(60,55)(70,15){4}{4}
		\Photon(62,55)(110,55){4}{5}
		\Photon(62,55)(110,55){-4}{5}
		\Line(60,55)(62,55)
		\Line(110,55)(112,55)
		\Vertex(60,55){1.5}
		\Text(72,100)[b]{a,$\mu$}
		\Text(10,87)[rb]{b,$\nu$}
		\Text(10,28)[rt]{c,$\rho$}
		\Text(72,10)[t]{d,$\sigma$}
		\Text(117,55)[l]{$\alpha,\beta$}
	\end{picture}
\end{center}
\end{minipage}
\begin{minipage}{10cm}
	\begin{eqnarray}
		- i g^2 \:\frac{\kappa}{2} \!\!\! & ( \!\!\! & \mbox{}
			+f^{abe}f^{cde}G_{\alpha\beta,\mu\rho\nu\sigma}
			+f^{ace}f^{bde}G_{\alpha\beta,\mu\nu\rho\sigma}
					\nonumber \\ && \mbox{}
			+f^{ade}f^{bce}G_{\alpha\beta,\mu\nu\sigma\rho}\ \ )
	\end{eqnarray}
\end{minipage}

\noindent
\begin{minipage}{6cm}
\begin{center}
          \SetWidth{1}
	\begin{picture}(130,90)(0,0)
		\DashArrowLine(60,45)(20,80){4}
		\DashArrowLine(20,10)(60,45){4}
		\Photon(62,45)(110,45){4}{5}
		\Photon(62,45)(110,45){-4}{5}
		\Line(60,55)(62,55)
		\Line(110,45)(112,45)
		\Vertex(60,45){1.5}
		\Text(15,80)[r]{b}
		\Text(45,68)[lb]{$p_2^{}$}
		\Text(45,25)[lt]{$p_1^{}$}
		\Text(15,10)[r]{a}
		\Text(117,45)[l]{$\alpha,\beta$}
	\end{picture}
\end{center}
\end{minipage}
\begin{minipage}{10cm}
	\begin{eqnarray}
		- i \:\frac{\kappa}{2}\: \delta^{ab}\, C_{\alpha\beta,\mu\nu}\,
            {p_1^{}}^\mu{p_2^{}}^\nu
	\end{eqnarray}
\end{minipage}

\noindent
\begin{minipage}{6cm}
\begin{center}
          \SetWidth{1}
	\begin{picture}(140,110)(0,0)
		\DashArrowLine(20,55)(70,55){4}
		\DashArrowLine(70,55)(70,95){4}
		\Gluon(70,55)(70,15){4}{4}
		\Photon(72,55)(120,55){4}{5}
		\Photon(72,55)(120,55){-4}{5}
		\Line(70,55)(72,55)
		\Line(120,55)(122,55)
		\Vertex(70,55){1.5}
		\Text(70,100)[b]{a}
		\Text(15,55)[r]{b}
		\Text(75,75)[l]{$p_1$}
		\Text(45,60)[b]{$p_2$}
		\Text(70,10)[t]{c,$\mu$}
		\Text(127,55)[l]{$\alpha,\beta$}
	\end{picture}
\end{center}
\end{minipage}
\begin{minipage}{10cm}
	\begin{eqnarray}
		- g \:\frac{\kappa}{2}\: f^{abc} C_{\alpha\beta,\mu\nu}\,
           {p_1^{}}^\nu
	\end{eqnarray}
\end{minipage}


\vspace*{5mm}
\noindent
The tensors $C,D,E,F$ and $G$ in Eqs.~(C.10) -- (C.14) are defined by
\begin{eqnarray}
	C_{\alpha\beta,\mu\nu} & = &
			\delta_{\alpha\mu}\delta_{\beta\nu}
			+\delta_{\alpha\nu}\delta_{\beta\mu}
			-\delta_{\alpha\beta}\delta_{\mu\nu} 
\:\: , \nn \\[1mm]
	D_{\alpha\beta,\mu\nu}(p_1^{},p_2^{}) & = & \mbox{}
			\delta_{\alpha\beta}{p_1^{}}_\nu{p_2^{}}_\mu
				-\delta_{\alpha\nu}{p_1^{}}_\beta{p_2^{}}_\mu
				-\delta_{\alpha\mu}{p_1^{}}_\nu{p_2^{}}_\beta
				+\delta_{\mu\nu}{p_1^{}}_\alpha{p_2^{}}_\beta
					\nonumber \\ && \mbox{}
				-\delta_{\beta\nu}{p_1^{}}_\alpha{p_2^{}}_\mu
				-\delta_{\beta\mu}{p_1^{}}_\nu{p_2^{}}_\alpha
				+\delta_{\mu\nu}{p_1^{}}_\beta{p_2^{}}_\alpha
\:\: , \nn \\[1mm]
	E_{\alpha\beta,\mu\nu}(p_1^{},p_2^{}) & = &
				\delta_{\alpha\beta}({p_1^{}}_\mu{p_1^{}}_\nu
					+{p_2^{}}_\mu{p_2^{}}_\nu+{p_1^{}}_\mu{p_2^{}}_\nu)
					\nonumber \\ && \mbox{}
				-\delta_{\beta\nu}{p_1^{}}_\alpha{p_1^{}}_\mu
				-\delta_{\beta\mu}{p_2^{}}_\alpha{p_2^{}}_\nu
				-\delta_{\alpha\nu}{p_1^{}}_\beta{p_1^{}}_\mu
				-\delta_{\alpha\mu}{p_2^{}}_\beta{p_2^{}}_\nu
\:\: ,  \nn \\[1mm]
	F_{\alpha\beta,\mu\nu\rho}(p_1^{},p_2^{},p_3^{}) & = & \mbox{}
				+\delta_{\alpha\mu}\,\delta_{\nu\rho}(p_2^{}-p_3^{})_\beta
				+\delta_{\alpha\nu}\,\delta_{\mu\rho}(p_3^{}-p_1^{})_\beta
					\nonumber \\ && \mbox{}
				+\delta_{\alpha\rho}\,\delta_{\mu\nu}(p_1^{}-p_2^{})_\beta
				+\delta_{\beta\mu}\,\delta_{\nu\rho}(p_2^{}-p_3^{})_\alpha
					\nonumber \\ && \mbox{}
				+\delta_{\beta\nu}\,\delta_{\mu\rho}(p_3^{}-p_1^{})_\alpha
				+\delta_{\beta\rho}\,\delta_{\mu\nu}(p_1^{}-p_2^{})_\alpha
\:\: , \nn \\[1mm]
	G_{\alpha\beta,\mu\nu\rho\sigma} & = &
			\delta_{\alpha\beta}\,(\delta_{\mu\nu}\,\delta_{\rho\sigma}
						-\delta_{\mu\sigma}\,\delta_{\nu\rho})
					\nonumber \\ && \mbox{}
			+\delta_{\alpha\mu}\,\delta_{\beta\sigma}\,\delta_{\nu\rho}
			+\delta_{\alpha\rho}\,\delta_{\beta\nu}\,\delta_{\mu\sigma}
			-\delta_{\alpha\mu}\,\delta_{\beta\nu}\,\delta_{\rho\sigma}
			-\delta_{\alpha\rho}\,\delta_{\beta\sigma}\,\delta_{\mu\nu}
					\nonumber \\ && \mbox{}
			+\delta_{\beta\mu}\,\delta_{\alpha\sigma}\,\delta_{\nu\rho}
			+\delta_{\beta\rho}\,\delta_{\alpha\nu}\,\delta_{\mu\sigma}
			-\delta_{\beta\mu}\,\delta_{\alpha\nu}\,\delta_{\rho\sigma}
			-\delta_{\beta\rho}\,\delta_{\alpha\sigma}\,\delta_{\mu\nu}
\:\: . \quad
\end{eqnarray}

In addition we need a ghost contribution in the graviton for the unpolarized 
calculations. We call this particle the g-ghost and we need the vertices
($\,\omega = \sqrt{\frac{2}{3(D-2)}}\,$):

\noindent
\begin{minipage}{6cm}
\begin{center}
          \SetWidth{1}
	\begin{picture}(130,90)(0,0)
		\ArrowLine(60,45)(20,80)
		\ArrowLine(20,10)(60,45)
		\DashLine(60,47)(110,46.5){4}
		\DashLine(60,43)(110,43.5){4}
		\Vertex(60,45){2}
		\Text(15,80)[r]{i}
		\Text(45,68)[lb]{$p_2^{}$}
		\Text(45,25)[lt]{$p_1^{}$}
		\Text(15,10)[r]{j}
	\end{picture}
\end{center}
\end{minipage}
\begin{minipage}{10cm}
	\begin{eqnarray}
		i\, \omega \kappa \,\delta_{ij}\: \frac{3}{4}\: \gamma_\mu 
          (p_1^{}+p_2^{})^\mu
	\end{eqnarray}
\end{minipage}

\noindent
\begin{minipage}{6cm}
\begin{center}
          \SetWidth{1}
	\begin{picture}(140,110)(0,0)
		\ArrowLine(20,55)(70,55)
		\ArrowLine(70,55)(70,95)
		\Gluon(70,55)(70,15){-4}{4}
		\DashLine(70,57)(120,56.5){4}
		\DashLine(70,53)(120,53.5){4}
		\Vertex(70,55){2}
		\Text(70,100)[b]{i}
		\Text(15,55)[r]{j}
		\Text(70,10)[t]{a,$\mu$}
	\end{picture}
\end{center}
\end{minipage}
\begin{minipage}{10cm}
	\begin{eqnarray}
		- i \:\frac{3}{2}\: \omega g \kappa\, T_{ij}^{\,a}\, \gamma_\mu
	\end{eqnarray}
\end{minipage}

\noindent
\begin{minipage}{6cm}
\begin{center}
          \SetWidth{1}
	\begin{picture}(130,90)(0,0)
		\Gluon(20,80)(60,45){-4}{5}
		\Gluon(20,10)(60,45){4}{5}
		\DashLine(60,47)(110,46.5){4}
		\DashLine(60,43)(110,43.5){4}
		\Vertex(60,45){2}
		\Text(15,80)[r]{a,$\mu$}
		\Text(45,68)[lb]{$p_1^{}$}
		\Text(45,25)[lt]{$p_2^{}$}
		\Text(15,10)[r]{b,$\nu$}
		\Text(115,45)[l]{Q}
	\end{picture}
\end{center}
\end{minipage}
\begin{minipage}{10cm}
	\begin{eqnarray} &&
		i \,\omega \kappa \,\delta^{ab}\: \frac{1}{1-\xi}\ 
          ({p_1^{}}_\mu Q_\nu +{p_2^{}}_\nu Q_\mu)
	\end{eqnarray}
\end{minipage}

\noindent
\begin{minipage}{6cm}
\begin{center}
          \SetWidth{1}
	\begin{picture}(140,110)(0,0)
		\Gluon(20,55)(70,55){4}{5}
		\Gluon(70,15)(70,55){4}{4}
		\Gluon(70,55)(70,95){4}{4}
		\DashLine(70,57)(120,56.5){4}
		\DashLine(70,53)(120,53.5){4}
		\Vertex(70,55){2}
		\Text(70,100)[b]{a,$\mu$}
		\Text(15,55)[r]{b,$\nu$}
		\Text(70,10)[t]{c,$\rho$}
		\Text(77,75)[l]{$p_1^{}$}
		\Text(45,62)[b]{$p_2^{}$}
		\Text(77,35)[l]{$p_3^{}$}
	\end{picture}
\end{center}
\end{minipage}
\begin{minipage}{10cm}
	\begin{eqnarray}
		0
	\end{eqnarray}
\end{minipage}

\noindent
\begin{minipage}{6cm}
\begin{center}
          \SetWidth{1}
	\begin{picture}(130,110)(0,0)
		\Gluon(15,85)(60,55){-4}{5}
		\Gluon(15,30)(60,55){4}{5}
		\Gluon(60,55)(70,95){-4}{4}
		\Gluon(60,55)(70,15){4}{4}
		\DashLine(60,57)(110,56.5){4}
		\DashLine(60,53)(110,53.5){4}
		\Vertex(60,55){2}
		\Text(72,100)[b]{a,$\mu$}
		\Text(10,87)[rb]{b,$\nu$}
		\Text(10,28)[rt]{c,$\rho$}
		\Text(72,10)[t]{d,$\sigma$}
	\end{picture}
\end{center}
\end{minipage}
\begin{minipage}{10cm}
	\begin{eqnarray}
			0
	\end{eqnarray}
\end{minipage}

\vspace{2mm}
\noindent
Vertices involving both the standard ghost and the g-ghost were not 
required in our calculation.

We now turn to the projection operators for which we sometimes have more than 
one choice. 
The physical operator for the unpolarized gluon is given by$\,$%
\footnote{Here we use $Q$ for the momentum of the probe. Often $q$ is used 
after which $Q^2 = -q\cdot q$. In the following part $Q\cdot Q$ is just the 
square of the 4-vector $Q$, which keeps the notation in line with the computer 
programs.}
\begin{eqnarray}
	\Pi_{\kappa\lambda}(Q,P) & = &
		\delta_{\kappa\lambda}
			-Q_\kappa P_\lambda^{} / Q\cdot P
			-Q_\lambda^{} P_\kappa / Q\cdot P
			+P_\kappa P_\lambda^{} Q\cdot Q / Q\cdot P^2
\end{eqnarray}
in which $P\cdot P = 0$.
One can replace this by $\delta_{\kappa\lambda}$ and a ghost contribution 
in the regular way. This gives more diagrams, but they are easier to 
compute.
For the polarized gluon we use
\begin{eqnarray}
	\Pi_{\kappa\lambda}(Q,P) & = & \epsilon_{P Q \kappa \lambda} / Q\cdot P
\:\: .
\end{eqnarray}
For the unpolarized and polarized quark the projection operators are
\begin{eqnarray}
	\Pi(P) & = & \gamma^{\,\mu} P_\mu
\end{eqnarray}
and
\begin{eqnarray}
	\Pi(P) \equal  \gamma_5^{} \gamma^{\,\mu} P_\mu 
	       \equal \frac{1}{6}\: \epsilon_{\kappa \lambda \nu P}\,
				\gamma^{\,\kappa} \gamma^{\,\lambda} \gamma^{\,\nu}
\:\: .
\end{eqnarray}
The last form of the operator is necessary to deal with the issue of 
$\gamma_5$ in $D$ dimensions. At a later stage we then contract the 
Levi-Civita tensors in terms of the $D$-dimensional metric.

For the graviton the situation is more complicated as there are several
possible currents.  We follow Ref.~\cite{LamLi}, assuming a target mass of 
zero, and add the D-dimensional effects as given in Ref.~\cite{SVgDIS}.
Then for unpolarized scattering we have
\begin{eqnarray}
	W_{\alpha_1\beta_1,\alpha_2\beta_2} & = &
		F_1\, A^{(1)}_{\alpha_1\beta_1\alpha_2\beta_2}
		+F_2\, A^{(2)}_{\alpha_1\beta_1\alpha_2\beta_2}
		+F_3\, A^{(3)}_{\alpha_1\beta_1\alpha_2\beta_2}
\:\: ,
\end{eqnarray}
and for polarized scattering
\begin{eqnarray}
	W_{\alpha_1\beta_1,\alpha_2\beta_2} & = &
		F_4\, A^{(4)}_{\alpha_1\beta_1\alpha_2\beta_2}
		+F_6\, A^{(6)}_{\alpha_1\beta_1\alpha_2\beta_2}
\end{eqnarray}
with
\begin{eqnarray}
	A^{(1)}_{\alpha_1\beta_1\alpha_2\beta_2} & = &
		\overline{\pi}_{\alpha_1\beta_1}\overline{\pi}_{\alpha_2\beta_2}
\:\: , \nn \\[1mm]
	A^{(2)}_{\alpha_1\beta_1\alpha_2\beta_2} & = &
		{\P}_{\alpha_1}{\P}_{\alpha_2}{\GG}_{\beta_1\beta_2}
		+\P_{\alpha_1}\P_{\beta_2}\GG_{\beta_1\alpha_2}
		+\P_{\beta_1}\P_{\alpha_2}\GG_{\alpha_1\beta_2}
		+\P_{\beta_1}\P_{\beta_2}\GG_{\alpha_1\alpha_2} 
\nonumber \\ && \hspn \mbox{}
		-\frac{4}{D-1}\:(\P_{\alpha_1}\P_{\beta_1}\GG_{\beta_2\alpha_2}
		+\P_{\alpha_2}\P_{\beta_2}\GG_{\beta_1\alpha_1})
		+\frac{4}{(D-1)^2}\:\GG_{\alpha_1\beta_1}\GG_{\alpha_2\beta_2}\P\cdot\P
\qquad \nn \\
	A^{(3)}_{\alpha_1\beta_1\alpha_2\beta_2} & = &
		\GG_{\alpha_1\alpha_2}\GG_{\beta_1\beta_2}
		+\GG_{\alpha_1\beta_2}\GG_{\alpha_2\beta_1}
		-\frac{2}{D-1}\:\GG_{\alpha_1\beta_1}\GG_{\alpha_2\beta_2}
\:\: , \nn \\[1mm]
	A^{(4)}_{\alpha_1\beta_1\alpha_2\beta_2} & = &
		\epsilon_{\alpha_1\alpha_2 Q P}\P_{\beta_1}\P_{\beta_2}
		+\epsilon_{\alpha_1\beta_2 Q P}\P_{\beta_1}\P_{\alpha_2}
		+\epsilon_{\beta_1\alpha_2 Q P}\P_{\alpha_1}\P_{\beta_2}
		+\epsilon_{\beta_1\beta_2 Q P}\P_{\alpha_1}\P_{\alpha_2}
\:\: , \nn \\[1mm]
	A^{(6)}_{\alpha_1\beta_1\alpha_2\beta_2} & = &
		\epsilon_{\alpha_1\alpha_2 Q P}\GG_{\beta_1 \beta_2}
		+\epsilon_{\alpha_1\beta_2 Q P}\GG_{\beta_1 \alpha_2}
		+\epsilon_{\beta_1\alpha_2 Q P}\GG_{\alpha_1 \beta_2}
		+\epsilon_{\beta_1\beta_2 Q P}\GG_{\alpha_1 \alpha_2}
\:\: .
\end{eqnarray}
Here we have used
\begin{eqnarray}
	\overline{\pi}_{\alpha \beta} & = &
		\P_\alpha \P_\beta 
	 	- \:\frac{1}{D-1}\: \GG_{\alpha \beta} \P\cdot\P 
\:\: , \nn \\
	\P_\alpha & = & P_\alpha - Q_\alpha \:\frac{Q\cdot P}{Q\cdot Q} 
\:\: , \nn \\
	\GG_{\alpha \beta} & = & \delta_{\alpha \beta}
			\,-\:\frac{Q_{\alpha} Q_{\beta}}{Q\cdot Q}
\:\: .
\end{eqnarray}

When we construct the projection operators we demand $\Pi_i\, A_j = 
\delta_{ij}$, and after also using the symmetry in the graviton indices we 
have for the unpolarized operators
\begin{eqnarray}
	\Pi_1 & = & 256\: \frac{(D+1)(D+3)}{D(D-2)}\:
		P_{\alpha_1} P_{\beta_1} P_{\alpha_2} P_{\beta_2}
				\:\frac{1}{Q\cdot Q^3} \nonumber \\ &&
		+ 1024\: \frac{D+1}{D(D-2)}\:
			 P_{\alpha_1} P_{\alpha_2}\delta_{\beta_1\beta_2}
				\:\frac{Q\cdot P^2}{Q\cdot Q^4}\, 
		+ 512\,
			\delta_{\alpha_1\alpha_2}\delta_{\beta_1\beta_2}
				\:\frac{Q\cdot P^4}{Q\cdot Q^5} 
\:\: , \nn \\[1mm]
	\Pi_2 & = & 64\: \frac{(D+1)}{D(D-2)}
		P_{\alpha_1} P_{\beta_1} P_{\alpha_2} P_{\beta_2}
				\:\frac{1}{Q\cdot Q^3} 
		+ 32\: \frac{1}{D(D-3)}\:
			\delta_{\alpha_1\alpha_2}\delta_{\beta_1\beta_2}
				\frac{Q\cdot P^4}{Q\cdot Q^5} \nonumber \\ &&
		+ 64\: \frac{D^2-D-4}{D(D-2)(D-3)}\:
			 P_{\alpha_1} P_{\alpha_2}\delta_{\beta_1\beta_2}
				\:\frac{Q\cdot P^2}{Q\cdot Q^4} 
\:\: , \nn \\[2mm] 
	\Pi_3 & = & 16\: \frac{1}{D(D-2)}\:
		P_{\alpha_1} P_{\beta_1}^{} P_{\alpha_2} P_{\beta_2}^{}
				\:\frac{1}{Q\cdot Q^3} 
		+ 32\: \frac{1}{D(D-3)}\:
			 P_{\alpha_1} P_{\alpha_2}\delta_{\beta_1\beta_2}
				\:\frac{Q\cdot P^2}{Q\cdot Q^4} \nonumber \\ &&
		+ 16\: \frac{1}{D(D-3)}\:
			\delta_{\alpha_1\alpha_2}\delta_{\beta_1\beta_2}
				\:\frac{Q\cdot P^4}{Q\cdot Q^5}
\:\: .
\end{eqnarray}

For the polarized projection operators the situation is slightly more 
complicated. In principle we could work with $\Pi_4$ and $\Pi_6$ but we 
notice that, if both projections are needed, it is easier to work with 
the linear combinations $\Pi_D$ and $\Pi_F$. These are defined by
\begin{eqnarray}
	\Pi_4 & = & \frac{D+1}{D(D-2)(D-3)}\:\Pi_D 
            \,+\: \frac{1}{D(D-2)(D-3)}\:\Pi_F 
\:\: , \nn \\[1mm]
	\Pi_6 & = & \frac{1}{D(D-2)(D-3)}\:(\Pi_D + \Pi_F)
\:\: .
\end{eqnarray}
In any case we have a Levi-Civita tensor in the operator, and we contract this 
with the Levi-Civita tensor of the quark or the gluon. 
For the quark we obtain
\begin{eqnarray}
   \Pi^{\,q}_D \:\:=\:\: 
		4 \:\frac{P_{\alpha_1}P_{\alpha_2}}{Q\cdot Q^3}\:
		R^{\,q}_{\beta_1\beta_2} 
\;\; , \quad
   \Pi^{\,q}_F \:\:=\:\: 
		4 \:\frac{\delta_{\alpha_1\alpha_2} Q\cdot P^2}{Q\cdot Q^4}
		\: R^{\,q}_{\beta_1\beta_2} 
\end{eqnarray}
with
\begin{eqnarray}
	R^{\,q}_{\beta_1\beta_2} & = &
		\gamma^{\,\mu}\gamma^{\,\nu} P_\mu Q_\nu
			(\gamma_{\beta_1}^{} P_{\beta_2}^{}
               -\gamma_{\beta_2}^{} P_{\beta_1}^{})
		+\gamma^{\,\mu} P_\mu (\gamma_{\beta_1}^{}\gamma_{\beta_2}^{}
				-\delta_{\beta_1\beta_2}) Q\cdot P 
\nonumber \\[1mm] &&
		+\gamma^{\,\mu} P_\mu (P_{\beta_1}^{} Q_{\beta_2}^{}
                    -Q_{\beta_1}^{} P_{\beta_2})^{}
\:\: ,
\end{eqnarray}
and for the gluon we find
\begin{eqnarray}
	\Pi^{\,g}_D \:\:=\:\: 4 \:\frac{P_{\alpha_1}P_{\alpha_2}}{Q\cdot Q^3}\:
		R^{g}_{\beta_1\beta_2} 
\;\; , \quad
	\Pi^{\,g}_F \:\:=\:\: 4 \:\frac{\delta_{\alpha_1\alpha_2} Q\cdot P^2}
		{Q\cdot Q^4}\: R^{\,g}_{\beta_1\beta_2} 
\end{eqnarray}
with
\begin{eqnarray}
	R^{\,g}_{\beta_1\beta_2} & = &
			(\delta_{\kappa \beta_2}\delta_{\lambda \beta_1}
			-\delta_{\kappa \beta_1}\delta_{\lambda \beta_2})
					\:\frac{Q\cdot P^2}{Q\cdot Q}\,
			+(\delta_{\kappa \beta_2}P_{\beta_1}\P_\lambda
			 -\delta_{\kappa \beta_1}P_{\beta_2}\P_\lambda)
\nonumber \\ &&
			-(\delta_{\lambda \beta_2}P_{\beta_1}\P_\kappa
			 -\delta_{\lambda \beta_1}P_{\beta_2}\P_\kappa) 
\:\: , \nn \\[1mm]
	\P_\kappa & = & P_\kappa - Q_\kappa \:\frac{Q\cdot P}{Q\cdot Q}
\:\:,
\end{eqnarray}
where we have again used the symmetry in the graviton indices to simplify the 
expressions.

In the polarized case we do not need a ghost contribution, neither for the 
graviton nor for the gluon.
Propagators for the graviton and the corresponding ghost are not required 
since we do not consider internal gravitons.

{\small
\setlength{\baselineskip}{0.36cm}

}

\end{document}